\documentclass[10pt]{article}

\usepackage[twoside,nohead]{geometry}

\usepackage{amsmath, amsfonts, amsthm, amssymb,verbatim}
\usepackage{graphicx}
\usepackage{xspace}
\usepackage[mathscr]{euscript}
\usepackage{enumitem}
\usepackage{bm}
\usepackage{dsfont}
\usepackage{xspace,a4wide}
 \usepackage[perpage]{footmisc}

\newtheorem{theorem}{Theorem}[section]
\newtheorem*{theorem*}{Theorem}
\newtheorem{definition}[theorem]{Definition}
\newtheorem{lemma}[theorem]{Lemma}

\newtheorem{proposition}[theorem]{Proposition}

\numberwithin{equation}{section}

\usepackage[bookmarks=true,
           bookmarksopen=true,hidelinks]{hyperref}

\newcommand \bz {\begin{itemize}}
\newcommand \ez {\end{itemize}}
\newcommand \ben {\begin{enumerate}}
\newcommand\een {\end{enumerate}} 
\newcommand \R {\mathbb R}
\newcommand{\Eqref}[1]{Eq.~\eqref{#1}}
\newcommand{\Eqsref}[1]{Eqs.~\eqref{#1}}
\newcommand{\Sectionref}[1]{Section~\ref{#1}}  
\newcommand{\Sectionsref}[1]{Sections~\ref{#1}}  
\newcommand{\Defref}[1]{Definition~\ref{#1}}
\newcommand{\Propref}[1]{Proposition~\ref{#1}}
\newcommand{\Lemref}[1]{Lemma~\ref{#1}}
\newcommand{\Theoremref}[1]{Theorem~\ref{#1}}

\newcommand{\Conditionref}[1]{Condition~\ref{#1}}

\newcommand{\Figref}[1]{Figure~\ref{#1}}
\newcommand{\keyword}[1]{\textit{#1}}
\newcommand{\diag}{\text{diag}\xspace}

\newcommand \muh M 
\newcommand \vv V

\newcommand \del 		\partial
\newcommand \eps 		\epsilon
\newcommand \lam 		\lambda 
\newcommand \be 		{\begin{equation}}
\newcommand\ee 		{\end{equation}}

\newcommand{\symbolUL}[3]{\ensuremath{{#1}^{#2}_{#3}}\xspace}
\newcommand{\gravity}{\mathrm{\scriptscriptstyle{[G]}}}
\newcommand{\fluid}{\mathrm{\scriptscriptstyle{[F]}}}
\newcommand{\gravityvac}{\mathrm{\scriptscriptstyle{[GV]}}}
\newcommand{\gravityfluid}{\mathrm{\scriptscriptstyle{[GF]}}}
\newcommand{\truncated}{\mathrm{\scriptscriptstyle{\{T\}}}}
\newcommand{\vacuum}{\mathrm{\scriptscriptstyle{\{V\}}}}

\newcommand{\Vectorletter}{U}
\newcommand{\VectorSletter}{{U_*}}
\newcommand{\VectorRemletter}{W}

\newcommand{\UG}{\symbolUL{\Vectorletter}{}{\gravity}}
\newcommand{\UGcomp}[1]{\symbolUL{\Vectorletter}{#1}{\gravity}}
\newcommand{\UGT}{\symbolUL{\Vectorletter}{}{\truncated\gravity}}
\newcommand{\USG}{\symbolUL{\VectorSletter}{}{\gravity}}
\newcommand{\USGcomp}[1]{\symbolUL{\VectorSletter}{#1}{\gravity}}
\newcommand{\WG}{\symbolUL{\VectorRemletter}{}{\gravity}}

\newcommand{\WGT}{\symbolUL{\VectorRemletter}{}{\truncated\gravity}}

\newcommand{\UF}{\symbolUL{\Vectorletter}{}{\fluid}}
\newcommand{\UFcomp}[1]{\symbolUL{\Vectorletter}{#1}{\fluid}}
\newcommand{\UFT}{\symbolUL{\Vectorletter}{}{\truncated\fluid}}
\newcommand{\USF}{\symbolUL{\VectorSletter}{}{\fluid}}

\newcommand{\WF}{\symbolUL{\VectorRemletter}{}{\fluid}}

\newcommand{\WFT}{\symbolUL{\VectorRemletter}{}{\truncated\fluid}}

\newcommand{\UC}{\symbolUL{\Vectorletter}{}{}}

\newcommand{\UCT}{\symbolUL{\Vectorletter}{}{\truncated}}
\newcommand{\USC}{\symbolUL{\VectorSletter}{}{}}

\newcommand{\WC}{\symbolUL{\VectorRemletter}{}{}}

\newcommand{\WCT}{\symbolUL{\VectorRemletter}{}{\truncated}}
\newcommand{\WCV}{\symbolUL{\VectorRemletter}{}{\vacuum}}

\newcommand{\LOExpoletter}{\kappa}
\newcommand{\HOExpoletter}{\mu}
\newcommand{\HHOExpoletter}{\nu}
\newcommand{\LOExpoVecletter}{\hat\LOExpoletter}
\newcommand{\HOExpoVecletter}{\hat\HOExpoletter}
\newcommand{\HHOExpoVecletter}{\hat\HHOExpoletter}
\newcommand{\kappaVecG}{\symbolUL{\LOExpoVecletter}{}{\gravity}}

\newcommand{\kappaVecF}{\symbolUL{\LOExpoVecletter}{}{\fluid}}

\newcommand{\kappaVecC}{\symbolUL{\LOExpoVecletter}{}{}}
\newcommand{\muVecG}{\symbolUL{\HOExpoVecletter}{}{\gravity}}

\newcommand{\muVecF}{\symbolUL{\HOExpoVecletter}{}{\fluid}}

\newcommand{\muVecC}{\symbolUL{\HOExpoVecletter}{}{}}
\newcommand{\muGcomp}[1]{\symbolUL{\HOExpoletter}{#1}{\gravity}}
\newcommand{\muFcomp}[1]{\symbolUL{\HOExpoletter}{#1}{\fluid}}
\newcommand{\nuVecG}{\symbolUL{\HHOExpoVecletter}{}{\gravity}}
\newcommand{\nuVecF}{\symbolUL{\HHOExpoVecletter}{}{\fluid}}
\newcommand{\nuVecC}{\symbolUL{\HHOExpoVecletter}{}{}}

\newcommand{\operatordouble}[3]{\ensuremath{{#1}(#2)[#3]}\xspace}
\newcommand{\operatorsingle}[2]{\ensuremath{{#1}[#2]}\xspace}
\newcommand{\operatornull}[1]{\ensuremath{#1}\xspace}

\newcommand{\PPMatrixletter}{S}

\newcommand{\PPMatrixGnullcomp}[1]{\operatornull{\symbolUL{\PPMatrixletter}{#1}{\gravity}}}
\newcommand{\PPMatrixGcomp}[2]{\operatorsingle{\symbolUL{\PPMatrixletter}{#1}{\gravity}}{#2}}

\newcommand{\PPMatrixFnullcomp}[1]{\operatornull{\symbolUL{\PPMatrixletter}{#1}{\fluid}}}
\newcommand{\PPMatrixFcomp}[2]{\operatorsingle{\symbolUL{\PPMatrixletter}{#1}{\fluid}}{#2}}
\newcommand{\SZeroZeroF}{\symbolUL{\PPMatrixletter}{0}{0\fluid}}

\newcommand{\PPletter}{L}
\newcommand{\LG}[2]{\operatordouble{\symbolUL{\PPletter}{}{\gravity}}{#1}{#2}}
\newcommand{\LF}[2]{\operatordouble{\symbolUL{\PPletter}{}{\fluid}}{#1}{#2}}

\newcommand{\Sourceletter}{F}
\newcommand{\FGV}[1]{\operatorsingle{\symbolUL{\Sourceletter}{}{\gravityvac}}{#1}}
\newcommand{\FGF}[1]{\operatorsingle{\symbolUL{\Sourceletter}{}{\gravityfluid}}{#1}}
\newcommand{\FF}[1]{\operatorsingle{\symbolUL{\Sourceletter}{}{\fluid}}{#1}}
\newcommand{\FGVOne}[1]{\operatorsingle{\symbolUL{\Sourceletter}{(1)}{\gravityvac}}{#1}}
\newcommand{\FGVTwo}[1]{\operatorsingle{\symbolUL{\Sourceletter}{(2)}{\gravityvac}}{#1}}
\newcommand{\FGVTwoOne}[1]{\operatorsingle{\symbolUL{\Sourceletter}{(2,1)}{\gravityvac}}{#1}}
\newcommand{\FGVTwoTwo}[1]{\operatorsingle{\symbolUL{\Sourceletter}{(2,2)}{\gravityvac}}{#1}}

\newcommand{\Sourceredletter}{\mathcal{F}}
\newcommand{\FredGV}[1]{\operatorsingle{\symbolUL{\Sourceredletter}{}{\gravityvac}}{#1}}
\newcommand{\FredGVOne}[1]{\operatorsingle{\symbolUL{\Sourceredletter}{(1)}{\gravityvac}}{#1}}
\newcommand{\FredGVOneOne}[1]{\operatorsingle{\symbolUL{\Sourceredletter}{(1,1)}{\gravityvac}}{#1}}
\newcommand{\FredGVOneTwo}[1]{\operatorsingle{\symbolUL{\Sourceredletter}{(1,2)}{\gravityvac}}{#1}}

\newcommand{\FredGVTwoOne}[1]{\operatorsingle{\symbolUL{\Sourceredletter}{(2,1)}{\gravityvac}}{#1}}

\newcommand{\FredF}[1]{\operatorsingle{\symbolUL{\Sourceredletter}{}{\fluid}}{#1}}
\newcommand{\FredTF}[1]{\operatorsingle{\symbolUL{\Sourceredletter}{}{\truncated\fluid}}{#1}}

\newcommand{\Oletter}{\mathcal{O}}
\newcommand{\OGVOneOne}[1]{\operatorsingle{\symbolUL{\Oletter}{(1,1)}{\gravityvac}}{#1}}
\newcommand{\OGVOneTwo}[1]{\operatorsingle{\symbolUL{\Oletter}{(1,2)}{\gravityvac}}{#1}}
\newcommand{\OGVTwoOne}[1]{\operatorsingle{\symbolUL{\Oletter}{(2,1)}{\gravityvac}}{#1}}
\newcommand{\OGVTwoTwo}[1]{\operatorsingle{\symbolUL{\Oletter}{(2,2)}{\gravityvac}}{#1}}
\newcommand{\OGF}[1]{\operatorsingle{\symbolUL{\Oletter}{}{\gravityfluid}}{#1}}
\newcommand{\OF}[1]{\operatorsingle{\symbolUL{\Oletter}{}{\fluid}}{#1}}
\newcommand{\OOF}[1]{\operatorsingle{\symbolUL{\widehat{\Oletter}}{}{\fluid}}{#1}}
\newcommand{\OGPZero}[1]{\operatorsingle{\symbolUL{\Oletter}{(P,0)}{\gravity}}{#1}}
\newcommand{\OGPOne}[1]{\operatorsingle{\symbolUL{\Oletter}{(P,1)}{\gravity}}{#1}}
\newcommand{\OOGPOne}[1]{\operatorsingle{\symbolUL{\widehat{\Oletter}}{(P,1)}{\gravity}}{#1}}
\newcommand{\OFPZero}[1]{\operatorsingle{\symbolUL{\Oletter}{(P,0)}{\fluid}}{#1}}
\newcommand{\OFPOne}[1]{\operatorsingle{\symbolUL{\Oletter}{(P,1)}{\fluid}}{#1}}
\newcommand{\OTG}[1]{\operatorsingle{\symbolUL{\Oletter}{}{\truncated\gravity}}{#1}}
\newcommand{\OTF}[1]{\operatorsingle{\symbolUL{\Oletter}{}{\truncated\fluid}}{#1}}

\newcommand{\NMatrixletter}{N}
\newcommand{\NNG}{\symbolUL{\NMatrixletter}{}{\gravity}}
\newcommand{\NNF}{\symbolUL{\NMatrixletter}{}{\fluid}}
\newcommand{\NNFT}{\symbolUL{\NMatrixletter}{}{\truncated\fluid}}

\newcommand{\gsF}{\ensuremath{\mathcal F}\xspace}
\newcommand{\unitmatrix}{\mathds{1}}

\newcommand{\RR}[1]{\ensuremath{\mathcal{R}[#1]}\xspace}

\newcommand{\Sts}[1]{\ensuremath{S^j(#1)}\xspace}
\newcommand{\Sna}{\ensuremath{S^j}\xspace}
\newcommand{\St}[1]{\ensuremath{S^0(#1)}\xspace}
\newcommand{\Stna}{\ensuremath{S^0}\xspace}

\newcommand{\StLu}{\ensuremath{S^0_0(u_*)}\xspace}
\newcommand{\StLInv}[1]{\ensuremath{\left(S_0^0(#1)\right)^{-1}}\xspace}
\newcommand{\StH}[1]{\ensuremath{S^0_1(#1)}\xspace}
\newcommand{\StHu}[1]{\StH{u_*+#1}}
\newcommand{\Ssna}{\ensuremath{S^a}\xspace}
\newcommand{\Ss}[1]{\ensuremath{S^a(#1)}\xspace}

\newcommand{\N}[1]{\ensuremath{N(#1)}\xspace}
\newcommand{\Nna}{\ensuremath{N}\xspace}

\newcommand{\NLu}{\ensuremath{N_{0}(u_*)}\xspace}
\newcommand{\NODE}{\ensuremath{\mathcal N}\xspace}

\newcommand{\fna}{\ensuremath{f}\xspace}

\newcommand{\Fredu}[1]{\ensuremath{\mathscr F}(u_*)[#1]\xspace}

\let\oldmarginpar\marginpar
\renewcommand\marginpar[1]{\-\oldmarginpar[\raggedleft\footnotesize #1]%
{\raggedright\footnotesize #1}}

\title{Self--gravitating fluid flows with Gowdy symmetry
\\
near cosmological singularities}
\author{Florian Beyer$^1$ and Philippe G. LeFloch$^2$}
\footnotetext[1]{Department of Mathematics and Statistics, University of
  Otago, P.O.~Box 56, 
Dunedin 9054, New Zealand. Email: {\tt fbeyer@maths.otago.ac.nz}
\newline 
\indent \, $^2$Laboratoire Jacques-Louis Lions and Centre National de la Recherche Scientifique, 
Sorbonne Universit\'e, 4 Place Jussieu, 75252 Paris, France.
Email: {\tt contact@philippelefloch.org}}

\date{June 2017} 

\graphicspath{{.}{Figures/}}

\begin{document}

\maketitle

\begin{abstract} 
We consider self-gravitating fluids in cosmological spacetimes with Gowdy symmetry on the torus $T^3$ and, in this class, we solve the singular initial value problem for the Einstein-Euler system of general relativity, when an initial data set is prescribed on the hypersurface of singularity.
We specify initial conditions for the geometric and matter variables and  identify the asymptotic behavior of these variables near the cosmological singularity. Our analysis of this class of nonlinear and singular partial differential equations exhibits a condition on the sound speed, which leads us to the notion of sub-critical, critical, and super-critical regimes. Solutions to the Einstein-Euler systems when the fluid is governed by a linear equation of state are constructed in the first two regimes, while additional difficulties arise in the latter one. All previous studies on inhomogeneous spacetimes concerned vacuum cosmological spacetimes only. 
\end{abstract}

%
%
%
%


\section{Introduction} 
\label{sec:intro}

\paragraph{Objective.}

We present a mathematical analysis of a
class of {solutions to the Einstein-Euler equations} describing
inhomogeneous matter spacetimes, when the matter content is a {\sl
  perfect compressible fluid.}  We attempt to elucidate the
coupling between the spacetime geometry, which is determined by the
Einstein equations, and the matter content, whose evolution is governed by the Euler equations, in a situation when the
gravitational field diverges near a ``cosmological singularity'' or
``Big Bang''.

Fully nonlinear self-gravitating fluid models are the basis of modern
cosmology
\cite{Mukhanov:2005ts}. Our results are thus relevant for the early history of the
Universe just after it was born in the Big Bang. While the standard
model of cosmology is highly consistent with observations, the
underlying assumption of isotropy and spatial homogeneity (and
linearized perturbations thereof) has raised concerns in the
scientific community in recent years (see \cite{Buchert:2015tx} and
references therein). Our results strongly suggest that the early history of
more realistic cosmological models is inconsistent with this assumption due
to highly anisotropic and inhomogeneous effects associated with the so-called
\keyword{velocity term dominance} discussed in more detail below. It
is interesting to observe that the situation may be fundamentally different in the
presence of certain ``extreme'' matter fields. In the case of a
massless scalar field, for instance, it was recently proven
\cite{Rodnianski:2014tk} that the dynamics at the singularity is
indeed consistent with the standard model. This paramount difference
of ``extreme'' matter models (as for example scalar fields or stiff
fluid models) and ``ordinary'' matter models (as for example fluids
with an ``ordinary'' equation of state, which are the subject of our
investigation) demonstrates the significance of the so-called
\keyword{matter does not matter} paradigm which we will put
particular emphasis on in this paper.

We restrict our attention to Gowdy symmetry, that is, we
assume that the spacetimes admit two commuting, spacelike Killing
fields with vanishing twist and that the spatial topology is the
$3$-torus $T^3$. A particular motivation for this is evidence from
earlier research \cite{Garfinkle:2004bq,Lim:2009fz} that the
singular dynamics of models in less symmetric cases can be much more difficult without the taming properties of ``extreme'' matter fields and
hence would be far beyond the applicability of current mathematical
techniques available for the Einstein-matter equations.

\newpage 

\paragraph{Main result.} 

We establish here an existence theory
for the Einstein-Euler system, which can be formulated as a nonlinear
system of quasilinear hyperbolic equations. This system is analyzed 
in the neighborhood of the cosmological singularity in the 
Gowdy symmetry class. We work with a time variable $t>0$, normalized
such that the spacetime becomes singular in the limit $t\searrow 0$.
By prescribing a suitable data set for the geometry and matter
variables \textit{on} the singularity at $t=0$ in the sense of a
\keyword{singular initial value problem} (see below), we are able to
prove the existence of a broad class of spacetimes, having
a well-specified asymptotic behavior as the singularity is approached. 
A preliminary version of our main result can be stated as follows. 

\begin{theorem*}[Fluid flows near the cosmological singularity of a Gowdy-symmetric
  spacetime]
  Consider self-gravi\-tating perfect fluid flows in the $3$-torus $T^3$ characterized by an
  energy density $\rho$, a pressure $P$ and a $4$-velocity field
  $v^\alpha$ with a  linear equation of state
  \be
  \label{pressure}
  P = c_s^2 \, \rho, 
  \ee
  where the (constant) sound speed $c_s \in (0,1)$ is measured
  in units of the speed of light.   
  Then, the singular initial value problem with suitable data
  prescribed on the initial hypersurface of singularity admits a
  solution in wave coordinates with a time function $t>0$ normalized
  to vanish on the past singularity $t=0$. This solution has a
  well-defined asymptotic behavior\footnote{An expansion near the singularity will be provided below.}
and, in particular, it is consistent
  with the ``velocity term dominance'' and ``matter does not matter''
  paradigms. 
\end{theorem*}

All the assumptions and relationships relevant for this
theorem will be described in the rest of this text. The vacuum
case corresponding to $\rho \equiv 0$ has received much attention
previously and,
under the above symmetry assumption, the class of spacetimes under
consideration is known as the Gowdy spacetimes on $T^3$, first studied
in \cite{Gowdy:1974hv}. Later, a combination of theoretical and
numerical works has led to a clear picture of the behavior of
solutions to the vacuum Einstein
equations as one approaches the boundary of the spacetimes; see
\cite{Garfinkle:2004cl,Kichenassamy} 
and eventually to the resolution of the so--called \keyword{strong censorship
  conjecture} in this class \cite{Ringstrom:2006gy,Ringstrom:2009ji}.
Much less is known about the Einstein-Euler equations under Gowdy
symmetry which is therefore our main focus here.
Yet, the initial value problem was solved in recent years by LeFloch et al. in 
\cite{Grubic:2013,Grubic:2015gb,LeFloch:2011dt,LeFloch:2005wq}. 
On the other hand, when a positive cosmological constant is added to the Einstein equations 
\cite{LeFloch:2017Wei,Oliynyk:2016eo,Rendall:2004gm,Reula:1999kl,Rodnianski:2013kt}, the late-time
asymptotics of solutions without symmetry assumptions have been studied in the expanding time direction.

\paragraph{A critical phenomenon.}
We have discovered a new critical phenomenon and in order to
describe this further, let us continue the discussion in slightly more
technical terms. We are seeking for $(3+1)$--dimensional, matter spacetimes $(M,g)$ with spatial topology $T^3$, satisfying the Einstein--Euler system in Gowdy symmetry (see below). Recall that Einstein's field equations read  
\be
\label{Einstein}
G_{\alpha\beta} = \kappa \, T_{\alpha\beta}, 
\ee
where $\kappa>0$ is a constant normalized to unit from now on and, by convention, all Greek indices $\alpha, \beta, \ldots$ describe $0, \ldots, 3$. Here, $G_{\alpha\beta} := R_{\alpha\beta} - (R/2) g_{\alpha\beta}$ denotes the Einstein curvature, 
$R_{\alpha\beta}$ the Ricci curvature, and $R={R_\alpha}^\alpha$ the scalar curvature of the metric $g_{\alpha\beta}$.
The stress--energy tensor $T_{\alpha\beta}$ describes the matter content and, for perfect compressible fluids,
reads
\be
\label{momentum}
T_{\alpha\beta} = (\rho+P) \, u_\alpha u_\beta + P \, g_{\alpha\beta}.  
\ee  
The pressure $P = c_s^2 \, \rho$ is assumed to be a linear function and we also recall that the Euler equations read
\be
\nabla^\alpha T_{\alpha\beta}=0,
\ee
where $\nabla^\alpha$ is the covariant derivative operator associated
with $g_{\alpha\beta}$.  Several parameters are playing a key role in our analysis. First of all, the geometry is characterized by the so-called Kasner exponent (precisely defined below) in the direction of the fluid flow, which we denote as 
\be
\label{eq:p1interval}
p_1 \in [-1/3, 1).
\ee
This exponent determines the rate at which the spacetime is shrinking or expanding in the direction of the fluid flow (relatively to the volume of the spacetime slices, which tends to zero if the time variable is taken to decrease to $t=0$). In view of our assumption $P = c_s^2 \, \rho$, the fluid is characterized by the sound speed $c_s\in (0,1)$. Some remarks will be made below in the limiting cases of vanishing or unit sound speed. In the limit $c_s \nearrow 1$, we have a so-called stiff fluid and the sound speed and light speed coincide; this is an example of an ``extreme'' matter model mentioned earlier. On the other hand, the limit $c_s \searrow 0$ leads us to the so-called \keyword{zero-pressure
  model} -- a rather degenerate model exhibiting high concentration of
matter.  

The nonlinear coefficient 
\be
\label{eq:defGammaIntro}
\Gamma := {(c_s^2 - p_1)}/{(1 - p_1)}
\ee
can be interpreted as the discrepancy between the
(square of the) geometric speed $p_1$ and the fluid speed $c_s^2$, i.e., it is a measure of
how much the fluid is ``able to react'' by internal isotropic forces to
the external anisotropic gravitational strain.
As we will see, the analysis performed in the present paper suggests the following terminology: 

\begin{enumerate}

\item[] {\bf Sub-critical fluid flow $\Gamma>0$.} In this case
  the fluid comes to a rest asymptotically with respect to an observer moving orthogonally to the foliation slices, and the matter does not strongly interact with the geometry.  

\item[] {\bf Super-critical fluid flow $\Gamma<0$.} In this regime, the (un-normalized) fluid vector becomes asymptotically null as one approaches the singularity, and the fluid model breaks down eventually. The sound speed is smaller than the characteristic speed $\sqrt{p_1}$ associated with the geometry so that, at least at a heuristic level, the dynamics of the fluid is dominated by the geometry. 

\end{enumerate} 

As we will see, the coefficient $\Gamma$ appears naturally for the first
time in the analysis of the simplified setting in
\Sectionref{sec:FluidsKasnerNN} where it reveals its
criticality in a most explicit manner. When we then proceed to
the fully coupled self-gravitating
fluid model in \Sectionref{sec:coupledfluids}, it is of particular
interest whether this criticality is
retained. It is well known that in the generic Gowdy \textit{vacuum}
case, the coefficient $p_1$ takes values in the negative subinterval
$(-1/3,0)$ only. If ``matter really does not matter'', as one expects
for ``ordinary'' fluids having $c_s\in (0,1)$, the same should
be the case in the generic Gowdy Einstein-\textit{Euler} case.
\Eqref{eq:defGammaIntro} then suggests that the main case of interest
is the sub-critical case $\Gamma>0$. Our results in
\Sectionref{sec:coupledfluids}, in particular Theorems \ref{th:sub-criticalcoupledExistence} and \ref{th:sub-criticalcoupledEstimates}, and Theorems \ref{th:criticalcoupledExistence} and \ref{th:criticalcoupledEstimates}, indeed support this claim. In any case, the relevance of $\Gamma$ is obvious in less rigid problems, for example for fluids on a fixed background and for (half-)polarized Gowdy-matter spacetimes, which we, however, only discuss briefly in this paper.

\paragraph{Forward and backward evolution problems.} 

By convention, we always solve within the future of the cosmological singularity and we distinguish between two set-ups: a backward problem where we evolve toward the singularity, and a forward problem where we evolve toward the future. 
Since our method of proof will rely on
energy estimates, our first task is to formulate a fully hyperbolic set of evolution equations derived from the
Einstein-Euler system. For the Euler equations, we rely on the formalism
in \cite{Frauendiener:2003gb,Walton:2005vx} which yields quasilinear
symmetric hyperbolic evolution equations of the form
$$
{A^\delta}_{\alpha\beta} \nabla_\delta v^\beta = 0
$$
for a vector field $v=(v^\alpha)$ which fully describes the fluid evolution and for some coefficients $
{A^\delta}_{\alpha\beta}=
{A^\delta}_{\alpha\beta}(v)$. 
For the Einstein equations we use the
\keyword{(generalized) wave formalism} which leads to quasilinear
evolution equations of wave type for components of the Lorentzian
metric $g_{\alpha\beta}$ in the schematic form (cf.~\Sectionref{sec:2})
$$
\Box_g g_{\alpha\beta}=Q(\partial g,\partial g)+\text{matter
  terms}.
$$

Once these equations are properly formulated, we can introduce a \keyword{singular
  initial value problem}, also called a \keyword{Fuchsian problem}. Let us provide here a quick summary of the \keyword{forward problem} of interest in the present paper, 
and compare it to the more conventional \keyword{backward problem}. Further details of our approach will discussed later in
\Sectionref{sec:SIVPNN}. 

Consider any system of evolution
equations defined on $(t,x)\in (0,\delta]\times T^3$ ($\delta$ being a
possibly small positive constant) with a symbolic unknown $u=u(t,x)$.
Suppose that the Cauchy problem is well-posed when data from some
initial data function space $I$ are  prescribed at some time $t_0\in
(0,\delta]$
and when solutions are sought within some function space $S$. Moreover, suppose
that for each initial datum in $I$, the corresponding solution
  $u(t,x)$ in $S$ is
  always defined on the whole domain $(0,\delta]\times T^3$. The  \textit{backward
    problem} is the study of the behavior of these solutions in the, presumably singular,
  limit $t\searrow 0$. In our case, one would for example seek to establish that
  for some suitable norm $\|\cdot\|$ and some function $\lambda(x)$,
  each $u_*$ in $I$ can be associated with a function
  $u_0=u_0(t, x)$ in some \textit{asymptotic data space} $A$ so that
\be
\label{eq:convergencesymbolic}
  \lim_{t\searrow 0}
  \|t^{-\lambda}\left(u(t,\cdot)-u_0(t,\cdot)\right)\|=0.
\ee
The function $u$ here is the solution of the Cauchy problem in $S$
uniquely determined by $u_*$. 

In contrast to the above situation for the
\textit{forward problem} (i.e., the {singular initial value
  problem}), one seeks to establish that for
some suitable norm $\|\cdot\|$ and smooth function $\lambda(x)$,
each asymptotic data $u_0$ in $A$ gives rise to a unique solution
$u(t,x)$ in $S$ --- which henceforth determines data
$u_*(x):=u(t_0,x)$ in $I$ --- such that \Eqref{eq:convergencesymbolic}
holds. The backward problem can therefore be understood as a map
$I\rightarrow A$ while the forward problem as a map $A\rightarrow I$.
It is clear that these two problems contain quite different, rather
complementary information about the singular structure of the solution set of the evolution equations and hence about the
physical system they describe. Both kinds of information can be valuable (see
for example the discussion in Section~5.4 in \cite{Rodnianski:2014yaa}). 

In this paper, we focus on the forward problem, i.e., the
singular initial value problem. The asymptotic data, which describe the expected singular behavior of self-gravitating fluid models, are derived heuristically using particular key assumptions described in \Sectionref{sec:HeuristicAsymptotics}. In \Sectionsref{sec:coupledfluids} and \ref{sec:proofsN} we then formulate and analyze the singular initial value problem rigorously.
Our technique builds on earlier
investigations by Kichenassamy and Rendall and
co-authors on {Fuchsian} techniques in the real-analytic setting \cite{Kichenassamy,KR},
which were later
applied
\cite{Isenberg:1999ba,Isenberg:2002ku,ChoquetBruhat:2006fc,Andersson:2001fa,Damour:2002exa,lrr-2008-1}.
Let us mention the work Anguige \cite{ANguige} on perfect fluids  for the restricted class of polarised Gowdy symmetry and again under the assumption of analyticity of the data. 
The first attempt to overcome the analyticity restriction was made
in \cite{Claudel:1998tt,Rendall:2000ki} and, next, a series of papers was presented by Beyer and LeFloch \cite{Beyer:2010a,Beyer:2010b,Beyer:2010foa,Beyer:2011ce} and then extended 
in Ames et al. \cite{Ames:2012vz,Ames:2016uy}. This study 
led to a Fuchsian theory which applies to a quite general class of
quasilinear hyperbolic equations without the analyticity
restriction, but yet does not apply the coupling twith the Euler equations.

\paragraph{Overcoming a technical challenge.} In this paper,  
we introduce a further method for the class of partial differential equations under consideration,
which could in principle apply to a broader class of singular initial value problems for
nonlinear wave equations, well beyond the particular problem treated here. In
sharp contrast to the Fuchsian method in the real analytic setting
mentioned above, the Fuchsian method for quasilinear symmetric
hyperbolic equations (see \cite[Theorems~2.4 and 2.21]{Ames:2012vz} and  
\Theoremref{th:smoothexistenceN} below)  
does not always apply directly to the problems at hand. Let us
highlight an issue which is particularly relevant for
wave equations for which  
space and
time derivatives appear at the same order of differentiation in the energy estimates. In
all cases of study so far, this entails an unsatisfactorily weak control of the
behavior of spatial derivatives in the limit $t\searrow 0$ in
comparison to time derivatives\footnote{Typically, the unknowns to behave like
  $u(t,x)=c(x) t^{k(x)}$ for some functions $c$ and $k$, and one has 
  $\partial_t u=c k t^{k-1}$ and $\partial_x u=(c'+c k'\log t) t^k$.
  The standard wave energy is dominated by the time derivative
  $\partial_t u$ in the limit $t\searrow 0$ and provides only limited control 
  on the behavior of $\partial_x u$.}. In particular, the
above Fuchsian method only applies if one can establish sufficiently
\textit{strong estimates} for the source terms of the equations given
only such a \textit{weak control} over spatial derivatives.

This problem has a long history in Fuchsian studies of the
Einstein-vacuum equations for Gowdy symmetric spacetimes in the non-analytic
setting. Due to its fully explicit nature, the existing technique proposed in \cite{Rendall:2000ki,Stahl:2002bv} is, however, feasible only for problems which are as simple as the vacuum Gowdy equations in
areal gauge. It is hopeless for the significantly more complex equations we considered in the present paper. Furthermore, the use of an
iterative procedure involving the spatial derivative terms of the equations requires in the end the asymptotic data to be $C^\infty$-regular, while we also seek for $H^q$-regularity with for some finite $q$. The new
approach we introduce in the present work (and is presented in details in
\Sectionref{sec:applFuchsian}) is neither restricted to the
$C^\infty$-case nor to simple equations a priori. Our idea is natural, yet novel, namely in a first step we exploit the before-mentioned velocity term dominance by solving ``truncated evolution
equations'' (which by construction do not contain spatial
derivatives) and it is only in a second step that we solve the singular initial value problem of the full equations and, in this problem, we use the solutions in the first step as ``asymptotic data''. Our approach is both simple and natural and this gives hope that it will be used for more general problems in future work.

Our approach indeed applies, \textit{both}, to the vacuum case (but was not used in earlier works) and  to the fluid Gowdy case (which we treat here). A major technical difference between these two cases, however, is that while for vacuum one is allowed to choose the particular wave gauge called \keyword{areal gauge} \cite{Chrusciel:1990ti} for which Einstein's equations decouple significantly, the complicated structure of the principal part matrices in the presence of a fluid (see in particular \Eqref{eq:nnZO}) makes the additional ODE arguments in Section~2.4 of \cite{Ames:2012vz}  necessary to complete the proof. While in the vacuum case, our new approach therefore allows us to prove, for the first-order time, an existence result of the singular initial value problem with $H^q$-regularity for some finite $q$, we are still restricted to $C^\infty$-regularity in the fluid case. In any case, in the light of the above duality between the forward and the backward problem, this new result for the vacuum forward problem  therefore complements Ringstrom’s theory  regarding the vacuum backward problem \cite{Ringstrom:2006gy,Ringstrom:2009ji}.

 In summary, the present work provides the \textit{first} mathematically rigorous investigation of self--gravitating fluids in inhomogeneous spacetimes in a neighborhood of the cosmological singularity, while 
only the corresponding problem near {\sl isotropic} singularities was
studied in earlier works  
\cite{Tod:2002wd}. Our study has allowed us to identify specific
parameters (such as the exponent $\Gamma$) and their values. In future
studies, numerical experiments could be useful to further elucidate
the critical behavior we have uncover in this paper and possibly overcome some of the
restrictions of the existing theoretical techniques.


\section{Formulation of the Einstein-Euler system} 
\label{sec:2}

\subsection{The relativistic Euler equations}
\label{sec:generalEulerEqs}

We will use the 
symmetrization of the relativistic Euler equations which was independently introduced by Frauendiener \cite{Frauendiener:2003gb} and Walton \cite{Walton:2005vx}. (See also the alternative derivation by Beig and LeFloch~\cite{Beig:2014}.) 
The basic idea of this formulation is to work with a non-unit fluid velocity vector and to relate the norm of this vector field to the mass energy density of the fluid. 
It was shown therein that the divergence-free condition $\nabla_\beta {T_\alpha}^\beta=0$
of an arbitrary smooth $(0,2)$-tensor field of the form
\be
\label{momentum2}
T_{\alpha\beta} = f(x) \, v_\alpha v_\beta + g(x) \, g_{\alpha\beta}
\ee
implies a symmetric hyperbolic system of PDEs of the form
$
0={A^\delta}_{\alpha\beta} (v) \nabla_\delta v^\beta 
$
for the unknown (not necessarily normalized) timelike vector field $v=(v^\beta)$, provided the so far unspecified functions $f$ and $g$ satisfy
$
  f-x^3 g'=0.
$
with $1/x :=v:=\sqrt{-v_\alpha v^\alpha}$. 
Moreover, $T_{\alpha\beta}$ is the energy-momentum tensor of a perfect fluid as in \Eqref{momentum}, provided
\be 
g=P,\quad v^\alpha=v u^\alpha,\quad f/x^2=\rho+P.
\ee
In the case of the equation of state $P = c_s^2 \, \rho$ for some constant $c_s$, this leads to the following system of equations
\be
\label{eq:AAA1}
\aligned
0 &=\frac 1f{A^\delta}_{\alpha\beta}\nabla_\delta v^\beta,
\qquad \quad 
\frac 1f A^\delta_{\alpha\beta}
 =\frac{3\gamma-2}{\gamma-1} \frac{v_\alpha v_\beta}{v^2} v^\delta+v^\delta g_{\alpha\beta}
  +2{g^\delta}_{(\beta} v_{\alpha)}, 
\endaligned
\ee
which is therefore equivalent to the Euler equations. Here, we have introduced the parameter
\be
  \label{eq:defgamma}
  \gamma:=1+c_s^2,
\ee
in consistency with the literature. Our restriction $c_s\in (0,1)$ therefore translates to $\gamma\in (1,2)$. We note that in the zero-pressure case $\gamma=1$, \Eqref{eq:AAA1} becomes singular.

Let us also express the energy momentum tensor $T_{\alpha\beta}$ completely in terms of $v^\alpha$, $x$ and $\gamma$. 
For the equation of state $P = c_s^2 \, \rho$,
we thus find
$
g(x)=P_0 x ^{\frac {\gamma}{\gamma-1}}$
and 
$f(x)=P_0 \frac {\gamma}{\gamma-1} x^\frac{3\gamma-2}{\gamma-1}$ 
for some constant $P_0>0$. The energy-momentum tensor \Eqref{momentum2} therefore reads
\be
\label{momentum3}
T_{\alpha\beta} =P_0 \Big(
\frac {\gamma}{\gamma-1} v^\frac{2-3\gamma}{\gamma-1} \, v_\alpha v_\beta +  v^{-\frac {\gamma}{\gamma-1}} \, g_{\alpha\beta}
\Big).
\ee
Once the vector field $v^\alpha$ has been found as a solution of \eqref{eq:AAA1}, we can calculate the physical variables $\rho$, $P$ and $u^\alpha$ in \Eqref{momentum} from the following relationships:
\be
\label{eq:relationshipphysical}
u^\alpha=\frac{v^\alpha}{v},
\qquad P=P_0 v^{-\frac {\gamma}{\gamma-1}},
\qquad \rho=\frac{P_0}{\gamma-1} v^{-\frac {\gamma}{\gamma-1}}=T_{\alpha\beta}u^\alpha u^\beta.
\ee
Without loss of generality we set $P_0=1$ from now on. Observe that
while this formulation of the Euler equations also applies to the limiting case $\gamma=2$, it breaks down for $\gamma=1$ due to the presence of factors $1/(\gamma-1)$ in the formulas above. Note also that the vacuum case $\rho \to 0$ is recovered in the
limit $v\rightarrow + \infty$.


\subsection{The Einstein equations in generalized wave gauge}
\label{sec:waveformalism}

The technique in this section is standard and we  only sketch it while referring for instance to \cite{Ringstrom:2013} for the details. 
We start with Einstein's field equations
\be
  \label{eq:EFE}
  R_{\alpha\beta}=T_{\alpha\beta}-\frac 12 g_{\alpha\beta} T,
\ee
where $T:={T_{\alpha}}^{\alpha}$ is the trace of the energy momentum tensor $T_{\alpha\beta}$,
and we introduce the following \textit{generalized Einstein equations}
\be
\label{eq:EinstEvolEqsSHORT}
R_{{\alpha} {\beta}} + \nabla_{( {\alpha}} \mathcal D_{{\beta})} + {C_{{\alpha}{\beta}}}^{{\gamma}} \mathcal D_{{\gamma}} =T_{\alpha\beta}-\frac 12 g_{\alpha\beta} T,
\ee
where we have set 
\be
  \label{eq:defD}
\mathcal D_{{\alpha}} := \gsF_{{\alpha}}-\Gamma_{{\alpha}},
\qquad 
\Gamma_{{\gamma}{\delta} {\alpha}} 
:= \frac 12 \left( \partial_{{\gamma}} g_{{\delta} {\alpha}} + \partial_{{\alpha}} g_{{\delta} {\gamma}} - \partial_{{\delta}} g_{{\gamma}{\alpha}} \right),
\qquad 
\Gamma_{ {\delta}} := g^{{\gamma}{\alpha}}\Gamma_{{\gamma}{\delta} {\alpha}},
\ee
and  $g^{{\gamma}{\alpha}}$ are the components of the inverse metric.
The terms $\gsF_{{\beta}}$ are the \keyword{gauge source
  functions} which are (freely specifiable) sufficiently regular functions of the coordinates $x^\alpha$ and the unknown metric components $g_{\alpha\beta}$ (but not of derivatives). The coefficients ${C_{{\alpha}{\beta}}}^{{\gamma}}$ are assumed to be symmetric in the first two indices, but apart from that are free functions of $x^\alpha$, $g_{\alpha\beta}$ and first derivatives. Observe that none of the terms $\mathcal D_{\alpha}$, $\gsF_{{\alpha}}$ and ${C_{{\alpha}{\beta}}}^{{\gamma}}$  are components of a tensor in general. The expression $\nabla_{\beta} \mathcal D_{\alpha}$ is a short hand notation for
$\nabla_{\beta} \mathcal D_{\alpha}=\partial_{\beta} \mathcal D_{\alpha}-\Gamma_{{\beta}{\delta} {\alpha}} \mathcal D_{\gamma} g^{{\delta}{\gamma}}$. 

We interpret \Eqsref{eq:EinstEvolEqsSHORT} as ``evolution equations'' since they are equivalent to a system of quasilinear wave equations
\be
\label{eq:wgEinstEqs}
\aligned
&- \frac{1}{2} g^{{\gamma}{\epsilon}} \partial_{\gamma}\partial_{\epsilon} g_{{\alpha} {\beta}}
+ \nabla_{({\alpha}} \mathcal F_{{\beta} )} + g^{{\gamma}{\epsilon}} g^{{\delta} {\phi}} 
  \left( \Gamma_{{\gamma}{\delta} {\alpha}} \Gamma_{{\epsilon} {\phi} {\beta}} + 
         \Gamma_{{\gamma}{\delta} {\alpha}} \Gamma_{{\epsilon} {\beta} {\phi}} +
         \Gamma_{{\gamma}{\delta} {\beta}} \Gamma_{{\epsilon} {\alpha} {\phi}} 
  \right)
\\
& 
+ {C_{{\alpha}{\beta}}}^{{\gamma}} \mathcal D_{{\gamma}} -T_{\alpha\beta}+\frac 12 g_{\alpha\beta} T=0,
\endaligned
\ee
which, under suitable conditions, admits a 
  well-posed initial value formulation with 
 Cauchy data $g_{{\alpha}{\beta}}$ (a Lorentzian metric) and $\partial_t g_{{\alpha}{\beta}}$ (a symmetric two-tensor) prescribed on a spacelike hypersurface. The solution is thus Lorentzian metric defined in a neighborhood of the given initial hypersurface.  

Suppose that $g_{{\alpha}{\beta}}$ is any solution to the evolution equations \eqref{eq:EinstEvolEqsSHORT} for some chosen gauge source functions with $\mathcal D_\alpha$ of the form \eqref{eq:defD}. It is clear that $g_{{\alpha}{\beta}}$ is an actual solution to the Einstein equations \Eqref{eq:EFE} {\sl if and only if} $\mathcal D_{{\alpha}}$ all vanish identically.
Furthermore, assuming the energy momentum tensor $T_{\alpha\beta}$ is divergence free, we can derive a system of equations for
$\mathcal D_{{\alpha}}$, that is,  
\be
\label{eq:ConstraintPropagationEquation}
 \nabla^{\alpha} \nabla_{\alpha} \mathcal D_{{\beta}} +{R_{\beta}}^{\epsilon} \mathcal D_{{\epsilon}}
  +\left(2\nabla_{\alpha}C^{{\alpha}}{}_{\beta}{}^{{\gamma}} -\nabla_{{\beta}} {C_{\epsilon}}^{{\epsilon}{\gamma}}\right) \mathcal D_{\gamma}+
\left(2C^{{\alpha}}{}_{\beta}{}^{{\gamma}}- {C_{\epsilon}}^{{\epsilon}{\gamma}}{\delta^{\alpha}}_{\beta}\right) \nabla_{{\alpha}} \mathcal D_{\gamma}=0, 
\ee
which is a linear homogeneous system of wave equations and is referred to as the \keyword{constraint propagation equations}  or the \keyword{subsidiary system}.
Recall that $\nabla_{\alpha}$ is the Levi-Civita covariant derivative of $g_{{\alpha}{\beta}}$ and $R_{{\alpha}{\beta}}$ is the corresponding Ricci tensor. We thus conclude that the terms $\mathcal D_{\beta}$ are identically zero (and hence the solution $g_{{\alpha}{\beta}}$ of the evolution equations is a solution to Einstein's equations) if and only if the Cauchy data on the initial hypersurface satisfy $\mathcal D_{\beta}=0$ and $\partial_t \mathcal D_{\beta}=0$. Motivated by this observation, we refer to $\mathcal D_{\beta}$ as the \keyword{constraint violation functions} and to the conditions $\mathcal D_{\beta}=0$ and $\partial_t\mathcal D_{\beta}= 0$ at the initial time as the \keyword{constraints} of the Cauchy problem.

Let us make a few further remarks on the constraints. From initial data $g_{{\alpha}{\beta}}$ and $\partial_t g_{{\alpha}{\beta}}$ prescribed at the initial time $t_*$ we can calculate the terms $\Gamma_{\alpha}$ at $t_*$. The constraint $\mathcal D_{\beta}=0$ implies that these terms must match the initial values of the gauge source functions; cf.\ \Eqref{eq:defD}. It follows that this condition is not a restriction on the Cauchy data but rather on the gauge source functions because for \textit{any} Cauchy data we can find gauge source functions whose initial values match  the terms $\Gamma_{\alpha}$ at $t_*$.
This suggests that $\mathcal D_{\alpha}=0$ is not a \textit{physical} restriction but merely a  \keyword{gauge constraint}. In contrast to this, the constraint $\partial_t{\mathcal D}_{\alpha}=0$ turns out to be a restriction on the Cauchy data but \textit{not} on the gauge source functions. 
In order to see this, we first realize that the values of the terms $\partial_t\Gamma_{\alpha}$ at $t_*$ can be calculated from the sole Cauchy data (and hence it can be checked if this constraint is satisfied) if we assume that the evolution equations hold at $t_*$. This is so because the constraint $\partial_t{\mathcal D}_{\alpha}=0$ contains second-order time derivatives of the metric at $t_*$ which can only be computed via the evolution equations. However, when all these second-order time derivatives in the constraint are expressed using the evolution equations, it turns out that all terms involving the gauge source functions drop out completely.
In fact, we find that the relationship
\be
G^{{\alpha}0}=-\frac 12 g^{00}g^{{\alpha}{\beta}}\partial_t \mathcal D_{{\beta}}
\ee
is valid at $t_*$. Hence the constraints $\partial_t{\mathcal D}_{\alpha}=0$ are equivalent to the standard Hamiltonian and momentum constraints, and we therefore refer to them as the \keyword{physical constraints}, in order to distinguish them from the \textit{gauge constraints} above.

\subsection{Spacetimes with Gowdy symmetry}
\label{sec:GowdysymSpacetimes}

For the purpose of this paper, we restrict to spacetimes with $U(1)\times U(1)$-symmetry. A $4$-dimensional smooth oriented time-oriented Lorentzian manifold $(M,g_{\alpha\beta})$ with $M\cong \R\times T^3$ is said to be
$U(1)\times U(1)$-symmetric provided there is a smooth effective action of the group $U(1)\times U(1)$ generated by two linear independent smooth commuting spacelike Killing vector fields $\xi_1^\alpha$ and $\xi_2^\alpha$. It can be shown that we can identify these Killing vector fields with two of the three spatial coordinate vector fields everywhere, say, $\partial_y$ and $\partial_z$, if the gauge source functions $\gsF_{{\alpha}}$ and the terms ${C_{\alpha\beta}}^\gamma$ in \Eqref{eq:EinstEvolEqsSHORT} do not depend on the spatial coordinates $y$ and $z$ and if the fluid vector commutes with $\partial_y$ and $\partial_z$.

For Gowdy-symmetric matter spacetimes, we choose
\be
  \label{eq:asympwavegauge} 
    \mathcal F_{0}(t,x,g) = - \frac 1t, \quad \mathcal F_{1}(t,x,g) = \mathcal F_{2}(t,x,g)= \mathcal F_{3}(t,x,g) = 0. 
\ee
The foliation of Cauchy surfaces generated by these gauge source functions can be shown to agree with that of {wave coordinates} asymptotically in the limit $t\searrow 0$. A more detailed discussion can be found in \cite{Ames:2016uy}. 
As we will see later, it is very useful to also choose
\be
  \label{eq:choiceconstraintmultiples}
  \begin{split}
  {C_{00}}^0(t,x)&=\frac{3+k^2(x)}{2t}, \quad {C_{01}}^1(t,x)={C_{10}}^1(t,x)=\frac{1+k^2(x)}{4t},\\
 {C_{\alpha\beta}}^\gamma(t,x)&=0
\qquad \text{ for all other $\alpha,\beta,\gamma$,}
\end{split}
\ee
for some (so far unspecified) smooth function $k$.
One can then show easily that the following block diagonal form of the metric is preserved during the evolution of the Einstein-Euler equations.

\begin{definition}[Block diagonal coordinates for $U(1)\times U(1)$-symmetric spacetimes]
\label{def:blockdiagonalcoords}
    Let $(M,g_{\alpha\beta})$ be a $U(1)\times U(1)$-symmetric spacetime with $M=(0,\delta)\times T^3$ (for some fixed $\delta>0$). A coordinate chart with dense domain $U\subset M$ and range $(0,\delta)\times (0,2\pi)^3$ is called \textbf{block diagonal coordinates} provided the metric $g_{\alpha\beta}$ 
we can write on $U$
\be
  \label{eq:blockdiagonalcoords} 
    g=g_{00}(t,x) dt^2+2g_{01}(t,x) dtdx+g_{11}(t,x)dx^2+R(t,x)\left(E(t,x)(dy+Q(t,x)dz)^2+\frac1{E(t,x)} dz^2\right), 
\ee
for some metric coefficients $g_{00}$, $g_{01}$, $g_{11}$, $R$, $E$, and $Q$. 
\end{definition}

Without loss of generality we will always assume that these functions extend as smooth $2\pi$-periodic functions (in $x$) to the domain $(0,\delta]\times\R$ (the extended functions are denoted by the same symbols) such that
 $g_{00}<0$, $g_{11}>0$, $R>0$ and $E>0$ everywhere. Note in particular that 
for block diagonal coordinates 
\be
  \label{eq:6NN}
  g_{02}\equiv g_{03}\equiv g_{12}\equiv g_{13}\equiv 0.
\ee
In the following we often refer to such a coordinate chart as ``block diagonal coordinates $(t,x,y,z)$''.
It follows from the results in \cite{Chrusciel:1990ti} that one can only find such block diagonal coordinates globally on $U(1)\times U(1)$-symmetric solutions of the vacuum equations if 
the  \keyword{twists} associated with the two Killing vector fields vanish: 
$
\kappa_i:=\epsilon_{\alpha\beta\gamma\delta}\xi_1^\alpha \xi_2^\beta \nabla^\gamma\xi_i^\delta$
(with $i=1,2$). 
This condition defines the class of \keyword{Gowdy symmetric} spacetimes. If a $U(1)\times U(1)$-symmetric metric is given in the form \eqref{eq:blockdiagonalcoords}, then $\kappa_i=0$ follows and hence that the corresponding spacetime is Gowdy symmetric.

A particular example of block diagonal coordinates are: 
(1)  \keyword{areal coordinates} given by the additional condition
\be
  \label{eq:arealcond}
  g_{01}=0,\quad R=t,
\ee
or (2) \keyword{conformal coordinates} given by
\be
  \label{eq:confcond}
  g_{01}=0,\quad g_{00}=-g_{11}.
\ee
In the vacuum case, each of these two gauge choices implies the other and they form a special gauge condition within the family of gauges given by \Eqsref{eq:asympwavegauge} and \eqref{eq:blockdiagonalcoords}. In the non-vacuum case considered here however neither \Eqref{eq:arealcond} nor \eqref{eq:confcond} is preserved by the evolution, while the more general gauge condition \Eqref{eq:blockdiagonalcoords} always is. In particular,  non-vanishing values for $g_{01}$ are always generated even from zero initial data. In summary one can say that in the Einstein-Euler case, \Eqref{eq:blockdiagonalcoords} is the simplest form of the metric which is preserved by the evolution and which is as close as possible to standard forms of the Gowdy metric in the literature.


\section{Asymptotics of self-gravitating fluid models}
\label{sec:HeuristicAsymptotics}

\subsection{Velocity term dominance and  the motto ``matter does not matter''}
\label{sec:VTDMDNM}

Now, based on certain heuristic arguments, we derive here the expected singular asymptotics of self-gravitating fluid models. In \Sectionsref{sec:coupledfluids} and \ref{sec:proofsN}, these asymptotics will then be used as guides towards correct asymptotic data for our singular initial value problem for the Einstein-Euler equations.
Of particular importance for us are the \keyword{velocity term dominance} and \keyword{matter does not matter} paradigms. For now we only introduce these rather informally; precise notions will be given in \Sectionref{sec:coupledfluids}. The main idea of velocity term dominance \cite{Eardley:1972ig,Isenberg:1990gn} is that spatial derivative terms (which can be interpreted as ``gravitational potential terms'')
are expected to be {\sl negligible near a cosmological singularity} in comparison to time derivative terms (which can be interpreted as ``kinetic terms'' or ``velocity terms'') in the equations. The singular dynamics should therefore be governed by  \keyword{truncated equations} obtained from the full evolution equations by dropping all spatial derivative terms.  One then expects that, asymptotically close to the singularity and at each spatial point, any solution should behave like an independent spatially homogeneous universe. Such statements, however, have to be handled with great care. On the other hand, ``matter does not matter'' is the idea \cite{Lifshitz:1963hz,Belinskii:1970fu} that most matter fields (with the exception of ``extreme'' matter fields like scalar fields or stiff fluids) should not change the leading dynamics of the gravitational degrees of freedom close to the singularity.

These two paradigms suggest that the leading dynamics of the gravitational field should be described by the spatially homogeneous vacuum Einstein's equations -- which give rise to the so-called \keyword{Kasner spacetimes} -- or, more generally in view of the truncated Einstein's equations -- which give rise to \keyword{asymptotically local Kasner spacetimes} (\Sectionref{sec:KasnerALK}). We therefore stress that velocity term dominance \textit{implies} asymptotically local Kasner behavior, but not the other way around. Finally, the leading dynamics of the fluid should be described by the spatially homogeneous Euler equations on fixed Kasner backgrounds, as discussed in \Sectionref{sec:FluidsKasnerNN} below.

\subsection{Kasner and asymptotically local Kasner spacetimes}
\label{sec:KasnerALK}

\begin{definition}
A \keyword{Kasner spacetime} is a spatially homogeneous, but in general highly anisotropic solution $(M,g_{\alpha\beta})$ of Einstein's vacuum equation for $M=(0, + \infty)\times T^3$ and
\be
\label{Kasner-k}
g = t^{\frac{k^2-1}{2}} \big( - dt^2 + dx^2 \big) + t^{1-k} dy^2 + t^{1+k} dz^2, 
\ee
with $t \in (0, +\infty)$ and $x,y,z \in (0,2\pi)$. The free parameter $k\in\R$  is often referred to as the \keyword{asymptotic velocity}. 
\end{definition}

 With respect to the time coordinate 
 $\tau:=\frac{4}{k^2+3} t^{\frac{k^2+3}4}$,
and, by some additional rescaling of the spatial coordinate $x$, 
this metric takes the more conventional form 
$$
g = -d\tau^2 + \tau^{2p_1} dx^2  + \tau^{2p_2} dy^2  + \tau^{2p_3} dz^2.
$$
By definition, the \keyword{Kasner exponents} are\footnote{Observe that $p_1$ already appeared in \Eqref{eq:defGammaIntro}.}
\be
 \label{eq:kasnerexponents}
 p_1 :=(k^2-1)/(k^2+3), \quad
 p_2 :=2(1-k)/(k^2+3),\quad
 p_3 :=2(1+k)/(k^2+3).
\ee
Except for the three flat Kasner cases given by
$k=1$, $k=-1$, and (formally) $|k|\rightarrow + \infty$, the Kasner metric has a curvature singularity at $t=0$. 

In the light of velocity term dominance and its highly spatially inhomogeneous features we must modify the Kasner spacetimes such that, at each spatial coordinate point $x$ of our local coordinate system, the metric asymptotes to the metric \Eqref{Kasner-k} in the limit $t\searrow 0$ for some $x$-dependent value of the Kasner parameter $k$. The quantity $k$ thereby turns into an $x$-dependent function $k=k(x)$. In order to allow for further coordinate degrees of freedom, it turns out that certain additional transformations of \eqref{Kasner-k} are in general necessary, giving rise to other $x$-dependent functions in the following definition.  

\begin{definition}[Asymptotically local Kasner spacetimes]
  \label{def:asymptlocKasner}
  Suppose $(M,g_{\alpha\beta})$ is a smooth Gowdy symmetric spacetime and $(t,x,y,z)$ are  block diagonal coordinates (\Defref{def:blockdiagonalcoords}). 
Choose  functions $k>0$, $\Lambda_*>0$, $E_*>0$, $Q_*$, $Q_{**}$, $\muGcomp1>0$,\ldots,$\muGcomp6>0$ in $C^\infty(T^1)$. Then,
  $(M,g_{\alpha\beta})$ is called an \keyword{asymptotically local Kasner spacetime} with respect to data $k$, $\Lambda_*$, $E_*$, $Q_*$, $Q_{**}$
  and exponents $\muGcomp1$,\ldots,$\muGcomp6$ 
  provided that for each sufficiently large integer $q$ there exists a constant $C>0$ such that  
  for all sufficiently small $t>0$
  \begin{align*}
    &\left\|t^{-\muGcomp1}\left(g_{00}(t) t^{-(k^2-1)/2}+\Lambda_{*}\right)\right\|_{{H^q(T^1)}}
    +\left\|t^{-\muGcomp1}{t\partial_t}\left(g_{00}(t) t^{-(k^2-1)/2}\right)\right\|_{{H^q(T^1)}}\\
    +&\left\|t^{-\muGcomp2}\left(g_{11}(t) t^{-(k^2-1)/2}-\Lambda_{*}\right)\right\|_{{H^q(T^1)}}
    +\left\|t^{-\muGcomp2}{t\partial_t}\left(g_{11}(t) t^{-(k^2-1)/2}\right)\right\|_{{H^q(T^1)}}\\
    +&\left\|t^{-\muGcomp3}g_{01}(t) t^{-(k^2-1)/2}\right\|_{{H^q(T^1)}}
    +\left\|t^{-\muGcomp3}{t\partial_t}g_{01}(t) t^{-(k^2-1)/2}\right\|_{{H^q(T^1)}}\\
    +&\left\|t^{-\muGcomp4}\left(R(t) t^{-1}-1\right)\right\|_{{H^q(T^1)}}
    +\left\|t^{-\muGcomp4}{t\partial_t}\left(R(t) t^{-1}\right)\right\|_{{H^q(T^1)}} \\
    +&\left\|t^{-\muGcomp5}\left(E(t) t^{k}-E_*\right)\right\|_{{H^q(T^1)}}
    +\left\|t^{-\muGcomp5}{t\partial_t}\left(E(t) t^{k}\right)\right\|_{{H^q(T^1)}}\\
    +&\left\|t^{-\muGcomp6}\left((Q(t)-Q_*) t^{-2k}-Q_{**}\right)\right\|_{{H^q(T^1)}}
    +\left\|t^{-\muGcomp6}{t\partial_t}\left((Q(t)-Q_*) t^{-2k}\right)\right\|_{{H^q(T^1)}}\le C.
  \end{align*}
\end{definition}

In \cite{Ames:2016uy}, a
family of asymptotically local Kasner spacetimes was constructed as solutions of the \textit{vacuum}
Einstein's equations for many types of asymptotic wave gauges. The class of asymptotically local Kasner spacetimes is therefore certainly non-trivial. 
Moreover, the Kasner spacetime is a particular example of an asymptotically local Kasner spacetime. As for Kasner spacetimes, we expect that in general asymptotically local Kasner spacetimes have curvature singularities at $t\searrow 0$. 
Observe however that only first-order time derivatives of the metric variables are considered in the estimate above. Second-order time derivatives, which are necessary to calculate the curvature tensor, can then typically be estimated by using the field equations. 
In any case, it is the very nature of singular initial value problems for wave equations (as for example Einstein's equations) that estimates of the form above involve time derivatives up to first order; see \Sectionsref{sec:SIVPNN} and \ref{sec:proofsN}.


\subsection{Spatially homogeneous fluid flows on Kasner backgrounds}
\label{sec:FluidsKasnerNN}

We next provide the heuristic understanding of the singular behavior of the fluid. In particular, the coefficient $\Gamma$ in \Eqref{eq:defGammaIntro} will now emerge naturally for the first-order time. Based on the heuristic ideas in \Sectionref{sec:VTDMDNM}, we make a number of simplifications for the sole purpose of deriving the expected leading-order behavior close to the singularity. We will return to the full problem in \Sectionref{sec:coupledfluids}.
 
The first simplifying idea is to consider the Euler equations in the form \eqref{eq:AAA1} on the \textit{fixed} background spacetime \Eqref{Kasner-k}.
The second major simplification is to restrict attention to \textit{spatially homogeneous
    fluids} for which the fluid vector $v^\alpha$ is of the form
\be
\label{eq:spathomfluid}
v^\alpha=v^0(t)(\partial_t)^\alpha+v^1(t)(\partial_x)^\alpha+v^2(t)(\partial_y)^\alpha+v^3(t)(\partial_z)^\alpha.
\ee 
The third major simplification for this section is to impose, in \Eqref{eq:spathomfluid},  
\be
  \label{eq:specfluid}
  v^2=v^3=0.
\ee
 This last restriction is motivated by the fact that it is a consequence of the fully coupled Einstein-Euler case for Gowdy symmetry considered in \Sectionref{sec:coupledfluids}.  

Under these three restrictions, \Eqsref{eq:AAA1} are equivalent to the following system of ODEs:
\be
  \label{eq:SHKasnerFrauendiener}
  \begin{split}
  t\partial_t v^0=\Gamma\frac{((v^0)^2+(v^1)^2)v^0}{(v^0)^2-(\gamma-1)(v^1)^2},
  \qquad 
  t\partial_t v^1&=2\Gamma\frac{(v^0)^2v^1}{(v^0)^2-(\gamma-1)(v^1)^2},
\end{split}
\ee
where $\Gamma$ is defined in \Eqref{eq:defGammaIntro} which can be expressed in terms of $\gamma$ and $k$ as
\be
  \label{eq:exprGammaN}
  \Gamma=\frac 14\left(3 \gamma-2-(2-\gamma)k^2\right),
\ee
using \eqref{eq:defgamma} and \eqref{eq:kasnerexponents}. 
Throughout this paper, we assume that the fluid is \keyword{future directed}\footnote{In the {future directed} case, the fluid ``flows out'' of the Kasner singularity at $t=0$ which hence represents an \textit{initial} singularity. Note, however, that the Euler equations (and in fact also the coupled Einstein-Euler system) are invariant under the transformation $v^\alpha\mapsto -v^\alpha$. Hence, any solution $v^\alpha$ of the (Einstein-) Euler equations with $v^0>0$ gives rise to a solution $-v^\alpha$ of the (Einstein-) Euler equations with $v^0<0$. In the latter case, $t=0$ represents a \textit{future singularity} because the fluid ``flows into it''.}, i.e.,
$v^0>0$, which allows us to define
\[V:={v^1}/{v^0}.\]
Then \Eqsref{eq:SHKasnerFrauendiener} yield
\be
  \label{eq:EqVHomN}
  {t\partial_t}V
  =\frac{{t\partial_t}v^1}{v^0}-V \frac{{t\partial_t}v^0}{v^0}
  =\Gamma\frac{2 v^0 v^1-V( (v^0)^2+(v^1)^2)}{(v^0)^2-(\gamma-1)(v^1)^2}
  =\Gamma\frac{V( 1-V^2)}{1-(\gamma-1)V^2},
\ee
which can readily be integrated (for any $t_0>0$ and $V(t_0)\in (-1,1)$) 
\be
  \label{eq:solV}
  \frac{V(t)}{(1-V^2(t))^{(2-\gamma)/2}}=C_1 t^\Gamma,
  \qquad\
  C_1=
  \frac{V(t_0)}{(1-V^2(t_0))^{(2-\gamma)/2}}{t_0^{-\Gamma}}. 
\ee

Regarding the limiting cases for $\gamma$, observe that for each fixed $t>0$, it follows that $(1-V^2(t))^{(2-\gamma)/2}\rightarrow 1$ in the limit $\gamma\nearrow 2$ (which implies $\Gamma\nearrow 1$). The other limiting case $\gamma=1$ is however excluded as we have observed in \Sectionref{sec:generalEulerEqs}. \Eqref{eq:solV} is therefore the implicit representation of the solutions to \Eqref{eq:EqVHomN} for \textit{all} $\gamma\in (1,2]$. We easily note that $V$ is either (i) identically zero (if $C_1=0$) or (ii) strictly positive (if $C_1>0$) or (iii) strictly negative (if $C_1<0$).
In case (i), it follows that $v^1$ is identically zero, and hence \Eqref{eq:SHKasnerFrauendiener} implies that $v^0=C_2 t^\Gamma$ for some constant $C_2$. In both cases (ii) and (iii) we can eliminate $v^1$ from \Eqsref{eq:SHKasnerFrauendiener} using the definition of $V$ and \Eqref{eq:EqVHomN} to find
\be
  \label{eq:v0sol}
  \frac{{\partial_t}v^0}{v^0}=\frac{{\partial_t}V}V \frac{1+V^2}{1-V^2},
\quad
\text{implying } 
  v^0(t)=C_2 \frac{V(t)}{1-V^2(t)}\quad\text{with}\quad
  C_2=v^0(t_0)\frac{1-V^2(t_0)}{V(t_0)},
\ee
for any $v^0(t_0)>0$. The remaining term $v^1$ can then be expressed in terms of $V$ straightforwardly.

The implicit formula \Eqref{eq:solV} for $V$ can be exploited to derive expansions of $V$, and thereby of $v^0$ and $v^1$, about $t=0$. Here we need to distinguish between different signs of $\Gamma$. 
The following is a summary of the result.

\begin{theorem}[Homogeneous fluid flows on Kasner spacetimes] 
\label{th:fluidKasner}
Consider the Euler equations in the form \eqref{eq:AAA1} on the fixed Kasner spacetime \Eqref{Kasner-k} given by any value of the parameter $k$. Choose  any equation of state parameter $\gamma\in(1,2]$ and define $\Gamma$ by \Eqref{eq:exprGammaN}. For each solution $v^\alpha$ of the form \Eqsref{eq:spathomfluid}--\eqref{eq:specfluid}, there either exist constants $v_*>0$ and $v_{**}\in\R$ such that\footnote{
The symbols $o(\cdot)$ refer to the limit $t\searrow 0$.} 
\be
  \label{eq:nonspecialcase}
  \begin{pmatrix}
  v^0(t), &
  v^1(t)
\end{pmatrix}
=
\begin{cases}
  \begin{pmatrix}
  v_* t^\Gamma(1+o(1)), &
  v_{**} t^{2\Gamma}(1+o(1))
\end{pmatrix}, & \text{$\Gamma>0$},\\
\begin{pmatrix}
  v_*, &
  v_{**}
\end{pmatrix}, & \text{$\Gamma=0$},\\
\begin{pmatrix}
  v_* t^{-2|\Gamma|/(2-\gamma)}+v_{**}\frac{\gamma}{2(\gamma-1)}, &
  \pm (v_* t^{-2|\Gamma|/(2-\gamma)}+v_{**})
\end{pmatrix}+o(1), & \text{$\Gamma<0$},
\end{cases}
\ee 
or, there exists a constant $v_*>0$ such that
\be
\label{eq:specialcase}
\begin{pmatrix}
  v^0(t), &
  v^1(t)
\end{pmatrix}
=
\begin{pmatrix}
  v_* t^\Gamma, &
  0
\end{pmatrix}
\quad\text{for every $\Gamma\in\R$}.
\ee
\end{theorem}

Observe that the factor $1/(2-\gamma)$ in the case $\Gamma<0$ is bounded, since $\gamma=2$ is excluded if $\Gamma<0$.
Let us remark briefly that in the ``dynamical system language'' of \cite{Wainwright:2005wss}, \Eqref{eq:specialcase} corresponds to the ``non-tilted'' fluid case on a Bianchi I (Kasner) background while \Eqref{eq:nonspecialcase} corresponds to ``tilted'' fluids.


Next we calculate some relevant physical terms which further describe the fluids in \Theoremref{th:fluidKasner}. In all of what follows we ignore the case of a fluid which is ``identically at rest'', i.e., we focus on \Eqref{eq:nonspecialcase}.
As a reference frame let us fix the congruence of freely falling observers tangent to the future-pointing timelike unit vector 
\[e_0^\alpha=t^{(1-k^2)/4}\partial_t^\alpha\]
in the Kasner background spacetime.
Since $e_0$ is the future pointing normal to the homogeneous hypersurfaces, these observers can be interpreted as being ``at rest'' in the Kasner spacetimes. We  refer to these as ``Kasner observers'' in the following.
The energy density of the fluids in \Eqref{eq:nonspecialcase} measured by these observers is
\be
  \label{eq:energydensKK}
T_{\alpha\beta} e_0^\alpha e_0^\beta
=
\begin{cases}
  O\left(t^{-\frac{\gamma}{2-\gamma}(1-\Gamma)}\right) & \text{if
    $\Gamma\ge 0$},\\
  O\left(t^{-\frac{\gamma-2\Gamma}{2-\gamma}}\right) & \text{if $\Gamma<0$}.
\end{cases}
\ee
Hence, this energy density blows up for any choice of $\gamma\in (1,2]$ and $k\in\R$, in particular, irrespective of the sign of $\Gamma$, in the limit $t\searrow 0$. The rate of divergence however is different for different signs of $\Gamma$ which suggests that different physical processes lead the dynamics of the fluid at the singularity.  

Another interesting quantity is the relative velocity of the fluid and the Kasner observers. To this end we fix 
\[e_1^\alpha=t^{(1-k^2)/4} \partial_x^\alpha\]
which is a spacelike unit vector field parallel to the flow of the fluid and which is orthogonal to $e_0^\alpha$. This vector field can be interpreted as a natural spatial unit length scale for the Kasner observers. The  relative velocity is then given by
\be
  \label{eq:fluidvelocity}
  V= 
-\frac{g_{\alpha\beta}e_1^\alpha v^\beta}{g_{\alpha\beta} e_0^\alpha v^\beta}
=
\begin{cases}
  \frac{v_{**}}{v_*} t^\Gamma(1+o(1))& \text{if $\Gamma\ge 0$},\\
  \pm 1+o(1) & \text{if $\Gamma<0$}.
\end{cases}
\ee
Hence, the fluid ``slows down'' relatively to the Kasner observers in the limit $t\searrow 0$ when $\Gamma>0$ while it accelerates towards the maximal possible velocity relative to the Kasner observers, i.e., the speed of light, in the case $\Gamma<0$ (unless it is at rest identically, see \Eqref{eq:specialcase}). 

Let us also consider the energy density of the fluid measured by observers who are co-moving with the fluid. This is the quantity $\rho$ in \Eqref{eq:relationshipphysical} for which we find
\be
  \label{eq:fluiddensityNN}
  \rho=
\begin{cases}
  O(t^{-\gamma(3+k^2)/4})=O\left(t^{-\frac{\gamma}{2-\gamma}(1-\Gamma)}\right) & \text{if $\Gamma\ge 0$},\\
  O(t^{-\gamma/(2-\gamma)}) & \text{if $\Gamma<0$}.
\end{cases}
\ee
We emphasize that for $\Gamma>0$ the terms $\rho$ and $T_{\alpha\beta} e_0^\alpha e_0^\beta$ blow up with the same rate as a consequence of the fact that the two families of observers are parallel in the limit $t\searrow 0$ (which is not the case for $\Gamma<0$). We can show that if we fix a small spatial volume element orthogonal to the fluid at some event in the Kasner spacetime, e.g., some $3$-space spanned by a basis of spacelike vectors orthogonal to $u^\alpha$ at that event, and let this volume element flow together with the fluid towards the singularity, then $\rho(t)=C \textrm{Vol}^{-\gamma}(t)$ for some constant $C>0$ irrespective of the sign $\Gamma$. Here $\textrm{Vol}(t)$ is the $3$-dimensional volume of this co-moving $3$-space which approaches zero in the limit $t\searrow 0$ irrespective of the sign of $\Gamma$. This shows that the blow up of the fluid energy density is caused by the shrinking of ``space'' in the Kasner spacetime as one approaches the singularity. The different blow-up rates for different signs of $\Gamma$ in \Eqref{eq:fluiddensityNN} can hence be understood as a consequence of the fact that observers co-moving with the fluid measure spatial volumes differently depending on whether they approach the speed zero with respect to Kasner observers in the limit $t\searrow 0$ (for $\Gamma>0$) or the speed of light (for $\Gamma<0$).

In summary, we have now provided ample justification for the interpretation of $\Gamma$ as a critical parameter as outlined in \Sectionref{sec:intro}.
\begin{figure}[t]
  \centering
  \includegraphics[width=0.6\textwidth]{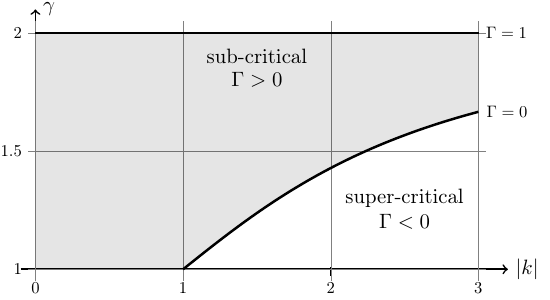}
  \caption{The parameter state space of homogeneous fluids on Kasner backgrounds.}
  \label{fig:criticalGamma}
\end{figure}

The ``state space'' of homogeneous fluids on Kasner backgrounds is illustrated in \Figref{fig:criticalGamma}.


\section{Self-gravitating fluids near the cosmological singularity}
\label{sec:coupledfluids}

\subsection{The sub-critical regime \texorpdfstring{$\Gamma>0$}{(Gamma>0)}}
\label{sec:coupledfluidssuper}

Now we turn our attention again to general Gowdy symmetric spacetimes
as described in \Sectionref{sec:GowdysymSpacetimes}.
We call a fluid \keyword{$U(1)\times U(1)$ symmetric, or, Gowdy symmetric} if the fluid vector field $v^\alpha$ commutes with the Killing vector fields $\partial_y^\alpha$ and $\partial_z^\alpha$ in block diagonal coordinates $(t,x,y,z)$ (see \Defref{def:asymptlocKasner}). We will continue to make the restriction $v^2=v^3=0$ motivated in \Sectionref{sec:FluidsKasnerNN}, and hence focus on fluids of the form
\be
  \label{eq:Gowdysymmfluid}
  v^\alpha=v^0(t,x)(\partial_t)^\alpha+v^1(t,x)(\partial_x)^\alpha.
\ee
We are now in a position to state the main results of the present paper about compressible perfect fluids in Gowdy-symmetric spacetimes near the cosmological singularity. 
 We begin with the sub-critical case\footnote{Throughout, periodicty conditions are imposed, so thet regularity of the solution can be ensured; see the discussion of the constraint equation in Section~\ref{sec:constraints} below.} 

\begin{theorem}[Sub-critical regime for self-gravitating fluid flows. Existence statement]
  \label{th:sub-criticalcoupledExistence}
  Suppose that $\Gamma>0$. 
Choose  fluid data
$v_*^0>0$ and $v_*^1$ in $C^\infty(T^1)$, an equation of state with adiabatic exponent $\gamma\in (1,2)$, and geometric data $k \in (0,1)$, $E_*>0$, $Q_*$, and $Q_{**}$ in $C^\infty(T^1)$ 
as well as a constant $\Lambda_{**}>0$ such that  the following functions are in $C^\infty(T^1)$: 
\begin{align}
  \label{eq:LambdaSconstr}
  \Lambda_*(x)&:=\Lambda_{**} \exp \Bigg( 
  \int_0^{x}\Bigl(
    -k(\xi)\frac{ E_*'(\xi)}{E_*(\xi)}+2 k(\xi) E_*^2(\xi) Q_{**}(\xi) Q_{*}'(\xi) -\frac{2\gamma v^1_*(\xi) (v^0_*(\xi))^{\frac{1-2\gamma}{\gamma-1}}}{\gamma -1}\Bigr)d\xi\Bigg),
    \\
  \label{eq:v1constr}
  \hat v^1_*(x)&:= v^1_*(x) (\Lambda_*(x))^{\frac{2-\gamma}{2(\gamma-1)}}. 
\end{align}
Then, there exists a constant $\delta>0$ and a solution to the Einstein-Euler equations (\Eqsref {eq:AAA1}, \eqref{momentum3}, \eqref{eq:EFE}) in the gauge given by the gauge source functions \Eqref{eq:asympwavegauge} which, for some choice of positive exponents $\muGcomp1$, \ldots, $\muGcomp6$, $\muFcomp1$ and $\muFcomp2$, is determined by the following conditions:
\begin{enumerate}[label=\textit{(\roman{*})}, ref=(\roman{*})] 

\item The metric admits the form \Defref{def:blockdiagonalcoords} for some functions $g_{00}$, $g_{01}$, $g_{11}$, $R$, $E$, and $Q$ in $C^\infty((0,\delta]\times T^1)$, and is an asymptotically local Kasner spacetime with respect to data $k$, $\Lambda_*$, $E_*$, $Q_*$, $Q_{**}$
  and exponents $\muGcomp1$, \ldots, $\muGcomp6$  (\Defref{def:asymptlocKasner}).

\item The fluid flow has the form \eqref{eq:Gowdysymmfluid} for some $v^0, v^1$ in $C^\infty((0,\delta]\times T^1)$, and for any (sufficiently large) integer $q$ 
there exists a constant $C_q>0$ such that, for all $t\in (0,\delta]$ with $\Gamma$ given in \Eqref{eq:exprGammaN}, 
 these functions satisfy
\be
  \label{eq:fluidestimate}
  \left\|t^{-\muFcomp1}\left(v^0 (t) t^{-\Gamma}-v^0_*\right)\right\|_{{H^q(T^1)}}
  +\left\|t^{-\muFcomp2}\left(v^1 (t) t^{-2\Gamma}-\hat v^1_*\right)\right\|_{{H^q(T^1)}}
  \le C_q. 
\ee
\end{enumerate}
\end{theorem}

Before we can state a result on the asymptotic properties, let us introduce some further notions.
Suppose two metrics $g$ and $h$ are given which are both Gowdy-symmetric and of the form \eqref{eq:blockdiagonalcoords}. We say that they \keyword{agree at order 
$(\muGcomp1,\muGcomp2,\muGcomp3,\muGcomp4,\muGcomp5,\muGcomp6)$} in the limit $t\searrow 0$ provided that 
for any smooth exponents $\widetilde{\muGcomp{i}}<\muGcomp{i}$, $i=1,\ldots,6$, and 
for each sufficiently large integer $q$,
there exists a constant $C_q>0$ such that, 
for all $t\in (0,\delta]$, 
\begin{align*}
  &\left\|t^{-(k^2-1)/2-\widetilde{\muGcomp1}}\left(g_{00}(t)-h_{ 00}(t)\right)\right\|_{{H^q(T^1)}}
    +\left\|t^{-(k^2-1)/2-\widetilde{\muGcomp1}}{t\partial_t}\left(g_{00}(t)-h_{ 00}(t)\right)\right\|_{{H^q(T^1)}}
    \\
  +&\left\|t^{-(k^2-1)/2-\widetilde{\muGcomp2}}\left(g_{11}(t)-h_{ 11}(t)\right)\right\|_{{H^q(T^1)}}
    +\left\|t^{-(k^2-1)/2-\widetilde{\muGcomp2}}{t\partial_t}\left(g_{11}(t)-h_{ 11}(t)\right)\right\|_{{H^q(T^1)}}
    \\
  +&\left\|t^{-(k^2-1)/2-\widetilde{\muGcomp3}}\left(g_{01}(t)-h_{ 01}(t)\right)\right\|_{{H^q(T^1)}}
    +\left\|t^{-(k^2-1)/2-\widetilde{\muGcomp3}}{t\partial_t}\left(g_{01}(t)-h_{ 01}(t)\right)\right\|_{{H^q(T^1)}}
    \\    
    +&\left\|t^{-1-\widetilde{\muGcomp4}}\left(R_g(t)-R_h(t)\right)\right\|_{{H^q(T^1)}}
    +\left\|t^{-1-\widetilde{\muGcomp4}}{t\partial_t}\left(R_g(t)-R_h(t)\right)\right\|_{{H^q(T^1)}} \\
    +&\left\|t^{k-\widetilde{\muGcomp5}}\left(E_g(t)-E_h(t)\right)\right\|_{{H^q(T^1)}}
    +\left\|t^{k-\widetilde{\muGcomp5}}{t\partial_t}\left(E_g(t)-E_h(t)\right)\right\|_{{H^q(T^1)}}\\
    +&\left\|t^{-2k-\widetilde{\muGcomp6}}\left(Q_g(t)-Q_h(t)\right)\right\|_{{H^q(T^1)}}
    +\left\|t^{-2k-\widetilde{\muGcomp6}}{t\partial_t}\left(Q_g(t)-Q_h(t)\right)\right\|_{{H^q(T^1)}}  
    \le C_q.
\end{align*}
Correspondingly, we say that two fluid
vectors $(v^0,v^1)$ and $(\tilde v^0,\tilde v^1)$ of the form \eqref{eq:Gowdysymmfluid}
\keyword{agree at order 
$(\muFcomp1,\muFcomp2)$} at $t\searrow 0$ provided that for any smooth exponents $\widetilde{\muFcomp1}<\muFcomp1$ and $\widetilde{\muFcomp2}<\muFcomp2$ and for each sufficiently large integer $q$,
there exists a constant $C_q>0$ such that
\be
  \label{eq:shortestimatefluid2}
  \left\|t^{-\Gamma-\widetilde{\muFcomp1}}\left(v^0 (t)-\tilde v^0 (t)\right)\right\|_{{H^q(T^1)}}
  +\left\|t^{-2\Gamma-\widetilde{\muFcomp2}}\left(v^1 (t)-\tilde v^1 (t)\right)\right\|_{{H^q(T^1)}}
  \le C_q.
\ee

\begin{theorem}[Sub-critical regime for self-gravitating fluid flows. Asymptotic properties]
\label{th:sub-criticalcoupledEstimates}
The solutions to the Einstein-Euler equations constructed in
\Theoremref{th:sub-criticalcoupledExistence} satisfy the following
properties as one approaches $t=0$:  
\begin{enumerate}[label=\textit{(\Roman{*})}, ref=(\Roman{*})] 

\item \label{statement:singular}  \textbf{Cosmological singularity}:  The metric is singular in the sense that for any sufficiently small $\epsilon>0$ and sufficiently large integer $q$ there exists a constant $C_q>0$ such that, for all $t\in (0,\delta]$, 
\[
\left\|t^{\frac{2\gamma}{2-\gamma}(1-\Gamma)}\mathrm{Ric}^2(t)-\frac{1+3(\gamma-1)^2}{(\gamma -1)^2}{\left((v^0_*)^2 \Lambda_*\right)^{\frac{2\gamma }{2-2 \gamma
   }}}\right\|_{H^q(T^1)}\le C_q t^\epsilon
\]
with $\mathrm{Ric}^2:=R_{\alpha\beta}R^{\alpha\beta}$. Due to the identity $\mathrm{Ric}^2=(1+3(\gamma-1)^2)\rho^2$ implied by Einstein's equations, 
the fluid energy density $\rho$ also blows-up.

\item \label{statement:improvedshift} \textbf{Improved decay of the shift}: For any sufficiently small $\epsilon>0$ and sufficiently large integer $q$ there exists a constant $C_q>0$ such that, 
for the ``shift'' $g_{01}$ for all $t\in (0,\delta]$, 
\[
\left\|t^{-(k^2+1)/2}g_{01}\right\|_{H^q(T^1)}+\left\|t^{-(k^2+1)/2}Dg_{01}\right\|_{H^q(T^1)}\le C_q t^\epsilon.
\]

\item \label{statement:coupledsub2} 
\textbf{Velocity term dominance}: Consider the ``truncated Einstein-Euler evolution equations'' in the gauge  \eqref{eq:asympwavegauge}; these are obtained from \Eqsref {eq:AAA1}, \eqref{momentum3} and \eqref{eq:wgEinstEqs} with \Eqsref{eq:asympwavegauge} and \eqref{eq:choiceconstraintmultiples} by dropping all $x$-derivatives of the metric and the fluid variables.
  These equations admit a solution $(g_\truncated, v_\truncated)$ in the form stated in  \Defref{def:blockdiagonalcoords} and \Eqref{eq:Gowdysymmfluid} 
such that $g$ and $g_\truncated$ agree at order 
$(\sigma,\sigma,\sigma,\sigma,\sigma,\sigma)$ 
and $(v^0,v^1)$ and $(v^0_\truncated,v^1_\truncated)$ agree at order
$(\sigma,\max\{0,\sigma-\Gamma\})$ for $\sigma=\min\{1,2(1-k)\}$ in  the limit $t\searrow 0$.
\item \label{statement:coupledsub3} 
\textbf{Matter matters at higher order}: 
  There exists a solution $g_\vacuum$ of the vacuum Einstein evolution equations in the form of \Defref{def:blockdiagonalcoords} in the gauge given by \Eqref{eq:asympwavegauge} (i.e., \Eqref{eq:wgEinstEqs} with $T_{\alpha\beta}=0$, \Eqsref{eq:asympwavegauge} and \eqref{eq:choiceconstraintmultiples})
such that $g$ and $g_\vacuum$ agree at order 
$(1-\Gamma,1-\Gamma,1-\Gamma,1-\Gamma,1-\Gamma,\min\{1-\Gamma,2(1-k)\})$ in  the limit $t\searrow 0$.
\end{enumerate}
\end{theorem}

\Sectionref{sec:proofsN} is devoted to the proofs of both theorems. Let us proceed with some remarks.
We point out that our method of proof here introduces new ideas which, for instance, would make it possible to circumvent some cumbersome arguments which have been necessary to cover the full interval $(0,1)$ for $k$ in earlier treatments of (vacuum) Gowdy solutions with the Fuchsian method in the non-analytic setting \cite{Rendall:2000ki,Stahl:2002bv}.
Observe that the restriction to the sub-critical case $\Gamma>0$ follows from the restrictions $k\in(0,1)$ and $\gamma\in(1,2)$ (see \Figref{fig:criticalGamma}). The critical and super-critical cases $\Gamma\le 0$ are only possible if $|k|\ge 1$. In the same way as in vacuum, this however turns out to be possible only in the (half-)polarized case, i.e., when $Q_*=\text{const}$. The critical case considered in the next subsection is therefore restricted to the half-polarized case.  
 It is interesting to observe in \Sectionref{sec:proofsN} that the estimates in the proof break down in the limit $\Gamma\searrow 0$. Roughly speaking, any series expansion of the unknown variables formally break down in the limit $\Gamma\searrow 0$ as both the powers \textit{and} the coefficients of all terms simultaneously approach $0$. The critical case $\Gamma=0$ therefore has to be treated separately, which we do in the next subsection.  

The point of \Theoremref{th:sub-criticalcoupledExistence} is to establish the existence of singular solutions of the Einstein-Euler equations which are determined by (up to certain constraints) free data with the same degrees of freedom as for the Cauchy problem.
In \Sectionref{sec:proofsN} we find detailed estimates for the exponents  $\muGcomp1$, \ldots, $\muGcomp6$, $\muFcomp1$ and $\muFcomp2$. 
These estimates give us a more detailed description of the behavior of the solution in the limit $t\searrow0$, and also give rise to a non-trivial uniqueness statement for this singular initial value problem. For the sake of brevity we omit such details from \Theoremref{th:sub-criticalcoupledExistence}.
We emphasize the fact that the fluid data $\hat v^1_*$ in \Eqref{eq:fluidestimate} is not prescribed freely, but is instead given by \Eqref{eq:v1constr} in terms of another free function $v^1_*$.  
In \Sectionref{sec:constraints} we discuss the origin of this.

An interesting consequence of \eqref{eq:LambdaSconstr} is that spatially homogeneous solutions of \Theoremref{th:sub-criticalcoupledExistence}, namely solutions  where the components of the metric and the fluid only depend on $t$, only exist if the fluid $4$-velocity is orthogonal to the symmetry hypersurfaces. This is consistent with the remark in \Sectionref{sec:FluidsKasnerNN} that it is a consequence of Einstein's equations that Gowdy symmetry restricts the fluid to flow only into non-symmetry directions. If all spatial directions are symmetries, as in the spatially homogeneous case, then the fluid is not allowed to flow at all.

Let us recall that the block diagonal coordinates in the gauge \eqref{eq:asympwavegauge} are in general neither areal nor conformal coordinates \eqref{eq:arealcond}--\eqref{eq:confcond} unless we are in the vacuum case. In particular, the shift quantity $g_{01}$ does not vanish except in vacuum. The evolution equations in our gauge are significantly more complicated than the ones in areal or conformal coordinates and hence are significantly harder to analyze.

Let us now consider \Theoremref{th:sub-criticalcoupledEstimates}. Regarding statement~\ref{statement:singular} it is interesting to recall our comment after \Defref{def:asymptlocKasner}. Namely, the fact that the solution metric is asymptotically local Kasner, as asserted by \Theoremref{th:sub-criticalcoupledExistence}, is in general not sufficient to make a statement about the curvature tensor. It is necessary for this to derive estimates for second-order time derivatives of the metric components first. Indeed, such estimates follow almost directly from the evolution equations and the Fuchsian theory. 
We mention without proof that in the half-polarized case $Q_*=\textrm{const}$ we can choose $k$ to be an arbitrary positive function and that the same estimates regarding the blow up of the fluid density and curvature hold. In particular, the curvature blows up even when  $k=1$ which is not the case in vacuum. 

Part~\ref{statement:improvedshift} of \Theoremref{th:sub-criticalcoupledEstimates} yields a significantly more detailed description of the shift $g_{01}$ than  the asymptotically local Kasner property asserted by \Theoremref{th:sub-criticalcoupledExistence}. Recall that the latter states that $g_{01}\sim t^{(k^2-1)/2+\tau}$ for some $\tau>0$ while the former states that $g_{01}\sim t^{(k^2+1)/2+\tau}$. In the proofs in \Sectionref{sec:proofsN} we find an interesting technical relationship between the decay of the shift and the dynamics of the constraint propagation terms $\mathcal D_\alpha$. In fact, if the asymptotic constraints of \Theoremref{th:sub-criticalcoupledExistence} are violated then Part~\ref{statement:improvedshift} of \Theoremref{th:sub-criticalcoupledEstimates} does in general not hold. This relationship was discovered first in \cite{Ames:2016uy}. 

The content of statement~\ref{statement:coupledsub2} of \Theoremref{th:sub-criticalcoupledEstimates} is that all our solutions demonstrate velocity term dominance. Hence, they can be approximated by solutions of the truncated equations as discussed in \Sectionref{sec:VTDMDNM}. In addition to this very fact, our theorem provides an estimate for the ``truncation error'' in terms of the exponents provided in statement~\ref{statement:coupledsub2}. It is interesting that this truncation error is large the closer $k$ is to $1$. In the proof we observe that the most significant contributions to this error can come from the leading term of the quantity $Q$, i.e., the data $Q_*$. If this is constant, i.e., in the (half)-polarized case, then the quantity $\sigma$ in the theorem can be normalized to unit,  and hence the truncation error can be much smaller.

Of particular interest now is statement~\ref{statement:coupledsub3}. According to this, ``matter does not matter'' at the singularity as discussed in \Sectionref{sec:VTDMDNM}. The purpose of our result is to give a qualitative estimate which we rephrase as ``matter matters at higher order''. 
It is interesting to observe that the restriction $\gamma<2$, which implies $\Gamma<1$, is crucial because our result  suggests that ``matter \textit{matters at leading order}'' if $\gamma=2$ and hence $\Gamma=1$. In fact, this case of a stiff fluid (equivalent to a linear massless scalar field), which has been considered in ground-breaking works \cite{Andersson:2001fa,Rodnianski:2014tk}, has significantly different asymptotics. An interesting aspect of statement~\ref{statement:coupledsub3} is that $g_\vacuum$ is only assumed to be a solution of the vacuum \textit{evolution} equations and hence may in general violate the constraints. In fact it is easy to see that if any solution of the {fully coupled} Einstein-Euler equations is supposed to agree with a solution of the vacuum equations in the above sense then they must both be asymptotically local Kasner with respect to the same data for the metric. However, it is not possible that both asymptotic constraint equations, first, \Eqref{eq:LambdaSconstr} for the Einstein-Euler metric and, second, the corresponding equation for the vacuum metric which is obtained from \Eqref{eq:LambdaSconstr} by deleting the last term, are satisfied for the same data unless $v^1_*=0$. In particular the function $\Lambda_*$ in the Einstein-Euler case can in general not match the function $\Lambda_*$ in the vacuum case at \textit{every} spatial point. We can only match them at one single point unless the vacuum solution violates the constraints. 


Statements~\ref{statement:coupledsub2} and \ref{statement:coupledsub3} also allow us to consider the relative significance of the ``velocity term dominance'' and the ``matter does not matter'' properties. 
Our results suggest that if $1-\Gamma<2(1-k)$, then the solution metric of the full set of equations agrees better with the solution of the truncated equations than with the solution of the vacuum equations in the limit $t\searrow 0$, i.e., ``matter is less negligible than spatial derivatives''. For example, this is the case when $k$ is close to zero and $\Gamma$ is close to $1$, i.e., when $\gamma$ is close to $2$. If $1-\Gamma>2(1-k)$ on the other hand, then they agree at the same order. So in short, we could say that ``matter is never more negligible than spatial derivatives''.


\subsection{The critical regime \texorpdfstring{$\Gamma=0$}{(Gamma=0)}}
\label{sec:coupledpolcritical}

In this section now, we consider the case of self-gravitating critical fluids. Recall that $\Gamma=0$ implies that $k$ must have the constant value 
\be
  \label{eq:kcrit}
  k=\sqrt{\frac{3 \gamma-2}{2-\gamma}},
\ee
 which is always larger or equal $1$. In the coupled Einstein-Euler case now this makes the {(half-)}polarized condition $Q_*=\text{const}$ necessary for us.

\begin{theorem}[Compressible perfect fluids in Gowdy-symmetric spacetimes near the cosmological singularity. Critical self-gravitating fluid flow and existence statement]
\label{th:criticalcoupledExistence}
Choose  fluid data
$v_*^0>0$ and $v_*^1$  in $C^\infty(T^1)$ with
\be
  \label{eq:fluidtimelikecoupled}
  v_*^0>|v_*^1|,
\ee
an equation of state parameter $\gamma\in (1,2)$, and
spacetime data $Q_{**},\Lambda_*>0$ in $C^\infty(T^1)$ and constants $Q_*\in\R$ and $E_{**}>0$,
such that
\begin{align}
  \label{eq:ESconstr}
  E_*(x):=\frac{E_{**}}{(\Lambda_*(x))^{1/k}} e^{-\frac{2\gamma}{k(\gamma -1)}\int_0^{x}\Bigl(
    v^0_*(\xi) v^1_*(\xi) \left((v^0_*)^2(\xi)-(v^1_*)^2(\xi)\right)^{\frac{2-3\gamma}{2(\gamma-1)}}\left(\Lambda_*(\xi)\right) ^{-\frac{2-\gamma}{2(\gamma-1)}}\Bigr)d\xi}
\end{align}
is a function in $C^\infty(T^1)$ with $k$ given by \Eqref{eq:kcrit}
Then, there exists a constant $\delta>0$ and a solution to the Einstein-Euler equations (\Eqsref {eq:AAA1}, \eqref{momentum3}, \eqref{eq:EFE}) in the gauge given by the gauge source functions \Eqref{eq:asympwavegauge} which, for some choice of positive exponents $\muGcomp1$, \ldots, $\muGcomp6$, $\muFcomp1$ and $\muFcomp2$, is determined by the following conditions:
\begin{enumerate}[label=\textit{(\roman{*})}, ref=(\roman{*})] 

\item The metric admits the form \Defref{def:blockdiagonalcoords} for some functions $g_{00}$, $g_{01}$, $g_{11}$, $R$, $E$, and $Q$ in $C^\infty((0,\delta]\times T^1)$, and is an asymptotically local Kasner spacetime with respect to data $k$, $\Lambda_*$, $E_*$, $Q_*$, $Q_{**}$
  and exponents $\muGcomp1$, \ldots, $\muGcomp6$.

\item The fluid flow has the form \eqref{eq:Gowdysymmfluid} for some $v^0, v^1$ in $C^\infty((0,\delta]\times T^1)$, and for any (sufficiently large) integer $q$ 
there exists a constant $C_q>0$ such that these functions satisfy 
for all $t\in (0,\delta]$
\be
  \left\|t^{-\muFcomp1}\left(v^0 (t) -v^0_*\right)\right\|_{{H^q(T^1)}}
  +\left\|t^{-\muFcomp2}\left(v^1 (t) -\hat v^1_*\right)\right\|_{{H^q(T^1)}}
  \le C_q.
\ee
\end{enumerate}
\end{theorem}

\begin{theorem}[Compressible perfect fluids in Gowdy-symmetric spacetimes near the cosmological singularity. Critical self-gravitating fluid flow and asymptotic properties]
\label{th:criticalcoupledEstimates}
The solutions to the Einstein-Euler equations constructed in \Theoremref{th:criticalcoupledExistence} satisfy the following properties as one approaches the cosmological singularity $t=0$:  
\begin{enumerate}[label=\textit{(\Roman{*})}, ref=(\Roman{*})] 

\item \label{statement:singularcrit}  
\textbf{Cosmological singularity at $t=0$}:  The metric is singular in the sense that for any sufficiently small $\epsilon>0$ and sufficiently large integer $q$ there exists a constant $C_q>0$ such that
\[
\left\|t^{\frac{2\gamma}{2-\gamma}(1-\Gamma)}\mathrm{Ric}^2(t)-\frac{1+3(\gamma-1)^2}{(\gamma -1)^2}{\left(\left((v^0_*)^2-(v^1_*)^2\right) \Lambda_*\right)^{\frac{2\gamma }{2-2 \gamma
   }}}\right\|_{H^q(T^1)}
\le C_q t^\epsilon\]
for $\mathrm{Ric}^2:=R_{\alpha\beta}R^{\alpha\beta}$ for all $t\in (0,\delta]$.
Due to the identity $\mathrm{Ric}^2=(1+3(\gamma-1)^2)\rho^2$ implied by Einstein's equations, a corresponding blow-up result holds for the fluid energy density $\rho$.

\item \textbf{Improved decay of the shift}: For any sufficiently small $\epsilon>0$ and sufficiently large integer $q$ there exists a constant $C_q>0$ such that, 
for the ``shift'' $g_{01}$ and for all $t\in (0,\delta]$, 
\[
\left\|t^{-(k^2+1)/2}g_{01}\right\|_{H^q(T^1)}+\left\|t^{-(k^2+1)/2}Dg_{01}\right\|_{H^q(T^1)}\le C_q t^\epsilon.
\]

\item \label{statement:coupledsub2crit} 
\textbf{Velocity term dominance}: 
  Consider the ``truncated Einstein-Euler evolution equations'' in the gauge  \eqref{eq:asympwavegauge}; these are obtained from \Eqsref {eq:AAA1}, \eqref{momentum3} and \eqref{eq:wgEinstEqs} with \Eqsref{eq:asympwavegauge} and \eqref{eq:choiceconstraintmultiples} by dropping all $x$-derivatives of the metric and the fluid variables.
  These equations admit a solution $(g_\truncated, v_\truncated)$ in the form stated in \Defref{def:blockdiagonalcoords} and \Eqref{eq:Gowdysymmfluid} 
such that $g$ and $g_\truncated$ agree at order 
$(\sigma,\sigma,\sigma,\sigma,\sigma,\sigma)$ 
and $(v^0,v^1)$ and $(v^0_\truncated,v^1_\truncated)$ agree at order 
$(\sigma,\sigma)$ for $\sigma=\min\{1,2(1-k)\}$.

\item \label{statement:coupledsub3crit} 
\textbf{Matter matters at higher order}: 
  There exists a solution $g_\vacuum$ of the vacuum Einstein evolution equations in the form of \Defref{def:blockdiagonalcoords} in the gauge given by \Eqref{eq:asympwavegauge} (\Eqref{eq:wgEinstEqs} with $T_{\alpha\beta}=0$, \Eqsref{eq:asympwavegauge} and \eqref{eq:choiceconstraintmultiples})
such that $g$ and $g_\vacuum$ agree at order 
$(1,1,1,1,1,\min\{1,2(1-k)\})$. 
\end{enumerate}
\end{theorem}

As we explain briefly in \Sectionref{sec:proofsN} the proofs of these theorems are significantly simpler than the proofs of the theorems in \Sectionref{sec:coupledfluidssuper} mainly due to the (half-)polarization restriction.
Most of the remarks regarding the previous theorems  also apply here. Note however that the free data for the fluid are chosen differently and, in particular, the asymptotic constraint \Eqref{eq:ESconstr} is considered as an equation for the data $E_*$ here (while \Eqref{eq:LambdaSconstr} is an equation for $\Lambda_*$). 
In particular the free fluid data $v^0_*$ and $v^1_*$ determine the
leading-order of the fluid variables directly, in contrast to
\Theoremref{th:sub-criticalcoupledExistence}. Another difference is
the occurrence of the \keyword{timelike condition}
\Eqref{eq:fluidtimelikecoupled}.


\subsection{The super-critical regime \texorpdfstring{$\Gamma<0$}{(Gamma<0)}}
\label{sec:fixedbhgammaneg}

Let
us finally discuss the \textit{super-critical case} $\Gamma<0$ where the local sound speed of the solution is too small to compete with the gravitational dynamics at the singularity. 
It turns out that this problem  cannot be solved completely with our method and we want to use this subsection to explain the technical reason for this in the case of a fluid of the form \eqref{eq:Gowdysymmfluid} on a fixed \textit{exact} Kasner background \Eqref{Kasner-k}. 
The Euler equations take the form
\[S^0 {t\partial_t}
\begin{pmatrix}
  v^0\\
  v^1
\end{pmatrix}
+S^1 t\partial_x
\begin{pmatrix}
  v^0\\
  v^1
\end{pmatrix}
=f
\]
with
\begin{align}
  \label{eq:KasnerS0}
  S^0&=\begin{pmatrix}
 {v^0} \left((v^0)^2+3 (\gamma -1) (v^1)^2\right) &
   {v^1} \left((1-2 \gamma ) (v^0)^2-(\gamma -1)
   (v^1)^2\right) \\
 {v^1} \left((1-2 \gamma ) (v^0)^2-(\gamma -1)
   (v^1)^2\right) & {v^0} \left((\gamma -1)
   (v^0)^2+(2 \gamma -1) (v^1)^2\right)
\end{pmatrix},\\
\label{eq:KasnerS1}
S^1&=\begin{pmatrix}
 {v^1} \left((2 \gamma -1) (v^0)^2+(\gamma -1)
   (v^1)^2\right) & {v^0} \left((1-\gamma )
   (v^0)^2+(1-2 \gamma ) (v^1)^2\right) \\
 {v^0} \left((1-\gamma ) (v^0)^2+(1-2 \gamma )
   (v^1)^2\right) & {v^1} \left(3 (\gamma -1)
   (v^0)^2+(v^1)^2\right)
\end{pmatrix},\\
f&=\left(\Gamma (v^0)^2
   \left((v^0)^2-(v^1)^2\right),-\Gamma  {v^0}
   {v^1} \left((v^0)^2-(v^1)^2\right)\right)^T.
\end{align}
In order to construct fluid solutions with the leading-order behavior given by \Eqref{eq:nonspecialcase} for $\Gamma<0$ %
for arbitrary smooth data $v_*>0$ and $v_{**}$ with the Fuchsian theory (outlined in detail in \Sectionref{sec:SIVPNN}), it is an important condition that the matrix $S^0$ in \Eqref{eq:KasnerS0} is uniformly positive definite in the limit $t\searrow 0$ (possibly after a multiplication of the whole system with some power of $t$) when it is evaluated on fluid vector fields with the above leading order behavior, and that the matrices $S^0$ and $S^1$ above are symmetric. 
Without going into technical details, we find for $\Gamma<0$ and $\gamma\in (1,2)$:
\[
S^0=(3 \gamma -2) (v_*)^3 t^{\frac{6 {\Gamma} }{2-\gamma }}
\Biggl(
\begin{pmatrix}
 1 &  -1 \\
 -1 & 1 
\end{pmatrix}
+
\begin{pmatrix}
 \frac{3 v_{**} \left(5 \gamma ^2-8 \gamma +4\right)}{2 {v_*} (\gamma -1) (3 \gamma -2)} &
   \frac{ v_{**} \left(-7 \gamma
   ^2+10 \gamma -4\right)}{{v_*} (\gamma -1) (3 \gamma -2)} \\
 \frac{ v_{**} \left(-7 \gamma ^2+10
   \gamma -4\right)}{{v_*} (\gamma -1) (3 \gamma -2)} & \frac{v_{**}
   \left(13 \gamma ^2-16 \gamma +4\right) }{2 {v_*} (\gamma -1) (3 \gamma -2)}
\end{pmatrix}t^{-\frac{2 {\Gamma} }{2-\gamma }}
\Biggr) +\ldots.
\]
When the Euler system is divided by $(3 \gamma -2) (v_*)^3 t^{\frac{6 {\Gamma} }{2-\gamma }}$, the eigenvalues of the matrix resulting from $S^0$ are $2+O(t^{-\frac{2 {\Gamma} }{2-\gamma }})$ and $O(t^{-\frac{4 {\Gamma} }{2-\gamma }})$ and hence the before mentioned uniform positivity condition is violated. We could attempt to compensate this by multiplying the system with a suitable time dependent matrix and thereby obtain a new, now uniformly positive matrix $S^0$. This, however, would destroy the symmetry of the matrix resulting from $S^1$. Due to this, the Fuchsian theory in \Sectionref{sec:SIVPNN} does not apply. It is an interesting question whether the reason for this is that our Fuchsian method is not ``good enough'' or whether there is an actual physical phenomenon which prevents general super-critical inhomogeneous solutions of the Euler equations from existing. Surprisingly, though, if we restrict ourselves to the restricted class of analytic data, it turns out to be possible to solve this singular initial value problem in the super-critical inhomogeneous case. This is the content of the following theorem.

\begin{figure}[t]
  \centering
  \includegraphics[width=0.6\textwidth]{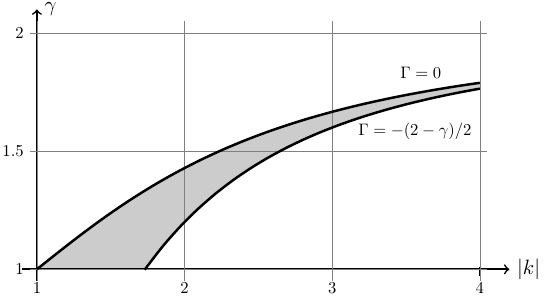}
\caption{The super-critical regime for inhomogeneous real-analytic data.}
\label{fig:super-criticalGamma}
\end{figure}

\begin{theorem}[Super-critical fluid flow on an (exact) Kasner spacetime for real-analytic data]
\label{th:super-criticalfixedbg}
Choose an equation of state parameter $\gamma\in (1,2)$ and a Kasner spacetime with parameter $k\in\R$ (recall \Eqref{Kasner-k}) such that
  \be
    \label{eq:negativelowerboundGamma}
    -\frac 12 (2-\gamma)<\Gamma<0.
  \ee
  Choose  fluid data $v_{*},v_{**}\in C^\omega(T^1)$ with $v_*(x)>0$ for all $x\in T^1$. Then 
for any exponent $\eta$ with
\be
  \label{eq:fluidineqNeg}
  0<\eta(x)<
\min\left\{1,-2\Gamma/(2-\gamma),(2(1+\Gamma)-\gamma)/(2-\gamma)\right\} \quad\text{for all $x\in T^1$,}
\ee
there exists some $\delta>0$ and a unique solution $v^\alpha$ of the form \eqref{eq:Gowdysymmfluid} of the Euler equations
with
\begin{equation*}
  \begin{split}
  v^0(t,x)&=v_*(x) t^{-2|\Gamma|/(2-\gamma)}+v_{**}(x)\frac{\gamma}{2(\gamma-1)}+W_0(t,x),\\ v^1(t,x)&=\pm (v_*(x) t^{-2|\Gamma|/(2-\gamma)}+v_{**}(x))+W_1(t,x),
\end{split}
\end{equation*}
for some remainders $W_0$, $W_1$ in $X_{\delta,\eta,\infty}$ which are continuous with respect to $t$ and real-analytic with respect to $x$ on $(0,\delta]\times T^1$.
\end{theorem}
We will not discuss the proof of this theorem and only mention that it is a direct application of Theorem~1 in \cite{Isenberg:1999ba}. The spaces $X_{\delta,\eta,\infty}$ are introduced in \Sectionref{sec:SIVPNN}. A particularly surprising outcome is that this existence result is subject to a  lower bound for $\Gamma$.
When this inequality  is violated, we find that the spatial derivative terms, i.e., the terms multiplied by $S^1$ in the Euler equations, cannot be neglected in leading order anymore and hence the assumption of velocity term dominance breaks down.  For inhomogeneous fluids, the super-critical case therefore applies only in the shaded region of \Figref{fig:super-criticalGamma}.


\section{Quasilinear symmetric hyperbolic Fuchsian systems}
\label{sec:SIVPNN}

A brief outline of the Fuchsian theory (which we will use in \Sectionref{sec:proofsN})
is now presented. Further details are available at \cite{Beyer:2010foa} which was later extended in  \cite{Ames:2016uy}.

\paragraph{Time-weighted Sobolev spaces.} In order to measure the regularity and the decay of certain kinds of functions near the ``singular time'' $t=0$, we introduce a family of time-weighted Sobolev spaces. Letting  $\mu: T^n \to \mathbb R^d$ be any smooth
 function, we define the $d\times d$-matrix 
\be
\label{eq:rmatrix}
\RR{\mu}(t,x) := \diag\left( t^{-\mu_1(x)},\ldots,t^{-\mu_d(x)} \right).
\ee
For functions
 $w: (0, \delta] \times T^n \rightarrow \R^d$ in $C^\infty((0,\delta]\times T^n)$ we set
\be
\label{eq:defnorm}
||w||_{\delta, \mu, q} := \sup_{t \in (0,\delta]} ||\RR{\mu} w ||_{H^q(T^n)},
\ee
whenever this expression is finite.
Here $H^q(T^n)$ denotes the usual Sobolev space of order $q$ on the $n$-torus $T^n$.
Based on this, we define $X_{\delta, \mu, q}$ to be the completion of the set of all smooth functions $w:(0, \delta] \times T^n \rightarrow \R^d$ 
for which \Eqref{eq:defnorm} is finite. Equipped with the norm \Eqref{eq:defnorm}, $X_{\delta, \mu, q}$ is therefore a Banach space. A closed ball of radius $r$ about $0$ in $X_{\delta, \mu, q}$  is denoted by $B_{\delta, \mu, q,r}$. 
To handle functions which are infinitely differentiable 
we also define $X_{\delta, \mu, \infty} := \cap_{q=0}^\infty X_{\delta, \mu, q}$.

A \keyword{function operator}  will be a map which assigns to any function $(0,\delta]\times T^n\rightarrow\R^d$ in some class a function $(0,\delta]\times T^n\rightarrow\R^m$ in some possibly different class. For all of what follows, $d$ and $m$ are positive integers. 
For our purposes we require precise control of the domain and range of our function operator.

\begin{definition}
  \label{def:functionoperatorsNN}
  Fix some positive integers $n$, $d$, $m$ and $q>n/2$. For any real number $s_0>0$ or $s_0=\infty$, set
\be
  \label{eq:Hspace}
  H_{\delta,q,s_0}:=
\Big\{
\text{$w: (0,\delta]\times T^n\rightarrow\R^d$ in $X_{\delta,0,q}$}\,\, \Bigl| \sup_{t\in
       (0,\delta]} \|w(t)\|_{L^\infty(T^n)}\le s_0
\Big\}
 \ee
and let $\nu$ be an exponent $m$-vector. 
A map $w\mapsto F(w)$ is called a \mbox{$(0,\nu,q)$-operator} provided:
 \begin{enumerate}[label=\textit{(\roman{*})}, ref=(\roman{*})] 
   \item \label{cond:functionoperatorsNN1} There exists a constant
     $s_0>0$ ($s_0=\infty$ is allowed) such that for each $\delta'\in (0,\delta]$ and $w\in
     H_{\delta',q,s_0}$, the image $F(w)$ is a well-defined function $(0,\delta']\times T^n\rightarrow\R^m$ in $X_{\delta',\nu,q}$.
  \item \label{cond:functionoperatorsNN2} For each $\delta'\in
    (0,\delta]$ and $q'=q,q-1$, there exists a constant $C_q>0$ such that the following local Lipschitz estimate holds for all $w,\tilde w\in   H_{\delta',q,s_0}$
\be
  \label{eq:Lipschitzfunctionoperator}
  \|F(w)-F(\tilde w)\|_{\delta',\nu,q'}\le C_q \left(1+\|w\|_{\delta',0,q'}+\|\tilde w\|_{\delta',0,q'}\right)\|w-\tilde w\|_{\delta',0,q'}.
\ee
  \end{enumerate}
  Now, let $\mu$ be an exponent $d$-vector. We call a map $w\mapsto
  F(w)$ a \mbox{$(\mu,\nu,q)$-operator} if the map $w\mapsto F(\RR{-\mu} w)$ is
 a $(0,\nu,q)$-operator.
 We call the map $w\mapsto F(w)$ a $(\mu,\nu, + \infty)$-operator if
 $w\mapsto F(w)$ is a \mbox{$(\mu,\nu,q)$-operator} for each $q\ge p$
 where $p$ is some integer with $p>n/2$ where the constant $s_0$ is supposed to be the same for all $q\ge p$. 
\end{definition}

 In the ``smooth case'' $q=\infty$, we do \textit{not} make any assumption about the dependence of the constant $C_q$ in \Conditionref{cond:functionoperatorsNN2} on $q$. Moreover, while we formally restrict $s_0$ to be the same for all $q$ in this case in \Defref{def:functionoperatorsNN}, this is not actually a restriction since $s_0$ is only a bound on the $L^\infty$-norm. In practice, the ``source'' exponent $\mu$ and the differentiability index $q$ are often clear from the context. Then we use the following simplified terminology: A \keyword{function operator $w\mapsto F(w)$ is $o(1)$} if there exists an exponent $\nu>0$ such that $F$ is a $(\mu,\nu, + \infty)$-operator.

Let us finally discuss a particularly important family of function operators which are induced by special functions $g$. First suppose that $m=1$ and that the function $g(t,x,u)$ is a polynomial with respect to the third argument where each coefficient function is of the type $(0,\delta]\times T^n\rightarrow\R$ in $X_{\delta,\nu,\infty}$ for some exponent scalar $\nu$. The induced function operator $w\mapsto g(w)$ given by $g(w)(t,x):=g(t,x,w(t,x))$ is called a \keyword{scalar polynomial function operator}. If $m>1$ and each component of $g$ induces a scalar polynomial function operator, then the induced function operator is called \keyword{vector (or matrix) polynomial function operator}.
Next suppose that $h_0$ is a scalar-valued function in
$X_{\delta,\eta,\infty}$ for some scalar exponent $\eta$ such that $1/h_0\in
X_{\delta,-\eta,\infty}$. Let $w\mapsto g_1(w)$ and $w\mapsto g_2(w)$ be two scalar polynomial
function operators and assume that $w\mapsto g_2(w)$ is a $(\mu,\zeta, + \infty)$-operator
for a scalar exponent $\zeta>0$. Then, the operator 
\be
  \label{eq:scalarrationalFOPs}
  w\mapsto h(w):={g_1(w)}/{ ((1+g_2(w)) h_0)}
\ee
is called a \keyword{scalar rational function operator}. Analogously we define \keyword{vector (or matrix) rational function operators}. Finally let us consider any constant $\gamma\in\R$ and set $g(t,x,u)=(1+u)^\gamma$. In this paper, we use the term \keyword{special function operator}\label{kw:special} to collectively refer to function operator induced by this function as well as to any  polynomial and rational function operator. It turns out that this class of function operators covers all function operators in this paper.

\paragraph{Quasilinear symmetric hyperbolic Fuchsian systems.} Let us now be specific about the most general class of equations for which our theory applies. 
Consider systems of quasilinear PDEs for the unknown $u: (0, \delta] \times T^n \to \mathbb R^d$: 
\be
  \label{eq:pde}
 \Stna(t,x,u(t,x)) Du(t,x) + \sum_{a=1}^n \Ssna(t,x,u(t,x)) t \partial_a u(t,x)  + \Nna(t,x,u(t,x)) u(t,x)  
= \fna(t,x,u(t,x)),
\ee
where each of the $n+1$ maps $\Stna,\ldots, S^n$ is a symmetric $d \times d$ matrix-valued function of the spacetime coordinates
$(t,x)\in (0, \delta] \times T^n$
 and of the unknown $u$, while
$\fna=\fna(t,x,u)$ is a prescribed $\R^d$--valued function of
$(t,x,u)$, and 
$\Nna$ is a $d\times d$-matrix-valued function of $(t,x,u)$. 
We set
\footnote{In all of what follows, indices $i,j,\ldots$ run over $0,1,\ldots,n$, while indices $a,b,\ldots$ take the values $1,\ldots,n$.}
$D := t \,  \partial_t$. 
At this point the reader may wonder why the term $\Nna(t,x,u) u$ is included in the principal part and not in the source $\fna(t,x,u)$. We leave these terms separate since, later on, $\fna(t,x,u)$ is considered as terms of ``higher order'' in $t$ at $t=0$ while the term $\Nna(t,x,u) u$  contains terms of the same order as the other terms in the ``principal part'' (see below) in $t$.
We list the precise requirements for $\Sna$, $\Nna$ and $\fna$ below.

\begin{definition}[(Special) quasilinear symmetric hyperbolic Fuchsian systems]
\label{def:quasilinearlimitN}
The PDE system of the type \Eqref{eq:pde} is called a \keyword{quasilinear symmetric hyperbolic Fuchsian system} around a specified smooth leading-order term 
$u_*: (0,\delta]\times T^n\rightarrow\R^d$
for parameters 
$\delta>0$ and a specified exponent $\mu$  
if there exists a positive-definite and symmetric matrix-valued function $\StLu\in C^{\infty}(T^n)$ 
and a matrix-valued function $\NLu \in C^{\infty}(T^n)$, 
such that 
all following function operators obtained from \Eqref{eq:pde} are $o(1)$:
\begin{align}
  \label{Ncond}
  w&\mapsto N(u_*+w)-\NLu,\\
  \label{S01}
  w&\mapsto \StHu{w} := \St{u_* + w} - \StLu,\\
  \label{Sa}
  w&\mapsto t\Ss{u_*+w},\\
  \begin{split}
  w&\mapsto \RR{\mu}\Fredu{w}\\
  &:=\RR{\mu}\left(\fna(u_*+w)-\St{u_* + w}Du_*-\sum_{a=1}^n\Ss{u_* + w} t\partial_x u_*-N(u_*+w) u_*\right), 
\end{split}
\end{align}
where by convention $\St{u_* + w}(t,x):=\Stna(t,x,u_*(t,x)+w(t,x))$, etc.
If all the function operators obtained from \Eqref{eq:pde}
are special (see \Sectionref{kw:special}), then the PDE system is labeled a \keyword{special quasilinear symmetric hyperbolic Fuchsian system}. 
\end{definition}

In order to formulate our Fuchsian theorem, we need to introduce some further technical concepts. Suppose that $M: (0,\delta]\times T^n\rightarrow\R^{d\times d}$ is any continuous $d\times d$-matrix-valued function.
  Suppose $\mu$ is some $d$-vector-valued exponent.
  A matrix-valued function $M$ is called \keyword{block diagonal with respect to $\mu$} provided 
\be
M(t,x)\RR{\mu}(t,x) - \RR{\mu}(t,x)M(t,x) = 0,
\ee
  for all $(t,x)\in(0,\delta]\times U$.
Let $\mu$ be a $d$-vector-valued exponent which is
\keyword{ordered}, i.e.,
\be
  \label{eq:orderedmu}
  \mu(x)=\Bigl(\underbrace{\mu^{(1)}(x),\ldots,\mu^{(1)}(x)}_{\text{$d_1$-times}},
  \underbrace{\mu^{(2)}(x),\ldots,\mu^{(2)}(x)}_{\text{$d_2$-times}},
  \ldots,
  \underbrace{\mu^{(l)}(x),\ldots,\mu^{(l)}(x)}_{\text{$d_l$-times}}\Bigr),
\ee
where $l\in\{1,\ldots,d\}$, $\mu^{(i)}\not=\mu^{(j)}$ for all $i\not=j\in\{1,\ldots,l\}$, and $d_1,\ldots,d_l$ are positive integers with $d_1+d_2+\ldots+d_l=d$.
It follows that any continuous $d\times d$-matrix-valued function $M$ is block
diagonal with respect to $\mu$ if and only if $M$ is of the form
  $M(t,x)=\mathrm{diag}\,\Bigl(M^{(1)}(t,x),\ldots,M^{(l)}(t,x)\Bigr)$,
where each $M^{(i)}(t,x)$ is a continuous $d_i\times
d_i$-matrix-valued function.

\begin{definition}
  \label{def:bdsystem}
  Choose any integer $q>n/2+2$ and a constant $\delta>0$.
  Suppose that $u_*$ is a given leading-order term and $\mu$ an
  exponent vector. The system (\ref{eq:pde}) is called  \keyword{block
    diagonal with respect to $\mu$} if, for each $u = u_*+w$ with $w\in
  X_{\delta,\mu,q}$ for which the following expressions are defined,
  the matrices $\Sts{u_*+w}$ and 
$\N{u_*+w}$ and all their spatial derivatives are block diagonal with respect to $\mu$.
\end{definition}

For all of the following we want to assume that the system \eqref{eq:pde} is  block
    diagonal with respect to $\mu$ and
  that $\mu$ is ordered. Hence all matrices in the principal part have the same block diagonal structure. In particular, the matrix
  \be
    \label{eq:defNODE}
    \NODE=\NODE(u_*):=\StLInv{u_*}\NLu
  \ee
  is block diagonal with respect to $\mu$; recall that by Definition \ref{def:quasilinearlimitN}
$\StLu$ is invertible.
Then we set
\be
  \label{eq:listoflambdas}
  \Lambda:=(\lambda_1,\ldots,\lambda_d)
\ee
as the vector of (possibly repeated) eigenvalues $\lambda_i$ of \NODE which is sorted by the blocks of \NODE.

\begin{theorem}
\label{th:smoothexistenceN}
Suppose that \Eqref{eq:pde} is a {quasilinear symmetric hyperbolic Fuchsian system} around $u_*$
with  the choice of the parameters $\delta$, $\mu$ as specified in
\Defref{def:quasilinearlimitN} and that  $\mu$ is ordered. Suppose that \Eqref{eq:pde}  is block
diagonal with respect to $\mu$ and that
\be
  \label{eq:eigenvaluecondition}
  \mu>-\mathrm{Re}\,\Lambda,
\ee
where $\Lambda$ is defined in \Eqref{eq:listoflambdas}. 
Then there exists a unique solution $u$  to \Eqref{eq:pde} with remainder $w:=u-u_*$ belonging to $X_{\widetilde\delta, \mu, \infty}$ for some $\widetilde \delta \in (0, \delta]$. Moreover, $w$ is differentiable with respect to $t$ and $Dw \in X_{\widetilde \delta, \mu, \infty}$.
\end{theorem}

The proof of this theorem has essentially been given in \cite{Ames:2012vz}; cf.\ Theorem~2.21 therein. The statement of the theorem therein significantly simplifies thanks to the restriction to \textit{special} function operators here. In fact, the additional technical requirements in the theorem in \cite{Ames:2012vz} hold for all members of this class of function operators. 


\section{Existence theory for self-gravitating fluids}
\label{sec:proofsN}

\subsection{First-order reduction of the Einstein-Euler system}
\label{sec:EinsteinEuler}

We now consider the Einstein evolution equations \eqref{eq:wgEinstEqs} with \Eqsref{eq:asympwavegauge}, \eqref{eq:choiceconstraintmultiples} and \eqref{momentum3}.
The function $k$ here is so far unspecified; later it will agree with the data $k$ in \Theoremref{th:sub-criticalcoupledExistence} in the sub-critical case, or the quantity $k$ in \Eqref{eq:kcrit} in the critical case, and, with the quantity $k$ in \Defref{def:asymptlocKasner}. 
These evolution equations are of the form 
\be
  \label{eq:wavepdeNGowdy}
      \sum_{\gamma,\delta=0}^1g^{\gamma \delta}\partial_{x^\gamma}\partial_{x^\delta}g_{{\alpha}{\beta}}= 2\hat H_{{\alpha}{\beta}},  
\ee
where
\be
\label{eq:wavepdeNGowdy2}
\hat H_{{\alpha}{\beta}}:=\nabla_{({\alpha}} \mathcal F_{{\beta} )} + g^{{\gamma}\delta} g^{{\epsilon} {\zeta}} 
  \left( \Gamma_{{\gamma}{\epsilon} {\alpha}} \Gamma_{\delta {\zeta} {\beta}} + 
         \Gamma_{{\gamma}{\epsilon} {\alpha}} \Gamma_{\delta {\beta} {\zeta}} +
         \Gamma_{{\gamma}{\epsilon} {\beta}} \Gamma_{\delta {\alpha} {\zeta}} 
  \right)
+ {C_{{\alpha}{\beta}}}^{{\gamma}} \mathcal D_{{\gamma}}-T_{\alpha\beta}+\frac 12 g_{\alpha\beta} T.
\ee
In consistency with \Defref{def:blockdiagonalcoords},
the unknown metric variables
in the parametrization given by \Eqref{eq:blockdiagonalcoords} 
are 
$g_{00}(t,x)$, $g_{11}(t,x)$, $g_{01}(t,x)$, $R(t,x)$, $E(t,x)$ and $Q(t,x)$.
The first step of our discussion 
is to convert our second-order evolution system \eqref{eq:wavepdeNGowdy}--\eqref{eq:wavepdeNGowdy2} to first-order symmetric hyperbolic form. To this end, we set
\be
  \label{eq:firstordervariablesG1}
    \UG:=(\UGcomp1,\ldots,\UGcomp6)^T,
  \ee
where, for each $i=1,\ldots,6$, we define
\be
  \UGcomp{i}:=(\UGcomp{i,-1},\UGcomp{i,0},
  \UGcomp{i,1})^T
\ee
with
\begin{align}
  \label{eq:firstordervariablesfirst}
  \UGcomp{1,-1}&:=g_{00},& \UGcomp{1,0}&:=Dg_{00}-\alpha g_{00},&
  \UGcomp{1,1}&:=t\partial_x g_{00},\\
  \UGcomp{2,-1}&:=g_{11},& \UGcomp{2,0}&:=Dg_{11}-\alpha g_{11},&
  \UGcomp{2,1}&:=t\partial_x g_{11},\\
  \UGcomp{3,-1}&:=g_{01},& \UGcomp{3,0}&:=Dg_{01}-\alpha g_{01},&
  \UGcomp{3,1}&:=t\partial_x g_{01},\\
  \UGcomp{4,-1}&:=R,& \UGcomp{4,0}&:=DR-\alpha R,&
  \UGcomp{4,1}&:=t\partial_x R,\\
  \UGcomp{5,-1}&:=E,& \UGcomp{5,0}&:=DE-\alpha E,&
  \UGcomp{5,1}&:=t\partial_x E,\\
  \UGcomp{6,-1}&:=Q-Q_*,& \UGcomp{6,0}&:=DQ-\alpha (Q-Q_*),&
  \label{eq:firstordervariablesLast}                                                           
  \UGcomp{6,1}&:=t\partial_x (Q-Q_*),  
\end{align}
with some constant $\alpha$ to be fixed later. $Q_*(x)$ is some (so far freely) specified  smooth function which will later be matched to the data in \Theoremref{th:sub-criticalcoupledExistence} and \Theoremref{th:criticalcoupledExistence}, respectively. 
\Eqsref{eq:wavepdeNGowdy}--\eqref{eq:wavepdeNGowdy2} imply the following first-order system for this vector $\UG$:
\be 
  \label{eq:firstordersystemGowdyN}
    \PPMatrixGnullcomp0 D\UG+\PPMatrixGnullcomp1 t\partial_{x} \UG 
    =f_{\gravity},
   \ee
 with
  \be
    \label{eq:block2fullGowdyN}
    \PPMatrixGnullcomp0:=\text{diag} (s^0,\ldots, s^0),\quad
    \PPMatrixGnullcomp1:=\text{diag} (s^1,\ldots, s^1),
  \ee
and
\be
  \label{eq:s1Gowdy}
  s^0
    :=
    \begin{pmatrix}
      1 & 0 & 0\\
      0 & 1 & 0\\
      0 & 0 & -\UGcomp{1,-1}/\UGcomp{2,-1}
    \end{pmatrix},
    \qquad   
  s^1
  :=
    \begin{pmatrix}
      0 & 0 & 0 \\
      0 & -2\UGcomp{3,-1}/\UGcomp{2,-1} & \UGcomp{1,-1}/\UGcomp{2,-1}\\
      0 & \UGcomp{1,-1}/\UGcomp{2,-1} & 0
    \end{pmatrix}.
\ee
The lengthy expression for $f_{\gravity}$ in \Eqref{eq:firstordersystemGowdyN} can be obtained explicitly from \Eqsref{eq:wavepdeNGowdy}--\eqref{eq:wavepdeNGowdy2}, but we refrain from writing it down here.

The Euler equations \eqref{eq:AAA1} are already in first-order form which we write symbolically as
\be 
  \label{eq:firstordersystemEulerN}
  \PPMatrixFnullcomp0 D\UF+\PPMatrixFnullcomp1 t\partial_{x} \UF
  =f_{\fluid},
\ee
with
$\UF:=(v^0,v^1)^T$. 
Again, the expression for $\PPMatrixFnullcomp0$, $\PPMatrixFnullcomp1$ and $f_{\fluid}$ can be derived explicitly.

For large parts of our discussion it is convenient to adopt the following operator notation: For any vectors $U$ and $\tilde U$ as above, we set
\be
  \label{eq:LG}
  \LG{{\tilde U}}{\UG}:=\PPMatrixGcomp0{\tilde U} D\UG
+\PPMatrixGcomp1{\tilde U} t\partial_x\UG +\NNG \UG
\ee
and
\be
  \label{eq:LF}
  \LF{{\tilde U},{\tilde V}}{\UF}
:=\PPMatrixFcomp0{{\tilde U},{\tilde V}} D\UF
+\PPMatrixFcomp1{{\tilde U},{\tilde V}} t\partial_x\UF
+\NNF\UF
\ee
where the matrices $\NNG$ and $\NNF$ are so far arbitrary.
The right-hand side of \eqref{eq:firstordersystemGowdyN} is written as
\be
  \label{eq:fgravity}
  f_{\gravity}+\NNG \UG=:\FGV{\UG}+\FGF{\UG,\UF}
\ee
where $\FGV{\UG}$ is the vacuum operator (obtained from \Eqref{eq:wavepdeNGowdy2} by setting $T_{\alpha\beta}=0$) and $\FGF{\UG,\UF}$ covers all the matter terms in \Eqref{eq:wavepdeNGowdy2}.
Finally, the right side of \eqref{eq:firstordersystemEulerN} is written as
\be
  \label{eq:ffluid} 
  f_{\fluid}+\NNF \UF=:\FF{\UG,\UF}.
\ee
The following systems will play a major role:
\begin{enumerate}
\item The \keyword{vacuum Einstein evolution system}:
$
  \LG{\UG}{\UG}=\FGV{\UG}
$
for the $18$-dimensional unknown $\UG$.
\item The \keyword{Einstein-Euler evolution system}:
\be
  \label{eq:EinsteinEuleroperator}
    \LG{\UG}{\UG}=\FGV{\UG}+\FGF{\UC},\quad
    \LF{\UC}{\UF}=\FF{\UC}
\ee
for the $20$-dimensional unknown $\UC:=(\UG,\UF)^T$.
\end{enumerate}

\subsection{The singular initial value problem}
\label{sec:SetupSIVP}

Next we formulate a singular initial value problem which matches the heuristic discussion in \Sectionref{sec:HeuristicAsymptotics} and the statements of \Theoremref{th:sub-criticalcoupledExistence} and \Theoremref{th:criticalcoupledExistence}.
The first step for this is to choose appropriate leading-order terms.
The choice of the first-order
  variables in \Eqsref{eq:firstordervariablesG1}--\eqref{eq:firstordervariablesLast} suggest 
\be
  \label{eq:firstordervariablesSG1}
    \USG:=(\USGcomp1,\ldots,\USGcomp6)^T
  \ee
where, for each $i=1,\ldots,6$, we define
\be
  \USGcomp{i}:=(\USGcomp{i,-1},\USGcomp{i,0},
  \USGcomp{i,1})^T
\ee
with
\begin{align}
  &\USGcomp{1,-1}:=-\Lambda_{*} t^{(k^2-1)/2},& 
  &\USGcomp{2,-1}:=\Lambda_{*} t^{(k^2-1)/2},&
&\USGcomp{3,-1}:=0,&\\
&\USGcomp{4,-1}:=t,&
&\USGcomp{5,-1}:=E_* t^{-k},&
&\USGcomp{6,-1}:=Q_{**}t^{2k},&
\end{align}
and, for each $i=1,\ldots,6$,
\be
  \label{eq:LOTderivatives}
  \USGcomp{i,0}:=D\USGcomp{i,-1}-\alpha\USGcomp{i,-1},\quad
  \USGcomp{i,1}:=0.
\ee
We stress that so far the data functions can be specified freely; in particular, there are no constraints for these data yet.
It turns out that the possibly more intuitive, but also more complicated choice $\USGcomp{i,1}=t\partial_x \USGcomp{i,-1}$ has no advantages over $\USGcomp{i,1}=0$ in \Eqref{eq:LOTderivatives} and in fact leads to the same results.

For later convenience, we also define the following $18$-dimensional vectors
  \be
    \label{eq:kappaG}
    \begin{split}
    \kappaVecG:=\Bigl(&(k^2-1)/2,(k^2-1)/2,(k^2-1)/2;
    (k^2-1)/2,(k^2-1)/2,(k^2-1)/2;\\
    &(k^2-1)/2,(k^2-1)/2,(k^2-1)/2;
    1,1,1;
    -k,-k,-k;
    2k,2k,2k\Bigr)
  \end{split}
\ee
  and
  \be
    \label{eq:muG}
    \begin{split}   
    \muVecG:=\Bigl(&\muGcomp1,\muGcomp1,\muGcomp1+\eta;
    \muGcomp1,\muGcomp1,\muGcomp1+\eta;
    \muGcomp1+\eta,\muGcomp1+\eta,\muGcomp1+2\eta;
    \muGcomp4,\muGcomp4,\muGcomp4+\eta;\\
    &\muGcomp5,\muGcomp5,\muGcomp5+\eta;
   \muGcomp6,\muGcomp6,\muGcomp6+\eta\Bigr)
 \end{split}
\ee
  for (so far unspecified) smooth scalar functions $\muGcomp{i}>0$ and $\eta\ge0$.
The particular structure and purpose of these exponent vectors and, in particular, the role of the function $\eta$ will be explained later.

For the leading-order term of the fluid, the results in \Sectionref{sec:HeuristicAsymptotics} suggest 
\be
  \label{eq:defUSF}
  \USF:=\Bigl(v_{*}^0t^{\Gamma},v_{*}^1t^{2\Gamma}\Bigr)^T
\ee
as the leading order term.
In analogy to the above we also define
\be
  \label{eq:muF}
  \kappaVecF:=(\Gamma,2\Gamma),\quad \muVecF:=(\muFcomp1,\muFcomp2-\Gamma).
\ee
We observe that the quantity $v^1_*$ in \Eqref{eq:defUSF} will later be called $\hat v^1_*$ to match the statement of \Theoremref{th:sub-criticalcoupledExistence} (this is not necessary for \Theoremref{th:criticalcoupledExistence}). The origin of this will become clear not before we incorporate the constraints in our analysis.

The next step in the proof of \Theoremref{th:sub-criticalcoupledExistence} and \Theoremref{th:criticalcoupledExistence} is to solve the singular initial value problem of \eqref{eq:EinsteinEuleroperator}
of the form
\be
  \label{eq:SIVP1}
  \UG=\USG+\WG, \quad \UF=\USF+\WF
\ee
for remainders 
\be
  \label{eq:SIVP2}
  \WG\in X_{\delta,\kappaVecG+\muVecG,\infty},\quad
  \WF\in X_{\delta,\kappaVecF+\muVecF,\infty}
\ee
for some constant $\delta>0$. 
With the short-hand notation
\be
  \label{eq:defcomb}
  \begin{split}
  \UC&:=(\UG,\UF)^T,\quad
  \USC:=(\USG,\USF)^T,\quad
  \WC:=(\WG,\WF)^T,\\
  \kappaVecC&:=(\kappaVecG,\kappaVecF),\quad
  \muVecC:=(\muVecG,\muVecF),
\end{split}
\ee
and the convention that we never write the leading-order term functions $\USC$, $\USG$ and $\USF$ explicitly unless they give rise to the only terms in some expression (as it is the case, e.g.,  for the second terms of \Eqsref{eq:FredGV} and \eqref{eq:FredF}),
we formally define {``reduced'' source term operators} 
  \be
    \label{eq:FredGV}
    \WG\mapsto\FredGV{\WG}:=\FGV{\WG}-\LG{\WG}{\USG}
  \ee
and
\be  
  \label{eq:FredF}
    \WC\mapsto\FredF{\WC}:=\FF{\WC}
    -\LF{\WC}{\USF}
\ee
from \Eqsref{eq:fgravity} and \eqref{eq:ffluid}.
In this notation the coupled Einstein-Euler evolution system \Eqref{eq:EinsteinEuleroperator} takes the form
  \be
  \label{eq:EinsteinEuleroperatorred}
    \LG{\WG}{\WG}=\FredGV{\WG}+\FGF{\WC},\quad
    \LF{\WC}{\WF}=\FredF{\WC}.
\ee

We remark 
that when we refer to the evolution equations in the form \eqref{eq:EinsteinEuleroperatorred} or to individual operators in \Eqref{eq:EinsteinEuleroperatorred}, we will always assume without further notice that the choices above have been made.
In particular, we will always consider $\kappaVecG$ and $\muVecG$ as given by \Eqsref{eq:kappaG} and \eqref{eq:muG} in terms of a smooth function $k$ and smooth exponents $\muGcomp{i}$ and $\eta$. In the same way we consider $\kappaVecF$ and $\muVecF$ as defined  by \Eqref{eq:muF}  from the function $\Gamma$ given by \Eqref{eq:exprGammaN}, $\gamma\in (1,2)$ and exponents $\muGcomp{i}$. 
We will also always consider $\USG$ and $\USF$ as defined in terms of smooth functions $\Lambda_*$, $E_*$, $Q_{**}$, $v_*^0$ and $v_*^1$ by \Eqsref{eq:firstordervariablesSG1}--\eqref{eq:LOTderivatives} and \Eqref{eq:defUSF}. 
In addition, the function $Q_*$ will always be considered as smooth.

\subsection{Estimates for our function operators}
\label{sec:foestimates}

In order to apply the Fuchsian theory in \Sectionref{sec:SIVPNN} to our singular initial value problem, \Theoremref{th:smoothexistenceN} requires that the function operators in our equations satisfy the estimates of the quasilinear symmetric hyperbolic Fuchsian property (recall \Defref{def:quasilinearlimitN}).
These estimates need to be proven under suitably general conditions in order to complete the arguments. In fact, we will see that the same estimates need to be applied at various, sometimes quite different stages of the proof and hence their hypotheses must be sufficiently general and flexible. On the other hand, however, the algebraic complexity of the expressions requires a certain degree of pragmatism which we aim for in our presentation. 

The main idea of the proofs of these estimates is to exploit the fact that all function operators which occur in the Einstein-Euler equations are \keyword{special} in the sense of \Sectionref{sec:SIVPNN} (see the end of the paragraph on function operators). Given leading-order terms and assumptions for the exponents, simple algebraic rules can be used to rigorously determine the leading terms and the estimates of interest. Because some of our function operators consist of hundreds of terms and sometimes subtle cancellations from all kinds of terms are crucial, we have programmed these algebraic rules into a computer algebra system. The computer is able to apply these rules repeatedly to all these terms efficiently. We stress that this yields fully rigorous estimates; no numerical approximations of any sort are used. The details of our computer algebra code are discussed in  
\cite{Ames:2016uy}.

We will present our estimates in the case $\Gamma>0$ only. The following lemmas hence lay the foundation for the proofs of \Theoremref{th:sub-criticalcoupledExistence} and \Theoremref{th:sub-criticalcoupledEstimates}. Regarding the case $\Gamma=0$ and \Theoremref{th:criticalcoupledExistence} and \Theoremref{th:criticalcoupledEstimates}, we will only make a few brief comments.

\paragraph{Principal part matrix operators} Let us start with the matrix operators which constitute the principal part of the evolution equations, i.e., $\PPMatrixGnullcomp{0}$, $\PPMatrixGnullcomp{1}$, $\PPMatrixFnullcomp{0}$ and $\PPMatrixFnullcomp{1}$, see \Eqsref{eq:block2fullGowdyN}--\eqref{eq:s1Gowdy}, 
and \Eqsref{eq:AAA1}. 

\begin{lemma}[Estimates for $\PPMatrixGnullcomp{0}$ and $\PPMatrixGnullcomp{1}$]
  \label{lem:princpartGN}
  Choose  functions $k$ and $\Lambda_*$ in $C^\infty(T^1)$ with $\Lambda_*>0$, and smooth exponent functions 
  $\muGcomp{i}>0$ and $\eta\ge0$.
  Then,  for any sufficiently small constant $\delta>0$, the function operator
$WG\mapsto \PPMatrixGcomp0{\WG}-\unitmatrix_{18}$
  is 
  $o(1)$, where $\unitmatrix_{18}$ represents the $18\times 18$-unit matrix. Moreover, 
  $\WG\mapsto t\PPMatrixGcomp1{\WG}$
  is a
  $(\kappaVecG+\muVecG,\zeta^{(1)}_{\gravity}, + \infty)$-operator with 
  \be
    \label{eq:xiG}
    \zeta^{(1)}_{\gravity}=(\infty,1,1,\ldots,\infty,1,1),
  \ee
  which is hence in particular $o(1)$.
\end{lemma}
Recall the paragraph after \Defref{def:functionoperatorsNN} for the definition of the $o(1)$-symbol for function operators.
In order to write an analogous result for the principal part matrices of the Euler equations, we  first define
\be
  \label{eq:SUTLF}
  \SZeroZeroF:=
  \diag\left(\frac{v_*^0\Lambda_*}{\gamma-1},v_*^0\Lambda_*\right).
\ee
This matrix is clearly  positive definite so long as $\Lambda_*,v^0_*>0$ and $\gamma>1$.

\begin{lemma}[Estimates for $\PPMatrixFnullcomp{0}$ and $\PPMatrixFnullcomp{1}$]
  \label{lem:princpartFN}
  Choose  functions $k$, $\Lambda_*$, $v^0_*$ and $v^1_*$ in $C^\infty(T^1)$ with $\Lambda_*,v^0_*>0$, a constant $\gamma\in (1,2)$ such that $\Gamma> 0$ (cf.\ \Eqref{eq:exprGammaN}), smooth exponent functions 
  $\muGcomp{i}>0$, $\eta\ge0$, and
  \[\muFcomp1\le\muFcomp2,\quad
 0<\muFcomp1<\min\{2\Gamma,\muGcomp1\}.\]
  Then,  for any sufficiently small constant $\delta>0$, the function operator
  \[\WC\mapsto \PPMatrixFcomp0{\WC}-\SZeroZeroF\]
  is 
  $o(1)$. Moreover, 
  $\WC\mapsto t\PPMatrixFcomp1{\WC}$
  is a
  $(\kappaVecC+\muVecC,\zeta^{(1)}_{\fluid}, + \infty)$-operator with
\be
    \label{eq:xiF}
    \zeta^{(1)}_{\fluid}=(1,1),
  \ee
  which is hence in particular $o(1)$.
\end{lemma}

\paragraph{Reduced source term operators} 
We continue with the reduced source term operators $\FredGV{\cdot}$, $\FGF{\cdot}$ and $\FredF{\cdot}$ defined in \Eqsref {eq:fgravity}, \eqref{eq:ffluid}, \eqref{eq:FredGV} and \eqref{eq:FredF}. 
First we specify the matrices $\NNG$ and $\NNF$ which have appeared the first-order time in \Eqsref{eq:LG} and \eqref{eq:LF}.
In agreement with \cite{Ames:2016uy}, we set
\be
  \label{eq:NNG}
  \NNG:=\mathrm{diag}\,(n_{01},n_R,n_E,n_Q),
\ee
where
\be
\label{eq:nnZO}
    n_{01}=\left(
\begin{array}{ccccccccc}
 -\alpha  & -1 & 0 & 0 & 0 & 0 & 0 & 0 & 0 \\
 \frac{1}{4} b^2 & -b-\alpha & 0 & -b & 2 & 0 & 0 & 0 & -4 \\
 0 & 0 & -\alpha -1 & 0 & 0 & 0 & 0 & 0 & 0 \\
 0 & 0 & 0 & -\alpha  & -1 & 0 & 0 & 0 & 0 \\
 0 & 0 & 0 & \frac{1}{4} b^2 &
   -b-\alpha & 0 & 0 & 0 & -2 \\
 0 & 0 & 0 & 0 & 0 & -\alpha -1 & 0 & 0 & 0 \\
 0 & 0 & 0 & 0 & 0 & 0 & -\alpha  & -1 & 0 \\
 0 & 0 & -\frac{3}{2} & 0 & 0 & -\frac{1}{2} & \frac{1}{4}
   b^2 & -b-\alpha & 0 \\
 0 & 0 & 0 & 0 & 0 & 0 & 0 & 0 & -\alpha -1 \\
\end{array}
\right),
\ee
with $b:=k^2-2 \alpha -1$, 
and
\begin{align*}
n_R&=\left(
\begin{array}{ccc}
 -\alpha  & -1 & 0 \\
 (\alpha -1)^2 & \alpha -2 & 0 \\
 0 & 0 & -\alpha -1 \\
\end{array}
\right),\quad
n_E=\left(
\begin{array}{ccc}
 -\alpha  & -1 & 0 \\
 (\alpha +k)^2 & \alpha +2 k & 0 \\
 0 & 0 & -\alpha -1 \\
\end{array}
\right),\\
&\qquad\qquad\qquad n_Q=\left(
\begin{array}{ccc}
 -\alpha  & -1 & 0 \\
 \alpha  (\alpha -2 k) & \alpha -2 k & 0 \\
 0 & 0 & -\alpha -1 \\
\end{array}
\right).
\end{align*}
In addition, we set
\be
\label{eq:defNNF}
\NNF:=
  \diag\left(-\Gamma\frac{ v_*^0\Lambda}{\gamma-1},-2\Gamma v_*^0\Lambda\right).
\ee

For the following it is a crucial observation that  $\FredGV{\cdot}$ is the only source term operator in \Eqref{eq:EinsteinEuleroperatorred} which depends on $Q$ (and hence in particular on $Q_*$) and derivatives. Indeed, $Q$ will play a distinguished role for the analysis. In order to anticipate this, we split this operator up as follows:
\be
  \FGV{\UG}=:\FGVOne{\UG}+\FGVTwo{\UG}
\ee
where $\FGVOne{\UG}$ is constructed from $\FGV{\UG}$ by dropping all terms proportional to $Q_*$ and derivatives. 
Then by \Eqref{eq:FredGV}, we have
\be
  \label{eq:FredGVsplit1}
  \begin{split}
  \FredGV{\WG}&=\FGV{\USG+\WG}-\LG{\USG+\WG}{\USG}\\
&=\underbrace{\FGVOne{\USG+\WG}-\LG{\USG+\WG}{\USG}}_{=:\FredGVOne{\WG}}+\FGVTwo{\USG+\WG}.
\end{split}
\ee
Observe that $\FredGVOne{\WG}$ therefore does not contain any $Q_*$ terms.
These operators are now decomposed further.
With $\Pi:=\diag\Bigl(\underbrace{1,\ldots,1}_{\text{$15$ entries}},0,0,0\Bigr)$, we set
\begin{align}
\FredGVOneOne{\WG}&:=\FGVOne{\Pi(\USG+\WG)}-\LG{\Pi(\USG+\WG)}{\USG},\\
\FredGVOneTwo{\WG}&:=\FredGVOne{\WG}-\FredGVOneOne{\WG}.
\end{align}
We note that $\FredGVOneOne{\WG}$ is \textit{completely} free of $Q$-terms while the second operator still contains higher-order contribution from the $Q$-variables (it does not contain any terms proportional to $Q_*$ and derivatives).
Analogously we set
\begin{align}
\FGVTwoOne{\UG}:=\FGVTwo{\Pi(\UG)},\quad
\FGVTwoTwo{\UG}:=\FGVTwo{\UG}-\FGVTwoOne{\UG}.
\end{align}
So, in total we have
\be
\label{eq:hsf}
\FredGV{\WG}=\FredGVOneOne{\WG}+\FredGVOneTwo{\WG}+\FGVTwoOne{\USG+\WG}+\FGVTwoTwo{\USG+\WG}.
\ee
We remark that in the half-polarized case $Q_*=\textrm{const}$, we have
\[\FredGV{\WG}=\FredGVOneOne{\WG}+\FredGVOneTwo{\WG},\]
while in the fully polarized case $Q=Q_*=\textrm{const}$, we have
$\FredGV{\WG}=\FredGVOneOne{\WG}$.
Even though the following results also hold in these special cases, the main focus is the general unpolarized case.
In consistency with our previous convention we will now often not write the leading term function $\USG$ explicitly in the last two terms of \eqref{eq:hsf}.
Recall the definition of $\RR{\cdot}$ in \Eqref{eq:rmatrix}.
\begin{lemma}[Estimates for  $\FredGV{\cdot}$]  
  \label{lem:sourceG1}
  Choose  functions $k, \Lambda_*,E_*, Q_*, Q_{**}\in C^\infty(T^1)$ with $\Lambda_*,E_*>0$ and $0<k<1$,
and smooth exponent functions 
  $\muGcomp{i}>0$ and $\eta\ge 0$.
Then for any sufficiently small constant $\delta>0$:
\begin{enumerate}[label=\textit{(\roman{*})}, ref=(\roman{*})] 
\item The operator
$\WG\mapsto \RR{\kappaVecG+\muVecG}
\FredGVOneOne{\WG}$
is 
$o(1)$ provided
\[\eta<1,\quad \muGcomp{5},\muGcomp{6}<1-\eta,\quad \muGcomp{1}<\min\{\muGcomp4,\muGcomp5\}.\]
\item The operator
$\WG\mapsto \RR{\kappaVecG+\muVecG}
\FredGVOneTwo{\WG}$
is a $(\kappaVecG+\muVecG,\zeta_\gravityvac^{(1,2)}, + \infty)$-operator for
\begin{align*}
  \zeta_\gravityvac^{(1,2)}=(&\infty,2 k-{\muGcomp1},\infty,
\infty,2 \eta +2 k-{\muGcomp1}+2
   {\muGcomp6},\infty,
\infty,2 k-{\muGcomp1}+{\muGcomp6},\infty,\\
&\infty,\infty,\infty,
\infty,2 k-{\muGcomp5},\infty,
\infty,\min\{2 \eta   +{\muGcomp1},\muGcomp5-\muGcomp6\},{\muGcomp1})>0
\end{align*}
provided
\[\muGcomp{1},\muGcomp{5}<2k,\quad \muGcomp{6}<\muGcomp5.\]
\item \label{part:estimate3} The operator
$\WG\mapsto \RR{\kappaVecG+\muVecG}
\FGVTwoOne{\WG}$
is a $(\kappaVecG+\muVecG,\zeta_\gravityvac^{(2,1)}, + \infty)$-operator for
\begin{align*}
  \zeta_\gravityvac^{(2,1)}=(&\infty,\infty,\infty,
\infty,2 (1-k)-{\muGcomp1},\infty,
\infty,\infty,\infty,\\
&\infty,\infty,\infty,
\infty,2 (1-k)-{\muGcomp5},\infty,
\infty,1+\eta -2 k+{\muGcomp1}-{\muGcomp6},\infty )
\end{align*}
 provided
\be
\label{eq:estimate3hypo}
\eta<1,\quad \muGcomp{1}<\min\{\muGcomp{5},1-\eta\},\quad \muGcomp{5},\muGcomp{6}<2(1-k).
\ee
We have $\zeta_\gravityvac^{(2,1)}>0$ under the additional restriction
\be
  \label{eq:restrictiveassumps}
  1+\eta -2 k+{\muGcomp1}-{\muGcomp6}>0.
\ee
\item The operator
$\WG\mapsto \RR{\kappaVecG+\muVecG}
\FGVTwoTwo{\WG}$
is a $(\kappaVecG+\muVecG,\zeta_\gravityvac^{(2,2)}, + \infty)$-operator for
\begin{align*}
  \zeta_\gravityvac^{(2,2)}=(&\infty,\infty,\infty,
\infty,\eta +1,\infty,
\infty,1-\eta  -{\muGcomp6},\infty,\\
&\infty,\infty,\infty,
\infty,1+\eta -{\muGcomp5}+{\muGcomp6},\infty,
\infty,\infty, \infty)>0
\end{align*}
 provided
\[\eta<1,\quad \muGcomp{1}=\muGcomp{6}<1-\eta,\quad \muGcomp{5}<1+\eta+\muGcomp{1}.\]
\end{enumerate}
\end{lemma}

A remarkable fact is that $\FGVTwoOne{\cdot}$ violates the $o(1)$-property required by the Fuchsian theorem (as part of \Defref{def:quasilinearlimitN}) unless \Eqref{eq:restrictiveassumps} is satisfied. This will indeed have important consequences below. By definition this operator vanishes if $Q_*=\textrm{const}$, and this issue therefore disappears in the half-polarized case and consequently the analysis becomes  significantly simpler. The terms  in the $17$th component of $\FredGVTwoOne{\cdot}$ which are responsible for this extra condition \Eqref{eq:restrictiveassumps} are
\be
  \label{eq:badterm}
  t Q_*' \frac{g_{01}}{g_{11}} \left(2 \frac{DE}E+1\right)-2t Q_*' \frac{g_{00}}{g_{11}}\frac{t\partial_x E}{E}.
\ee
We will use this later.

Next we discuss the operator which represents the matter terms in Einstein's equations, see \Eqref{eq:EinsteinEuleroperatorred}.
\begin{lemma}[Estimates for  $\FGF{\cdot}$]
  \label{lem:sourceG2}
  Choose  functions $k, \Lambda_*,E_*,v^0_*,v^1_*\in C^\infty(T^1)$ 
 with $\Lambda_*,E_*,v^0_*>0$, a constant $\gamma\in (1,2)$ such that $\Gamma> 0$ (cf.\ \Eqref{eq:exprGammaN}), and smooth exponent functions 
  $\muGcomp{i}>0$, $\muFcomp{i}>0$ and $\eta\ge 0$.
Then, for any sufficiently small constant $\delta>0$,
 the function operator
  $\WC\mapsto \RR{\kappaVecG+\muVecG}\FGF{\WC}$
  is a $(\kappaVecC+\muVecC,\zeta_\gravityfluid, + \infty)$-operator 
with
  \begin{align*}
    \zeta_\gravityfluid=(&\infty,1-\Gamma -{\muGcomp1},\infty,
\infty,1-\Gamma  -{\muGcomp1},\infty,\\
&\infty,\min\{1-\Gamma,1-\eta -{\muGcomp1},1-\Gamma -\eta
   +{\muFcomp2}-{\muGcomp1}\},\infty,\\
&\infty,1-\Gamma
   -{\muGcomp4},\infty,
\infty,\infty,\infty,
\infty,\infty,\infty )>0,
  \end{align*}
provided
\begin{gather*}
\eta<1,\quad\muGcomp{1}<\min\{1-\Gamma,1-\eta,1-\Gamma-\eta+\muFcomp2\},\quad\muGcomp{4}<1-\Gamma,\\
\muFcomp1<\min\{2\Gamma,\muGcomp1\},\quad \muFcomp1\le\muFcomp2.
\end{gather*}
\end{lemma}
We recall that $\Gamma$ is always smaller than $1$ as a consequence of the assumption $\gamma<2$.

Finally, we discuss the source term of the Euler equations. In addition to the operator $\FredF{\cdot}$ in \Eqref{eq:FredF}, we also consider
\be
  \label{eq:FredFTDef}
  \begin{split}
  \WC\mapsto\FredTF{\WC}
:=&(\PPMatrixFnullcomp0(\USC+\WC))^{-1}\left(\FredF{\WC}
    +\PPMatrixFnullcomp1(\USC+\WC) t\partial_{x} \USF\right)\\
&+
\left(\NNFT-(\PPMatrixFnullcomp0(\USC+\WC))^{-1}\NNF\right)\WF,
\end{split}
\ee
which we will use to study ``truncated versions'' of the Euler equations below.
Here,
\be
\label{eq:defNNFT}
\NNFT:=
  \diag\left(-\Gamma,-2\Gamma \right).
\ee
\begin{lemma}[Estimates for  $\FredF{\cdot}$]
  \label{lem:sourceF}
  Choose  functions $k, \Lambda_*,E_*,v^0_*,v^1_*\in C^\infty(T^1)$ 
  with $\Lambda_*,E_*,v^0_*>0$, a constant $\gamma\in (1,2)$ such that $\Gamma> 0$ (cf.\ \Eqref{eq:exprGammaN}), and smooth exponent functions 
  $\muGcomp{i}>0$, $\muFcomp{i}>0$ and $\eta\ge 0$.
Then, for any sufficiently small constant $\delta>0$,
 the function operator
  $\WC\mapsto \RR{\kappaVecF+\muVecF}\FredF{\WC}$
 is 
$o(1)$
provided
\be
\muFcomp1<\min\{1,2\Gamma,\muGcomp1,\muGcomp4\},\quad
\muFcomp1\le\muFcomp2<\min\{1,\eta+\muGcomp1,\Gamma+\muFcomp1\},
\ee
and the function operator
  $\WC\mapsto \RR{\kappaVecF+\muVecF}\FredTF{\WC}$
 is a $(\kappaVecC+\muVecC,\zeta_{\truncated\fluid}, + \infty)$-operator
for some $\zeta_{\truncated\fluid}>0$
provided
\be
\label{eq:conditionFFT}
\muFcomp1<\min\{2\Gamma,\muGcomp1,\muGcomp4\},\quad
\muFcomp1\le\muFcomp2<\min\{\eta+\muGcomp1,\Gamma+\muFcomp1\}.
\ee
\end{lemma}


\subsection{Solving the evolution equations: a new approach}
\label{sec:applFuchsian}

The next task in our discussion is to solve the singular initial value problem  \Eqsref{eq:SIVP1}, \eqref{eq:SIVP2} and \eqref{eq:EinsteinEuleroperatorred} using \Theoremref{th:smoothexistenceN} and the estimates obtained in the previous section. Before we do this, however, we want to give a quick argument why this can be done \textit{directly} (as opposed to our \textit{indirect} approach introduced below; see also the discussion in the last paragraph of \Sectionref{sec:intro}) {\sl 
only under quite restrictive conditions.} First we observe due to the coupled structure of, in particular,  \Eqref{eq:nnZO} that the block diagonal condition of \Theoremref{th:smoothexistenceN} requires $\eta=0$ (see \Eqref{eq:muG}). Part~\ref{part:estimate3} of \Lemref{lem:sourceG1} then yields the
condition
$1-2k+\muGcomp1-\muGcomp6>0$ which is necessary to guarantee that the operator $\WG\mapsto \FGVTwoOne{\WG}$ is $o(1)$. Since $\muGcomp6>0$, it is therefore necessary that
$\muGcomp1>2k-1$. This however is only compatible with the inequality $\muGcomp1<2(1-k)$ obtained from \Eqref{eq:estimate3hypo} if $0<k<3/4$. This is a disappointing result because one expects from earlier results in particular in the vacuum case \cite{Ringstrom:2006gy} that the permitted range for $k$ should be the interval $(0,1)$ in the general non-polarized case.

The basic idea of our new approach is very natural: roughly speaking it is to prove \Theoremref{th:sub-criticalcoupledExistence} and statement~\ref{statement:coupledsub2} of \Theoremref{th:sub-criticalcoupledEstimates} simultaneously 
--- as opposed to \textit{first} proving \Theoremref{th:sub-criticalcoupledExistence} and \textit{then} \Theoremref{th:sub-criticalcoupledEstimates}, as it has been done traditionally.
More specifically, we will \textit{not} solve the singular initial value problem outlined in \Sectionref{sec:SetupSIVP}
for the evolution equations \textit{directly} (this is why our new approach could be called ``indirect'').
Instead we will first construct solutions of the singular initial value problem in \Sectionref{sec:SetupSIVP} only to a truncated form of the evolution equations. These are (almost) the ``truncated equations'' considered in statement~\ref{statement:coupledsub2} of \Theoremref{th:sub-criticalcoupledEstimates}; see Step~1 below. 
Only after this has been achieved, we will consider the \textit{full} evolution equations in Step~2 below. The singular initial value problem, which we consider there, is defined by using the solutions in Step~1 as the leading-order term. It turns out that this  indeed resolves the technical problem above and allows us to consider the full interval $(0,1)$ for $k$. Roughly speaking, in this way we provide an ``improved leading-order term'' for the singular initial value problem in full analogy to the iterative approach by \cite{Rendall:2000ki,Stahl:2002bv} in Step~1 which is used then used in Step~2, but in a completely non-iterative fashion and without loss of regularity. 

Let us also briefly recall our previous claim that the analysis is significantly simpler in the half-polarized case, i.e., when $Q_*=const$. Now we can understand one particular reason for this claim. Since the restriction $k\in (0,3/4)$ found above is a consequence of the properties of the operator $\FGVTwoOne{\cdot}$, which is however identically zero in this case, the problem disappears when $Q_*=const$.

\paragraph{Step 1. Solving the partially truncated equations}
As in \Sectionref{sec:foestimates}, we continue to give details for the case $\Gamma>0$ and add only a few remarks regarding the case $\Gamma=0$.
As discussed above this step is only necessary in the non-polarized case $Q_*\not=\text{const}$. It is therefore essential for the proof of \Theoremref{th:sub-criticalcoupledExistence} (and \Theoremref{th:sub-criticalcoupledEstimates}) but not for \Theoremref{th:criticalcoupledExistence}. 

Let us recall the operator versions of the fully coupled Einstein-Euler equations \Eqref{eq:EinsteinEuleroperator} and their ``reduced version'' in \Eqref{eq:EinsteinEuleroperatorred}. The \keyword{partially truncated equations} 
are defined as
\be
  \label{eq:EinsteinEulerPartialTruncatedNonRed}
  \begin{split}
  \LG{\UG}{\UG}-\PPMatrixGcomp1{\UG} t\partial_{x} \UG&=\FGV{\UG}+\FGF{\UC},\\
  (\PPMatrixFcomp0{\UC})^{-1}\left(
\LF{\UC}{\UF}-\PPMatrixFcomp1{\UC} t\partial_{x} \UF\right)&=(\PPMatrixFcomp0{\UC})^{-1}\FF{\UC},
\end{split}
\ee
which yields
\be
  \label{eq:EinsteinEulerPartialTruncated}
  \begin{split}
    \LG{\WG}{\WG}-\PPMatrixFcomp1{\WG} t\partial_{x} \WG&=\FredGV{\WG}+\PPMatrixFcomp1{\WG} t\partial_{x} \USG+\FGF{\WC},\\
    D\WF+\NNFT\WF&=\FredTF{\WC},
  \end{split}
\ee
with \Eqsref{eq:FredFTDef} and \eqref{eq:defNNFT}.
Essentially, these partially truncated equations are derived from the full evolution equation by {\sl 
removing all those spatial derivative terms} which are multiplied with the matrices $\PPMatrixGnullcomp1$ and $\PPMatrixFnullcomp1$. Note however that this system still involves $Q_*$ and its derivatives as part of $\FGVTwo{\cdot}$ (see \Eqref{eq:FredGVsplit1}) --- this is why we refer to these equations as \textit{partially} reduced. 
The reason why we keep the derivatives of $Q_*$ will be explained below. Regarding the ``spatial derivative terms'' variables $\UGcomp{i,1}$ we find that the equations for the $6$ terms $\UGcomp{i,1}$ given by \Eqref{eq:EinsteinEulerPartialTruncated} are trivial and hence
\be
  \label{eq:truncatedsol}
  \UGcomp{i,1}\equiv 0
\ee
for all $i=1,\ldots,6$ is a solution which is compatible with \Eqref{eq:LOTderivatives}. With this the evolution equations of these terms and the terms themselves can be removed from our system completely which we assume now without further notice.

Let us also comment on the fact that  we multiply the second equation of \eqref{eq:EinsteinEulerPartialTruncatedNonRed} by $(\PPMatrixFnullcomp0)^{-1}$.
This would clearly be harmful for the \textit{original} equations because the matrix $(\PPMatrixFnullcomp0)^{-1}\PPMatrixFnullcomp1$ is in general not symmetric. Since this term is however not present in \Eqref{eq:EinsteinEulerPartialTruncated}, this conveniently decouples the principal parts of the Euler equations.

We see that \Eqref{eq:EinsteinEulerPartialTruncated} is a {\sl 
system of $x$-parametrized ODEs with respect to $t$.}
 The goal is now to show that we can pick the exponents $\muGcomp{i}$, $\muFcomp{i}$ and $\eta$ in \Sectionref{sec:SetupSIVP} so that the three conditions of \Theoremref{th:smoothexistenceN} are satisfied: (i) the system is a special quasilinear symmetric hyperbolic Fuchsian system (\Defref{def:quasilinearlimitN}), (ii) the block diagonal condition holds and (iii) the eigenvalue condition holds. This is achieved straightforwardly using the estimates in \Sectionref{sec:foestimates} and we obtain \Propref{prop:SIVPtruncated} below. The only non-trivial step is to satisfy \Eqref{eq:restrictiveassumps}, which with the judicious choice ${\muGcomp1}={\muGcomp6}$ leads to the condition $\eta>2k-1$, and the eigenvalue condition $\muFcomp2>\Gamma$ (see \Eqsref{eq:defNNFT} and \eqref{eq:muF}) which together with \Eqref{eq:conditionFFT} leads to the condition $\eta>\Gamma$. Moreover, we require $\eta<1$. It is important to note that the block diagonal condition is now essentially trivial and in particular $\eta$ does not need to vanish.

\begin{proposition}[Singular initial value problem for the partially truncated equations]
  \label{prop:SIVPtruncated}
  Choose  functions $k$, $\Lambda_*$, $E_*$, $Q_*$, $Q_{**}$, $v^0_*$ and $v^1_*$ in $C^\infty(T^1)$ such that $\Lambda_*,E_*,v^0_*>0$ and $1>k>0$, and a constant $\gamma\in (1,2)$. Choose  smooth functions $\muGcomp{i}$, $\muFcomp{i}$ and $\eta$ such that
  \begin{center}
    \begin{tabular}{rcccl}
      $\max\{\Gamma,2k-1\}$ & $<$ & $\eta$ & $<$ & $1$,\\
      $0$ & $<$ & $\muGcomp5$ & $<$ & $\min\{2k,2(1-k),1-\eta\}$,\\
      $0$ & $<$ & $\muGcomp4$ & $<$ & $1-\Gamma$,\\
      $0$ & $<$ & $\muGcomp1=\muGcomp6$ & $<$ & $\min\{\muGcomp4,\muGcomp5\}$,\\
      $0$ & $<$ & $\muFcomp1$ & $<$ & $\min\{\Gamma,\muGcomp1\}$,\\
      $\Gamma$ & $<$ & $\muFcomp2$ & $<$ & $\Gamma+\muFcomp1$.
    \end{tabular}
  \end{center}
  Then there exists some $\tilde\delta>0$, such that the partially truncated evolution equations,  \Eqref{eq:EinsteinEulerPartialTruncated}, 
  have a unique solution of the form
  $\UC=\USC+\WC$,
  for some 
  $\WC\in X_{\tilde\delta,\kappaVecC+\muVecC,\infty}$ with $\UGcomp{i,1}\equiv 0$ for all $i=1,\ldots,6$. The remainder $\WC$ is differentiable with respect to $t$ and $D\WC\in X_{\tilde\delta,\kappaVecC+\muVecC,\infty}$.
\end{proposition}
We recall again that the two restrictions $\gamma\in (1,2)$ and $k\in (0,1)$ imply $\Gamma\in (0,1)$. 
We remark without proof that exactly the same result holds for the \textit{fully} truncated equations, i.e., for \Eqref{eq:EinsteinEulerPartialTruncated} where the term $\FredGV{\WG}$ is replaced by $\FredGVOne{\WG}=\FredGVOneOne{\WG}+\FredGVOneTwo{\WG}$. As one would expect the hypothesis of that result is a little less restrictive than the hypothesis of \Propref{prop:SIVPtruncated} in as much as the second inequality in \Propref{prop:SIVPtruncated} can be replaced by
$0<\muGcomp5<\min\{2k,1-\eta\}$.

Before we proceed, let us make a general comment about inequalities for exponents in results obtained by the Fuchsian method, as for example the list of inequalities in \Propref{prop:SIVPtruncated}. On one hand, if one is interested in obtaining as much control as possible over the decay of the remainders, one chooses the exponents as large as allowed by these inequalities. If one is interested in the strongest uniqueness statement on the other hand, one chooses them as small as possible. Hence both upper and lower bounds characterize the singular initial value problem crucially and hence we list both of them whenever necessary.

\paragraph{Step 2. Modified singular initial value problem for the full equations}
 In the following we will refer to solutions of the partially truncated equations \eqref{eq:EinsteinEulerPartialTruncated}, in particular those given by \Propref{prop:SIVPtruncated}, as $\UGT$ and $\UFT$ with remainders $\WGT$ and $\WFT$; we will also write $\UCT=(\UGT,\UFT)$ and $\WCT=(\WFT,\WFT)$ as before. 
Let such a solution be given. As motivated at the beginning of \Sectionref{sec:applFuchsian}, the task of this step now is to solve the following ``modified'' singular initial value problem 
\be
  \label{eq:modsivp}
  \UC=\UCT+\WC=\USC+\WCT+\WC
\ee
for the \textit{full} equations, \Eqref{eq:EinsteinEuleroperatorred}, where $\UCT=\USC+\WCT$ is considered as the given leading-order term and $\WC=(\WG,\WF)$ is the unknown remainder in some to be specified space. To this end we rewrite the full equations as follows. Let us start with the Einstein part of the full equations in \Eqref{eq:EinsteinEuleroperator}:
\begin{align*}
  0=&\LG{\UG}{\UG}-\FGV{\UG}-\FGF{\UC}
\\
     =&\LG{\UG}{\WG}+\LG{\UG}{\UGT}-\FGV{\UG}-\FGF{\UC}\\
  =&\LG{\UG}{\WG}
    -(\LG{\UGT}{\UGT}-\LG{\UG}{\UGT})\\
&+\PPMatrixGcomp1 {\UGT} t\partial_{x} \UGT+\FGV{\UGT}+\FGF{\UCT}
-\FGV{\UG}-\FGF{\UC}.
\end{align*}
In this calculation we have assumed explicitly that $\WCT$ is a solution to the partially truncated equations \eqref{eq:EinsteinEulerPartialTruncatedNonRed}. Using \Eqref{eq:FredGV} for the definition of the reduced operators and \Eqref{eq:hsf}, and performing the same calculation for the Euler equations (and using the same short-hand notation for $\USC$ as before), we find the following system:
\be
  \label{eq:modeqs1}
  \begin{split}
  &\!\!\!\!\!\!\!\!\LG{\WGT+\WG}{\WG}\\
=&
            \underbrace{\FredGVOneOne{\WGT+\WG}-\FredGVOneOne{\WGT}}_{=:\OGVOneOne{\WG}}+\underbrace{\FredGVOneTwo{\WGT+\WG}-\FredGVOneTwo{\WGT}}_{=:\OGVOneTwo{\WG}}\\
&+\underbrace{\FGVTwoOne{\WGT+\WG}-\FGVTwoOne{\WGT}}_{=:\OGVTwoOne{\WG}}+\underbrace{\FGVTwoTwo{\WGT+\WG}-\FGVTwoTwo{\WGT}}_{=:\OGVTwoTwo{\WG}}\\
  &+\underbrace{\FGF{\WCT+\WC}-\FGF{\WCT}}_{=:\OGF{\WC}}   -\underbrace{(\PPMatrixGcomp0{\WGT+\WG}-\PPMatrixGcomp0{\WGT})D {\WGT}}_{=:\OGPZero{\WG}}\\
&-\underbrace{(t\PPMatrixGcomp1{\WGT+\WG}- t\PPMatrixGcomp1{\WGT}) \partial_x {\WGT}}_{=:\OGPOne{\WG}}
-\underbrace{t\PPMatrixGcomp1 {\WGT}\partial_{x} \UGT}_{=:\OTG{\WG}},
\end{split}
\ee
and
\be
  \label{eq:modeqs2}
\begin{split}
  \LF{\WCT+\WC}{\WF}
  &=
            \underbrace{\FredF{\WCT+\WC}-\FredF{\WCT}}_{=:\OF{\WC}} 
  -\underbrace{ (\PPMatrixFcomp0{\WCT+\WC}-\PPMatrixFcomp0{\WCT})D {\WFT}}_{=:\OFPZero{\WC}}\\
&-\underbrace{(t\PPMatrixFcomp1{\WCT+\WC}- t\PPMatrixFcomp1{\WCT})\partial_x {\WFT}}_{=:\OFPOne{\WC}}
-\underbrace{t\PPMatrixFcomp1 {\WCT}\partial_{x} \UFT}_{=:\OTF{\WC}}.
\end{split}
\ee
These equations are equivalent to \Eqref{eq:EinsteinEuleroperatorred} if $\WCT$ is the remainder of a solution to the partially truncated equations \eqref{eq:EinsteinEulerPartialTruncated}.
We will now allow $\WCT=(\WGT,\WFT)$ to be \textit{any} given function in\footnote{For simplicity we set $\delta=\tilde\delta$ without loss of generality; recall that $\delta$ is always considered as some sufficiently small positive quantity.} $X_{\delta,\kappaVecC+\muVecC,\infty}$ which is differentiable with respect to $t$ with $D\WCT\in X_{\delta,\kappaVecC+\muVecC,\infty}$ for some exponents $\muGcomp{i}>0$, $\muFcomp{i}>0$ and $\eta\ge 0$ assuming that \Eqsref{eq:kappaG}, \eqref{eq:muG} and \eqref{eq:muF} hold. Let us now focus on the singular initial value problem \Eqsref{eq:modsivp}, \eqref{eq:modeqs1} and \eqref{eq:modeqs2} for a remainder $\WC$ in $X_{\delta,\kappaVecC+\muVecC+\nuVecC,\infty}$ where
\be
  \label{eq:defnu1}
  \nuVecC:=(\nuVecG,\nuVecF),\quad
\nuVecG=(\nu_1,\ldots,\nu_1,\nu_2,\nu_2,\nu_2),\quad
\nuVecF=(\nu_1,\nu_1)
\ee
for some 
 scalar exponents $\nu_1, \nu_2> 0$; the particular structure of  \Eqref{eq:defnu1} anticipates the restrictions imposed by the block diagonal condition of \Theoremref{th:smoothexistenceN} for this singular initial value problem as we discuss below.
The following result will be proven below.

\begin{proposition}[Modified singular initial value problem for the full evolution equations]
    \label{prop:ModSIVP}
    Choose  functions $k$, $\Lambda_*$, $E_*$, $Q_*$, $Q_{**}$, $v^0_*$ and $v^1_*$ in $C^\infty(T^1)$ such that $\Lambda_*,E_*,v^0_*>0$ and $1>k>0$, and a constant $\gamma\in (1,2)$. Choose  a smooth function $\epsilon$ with
\[0<\epsilon<\min\left\{2\Gamma,\frac{1-\Gamma}4,\frac{2k}4,\frac{2(1-k)}4\right\}.\]
Set
\begin{equation*}
  \eta=0,\quad  \muFcomp1=\muFcomp2=\epsilon,\quad
  \muGcomp1=\muGcomp6=2\epsilon,\quad
  \muGcomp4=\muGcomp5=3\epsilon,\quad
\nu_1=1-4\epsilon,
\end{equation*}
and choose any smooth function $\nu_2$ such that
\[\max\{0,1-2k\}<\nu_2<\min\{1,2(1-k)\}-4\epsilon.\]
Choose  any function $\WCT$ in $X_{\delta,\kappaVecC+\muVecC,\infty}$ which is differentiable with respect to $t$ such that $D\WCT\in X_{\delta,\kappaVecC+\muVecC,\infty}$. Then, for some (sufficiently small) constant $\tilde\delta>0$ and some (sufficiently negative) constant $\alpha$, the singular initial value problem \Eqsref{eq:modsivp}, \eqref{eq:modeqs1} and \eqref{eq:modeqs2} has a unique solution for some remainder $\WC$ in $X_{\tilde\delta,\kappaVecC+\muVecC+\nuVecC,\infty}$ where $\nuVecC$ is given by \Eqref{eq:defnu1}. The remainder $W$ is differentiable with respect to $t$ and $D\WC$ is also in $X_{\tilde\delta,\kappaVecC+\muVecC+\nuVecC,\infty}$.
\end{proposition}

It is clear that any solution $\UCT=\USC+\WCT$ to the partially truncated equation given by \Propref{prop:SIVPtruncated} satisfies the hypothesis of \Propref{prop:ModSIVP}. The corresponding solution to \Propref{prop:ModSIVP} is therefore a solution to the \textit{original} singular initial value problem \Eqsref{eq:SIVP1} and \eqref{eq:SIVP2} of \eqref{eq:EinsteinEuleroperatorred}. We have therefore shown that the singular initial value problem of interest indeed has a solution (\textit{existence}). Note, however, that it is in principle possible that there are further solutions to the original singular initial value problem and hence uniqueness is not addressed by this. We are not going to address in this paper. In any case, note that \Propref{prop:ModSIVP} does not yet imply \Theoremref{th:sub-criticalcoupledExistence} since we have not yet imposed the constraints.
In any case, observe that we have written the hypothesis of \Propref{prop:ModSIVP} in terms of a single scalar quantity $\epsilon$. The loss of generality implied by this is insignificant and it also simplifies the statement of the proposition.

The proof of \Propref{prop:ModSIVP} makes heavy use of \Theoremref{th:smoothexistenceN} and of the following general lemma which can be proved with techniques presented in \cite{Ames:2012vz}.
\begin{lemma}
  \label{lem:hoc}
  Suppose $W\mapsto F[W]$ is any special $(\tilde\mu,\tilde\nu, + \infty)$-operator for any exponent vectors $\tilde\mu$ and $\tilde\nu$. Choose  any $W_0\in X_{\delta,\tilde\mu,\infty}$. Then
  \be
    \label{eq:diffop}
    W\mapsto F[W_0+W]-F[W_0]
  \ee
  is a $(\tilde\mu+\tau,\tilde\nu+\tau, + \infty)$-operator for any exponent scalar $\tau\ge0$.
\end{lemma}
First observe that all our function operators are special.
We stress however that it is a crucial assumption that the quantity $\tau$ is a \textit{scalar}. 
In our application, $\tau$ corresponds to $\hat \nu$ which by definition \eqref{eq:defnu1} can in general obviously not be identified with a scalar  (unless $\nu_1=\nu_2$). We can therefore only apply this lemma directly to operators which do not depend on $\UGcomp{6,-1}$, $\UGcomp{6,0}$ and $\UGcomp{6,1}$ related to the quantity $Q$ (see  \Eqref{eq:blockdiagonalcoords}). In \Sectionref{sec:foestimates} (cf.\ in particular the discussion of \eqref{eq:hsf}) we have seen that the only operators in our equations which depend on $\UGcomp{6,-1}$, $\UGcomp{6,0}$ and $\UGcomp{6,1}$ (and for which the lemma can therefore not be applied directly) are $\FredGVOneTwo{\cdot}$ and $\FGVTwoTwo{\cdot}$. For these two operators we will exploit a useful consequence of \Lemref{lem:hoc}, namely that the difference operator in \Eqref{eq:diffop} is in general at least a $(\tilde\mu+\tau,\tilde\nu+\min_{i\in\{1,\ldots,d\}}\tau_i, + \infty)$-operator if $\tau$ is an exponent \textit{vector}.

Now let us prove \Propref{prop:ModSIVP}. We assume that the data satisfy the hypothesis.
The main task is to apply \Theoremref{th:smoothexistenceN} to our modified singular initial value problem. The matrices $\NNG$ given by \Eqref{eq:NNG} and $\NNF$ given by \Eqref{eq:defNNF}  and the other matrices in the principal part are block diagonal (see the discussion before \Defref{def:bdsystem}) with respect to $\kappaVecC$, $\kappaVecC+\muVecC$ and $\kappaVecC+\muVecC+\nuVecC$ (we will make use of all three) provided $\nuVecC$ has the structure \Eqref{eq:defnu1} and
\be
  \label{eq:bddddd1}
  \eta=0,\quad \muFcomp1=\muFcomp2.
\ee
The eigenvalue condition of \Theoremref{th:smoothexistenceN}  is satisfied if 
\be
  \label{eq:bddddd2}
  \nu_1>\Gamma
\ee
and if we choose an arbitrary sufficiently negative constant $\alpha$ (see \Eqsref{eq:firstordervariablesfirst}--\eqref{eq:firstordervariablesLast}).
Since $\WCT+\WC$ is in  $X_{\delta,\kappaVecC+\muVecC,\infty}$, the principal part matrices of \Eqsref{eq:modeqs1} and \eqref{eq:modeqs2} satisfy the conditions for a special quasilinear symmetric hyperbolic Fuchsian system (\Defref{def:quasilinearlimitN}) provided, in addition to the above, we have
\be
  \label{eq:ccccccond1}
 0<\muFcomp1<\min\{2\Gamma,\muGcomp1\}
\ee
as a consequence of Lemmas~\ref{lem:princpartGN} and \ref{lem:princpartFN}. 

 Next we write down conditions for which the function operators on the right-hand side of \Eqsref{eq:modeqs1} and \eqref{eq:modeqs2}
satisfy the requirements of \Defref{def:quasilinearlimitN}. 
In the following, when we speak of a \textit{rescaled operator} we mean that a given operator has been multiplied with $\RR{\kappaVecG+\muVecG+\nuVecG}$ (for an operator on the right-hand side of Einstein's equations) or with $\RR{\kappaVecF+\muVecF+\nuVecF}$ (for an operator on the right-hand side of Euler's equations), respectively. If any such rescaled operator turns out to be a $(\kappaVecG+\muVecG+\nuVecG,\zeta, + \infty)$-, a $(\kappaVecF+\muVecF+\nuVecF,\zeta, + \infty)$-, or a $(\kappaVecC+\muVecC+\nuVecC,\zeta, + \infty)$-operator, respectively, for some $\zeta$ we say that \textit{its image exponent is $\zeta$}. We recall that a rescaled operator is $o(1)$ if its image exponent is positive. We have:
\begin{description}[labelsep=1ex, align=left,%
  labelwidth=0.08\textwidth,labelindent=0.02\textwidth,%
  font=\normalfont,leftmargin=0.112\textwidth]
\item [$\OGVOneOne\cdot$:] This operator does not depend on $\UGcomp{6,-1}$, $\UGcomp{6,0}$ and $\UGcomp{6,1}$. As a consequence of \Lemref{lem:sourceG1} and \Lemref{lem:hoc}, the rescaled operator is $o(1)$ provided, in addition to the above, we have
  \be
    \label{eq:ccccccond2}
    \nu_1\ge\nu_2,\quad \muGcomp{5},\muGcomp{6}<1,\quad \muGcomp{1}<\min\{\muGcomp4,\muGcomp5\}.
  \ee
\item [$\OGVOneTwo\cdot$:] This \textit{does} depend on $\UGcomp{6,-1}$, $\UGcomp{6,0}$ and $\UGcomp{6,1}$ and therefore the generalized version of \Lemref{lem:hoc} above must be used together with \Lemref{lem:sourceG1}. If, in addition to the above, we assume
  \be
    \label{eq:ccccccond3}
    \muGcomp{5}<2k,\quad \muGcomp{6}<\muGcomp5,
  \ee
  the image exponent of the rescaled operator is
  \begin{align*}
    (&\infty,2 k-{\muGcomp1}+\nu_2-\nu_1,\infty,
       \infty,2 k-{\muGcomp1}+2
       {\muGcomp6}+\nu_2-\nu_1,\infty,\\
     &\infty,2 k-{\muGcomp1}+{\muGcomp6}+\nu_2-\nu_1,\infty,
       \infty,\infty,\infty,\\
     &\infty,2 k-{\muGcomp5}+\nu_2-\nu_1,\infty,
       \infty,\min\{{\muGcomp1},\muGcomp5-\muGcomp6\},{\muGcomp1}).
  \end{align*}
  This is positive and hence the rescaled operator is $o(1)$ if, in addition to the above, we have
  \be
    \label{eq:ccccccond4}
    \nu_1-\nu_2<2k-\muGcomp5.
  \ee  
\item [$\OGVTwoOne\cdot$:] 
  This does \textit{not} depend on $\UGcomp{6,-1}$, $\UGcomp{6,0}$ and $\UGcomp{6,1}$. If if, in addition to the above, we have
  \be
    \label{eq:ccccccond5}
    \muGcomp{5}<2(1-k),
  \ee
  then the image exponent of the rescaled operator is
  \begin{align*}
    (&\infty,\infty,\infty,
       \infty,2 (1-k)-{\muGcomp1},\infty,
       \infty,\infty,\infty,\\
     &\infty,\infty,\infty,
       \infty,2 (1-k)-{\muGcomp5},\infty,
       \infty,1 -2 k+{\muGcomp1}-{\muGcomp6}+\nu_1-\nu_2,\infty ).
  \end{align*}
  This follows from \Lemref{lem:hoc} and \Lemref{lem:sourceG1}.
  This is positive and hence the rescaled operator is $o(1)$ if in addition to the above
  \be
    \label{eq:ccccccond6} 
    1 -2 k+{\muGcomp1}-{\muGcomp6}+\nu_1-\nu_2>0.
  \ee
\item  [$\OGVTwoTwo\cdot$:] 
  This \textit{does} depend on $\UGcomp{6,-1}$, $\UGcomp{6,0}$ and $\UGcomp{6,1}$. If, in addition to the above, we assume
  \be
    \label{eq:ccccccond7}
    \muGcomp{1}=\muGcomp{6},
  \ee
  then the image exponent of the rescaled operator is
  \begin{align*}
    (&\infty,\infty,\infty,
       \infty,1+\nu_2-\nu_1,\infty,
       \infty,1 -{\muGcomp6}+\nu_2-\nu_1,\infty,\\
     &\infty,\infty,\infty,
       \infty,1 -{\muGcomp5}+{\muGcomp6}+\nu_2-\nu_1,\infty,
       \infty,\infty, \infty).
  \end{align*}
  This follows from the generalized version of \Lemref{lem:hoc} and from \Lemref{lem:sourceG1}.
  This is positive and hence the rescaled operator is $o(1)$ if
  \be
    \label{eq:ccccccond8}
    \nu_1-\nu_2<1 -{\muGcomp5}.
  \ee
\item  [$\OGF\cdot$:]  This operator does not depend on $\UGcomp{6,-1}$, $\UGcomp{6,0}$ and $\UGcomp{6,1}$. If, in addition to the above, we assume
  \[\muGcomp{4}<1-\Gamma,\]
  then the rescaled operator is $o(1)$ as a consequence of \Lemref{lem:sourceG2} and \Lemref{lem:hoc}.
\item [$\OF\cdot$:] This operator does not depend on $\UGcomp{6,-1}$, $\UGcomp{6,0}$ and $\UGcomp{6,1}$. The above conditions suffice to show that
  the rescaled operator is $o(1)$ as a consequence of \Lemref{lem:sourceF} and \Lemref{lem:hoc}.
\item [$\OGPZero\cdot$,  $\OGPOne\cdot$, $\OFPZero\cdot$ and $\OFPOne\cdot$:] Here we make use of the fact that $D\WCT\in X_{\delta,\kappaVecC+\muVecC,\infty}$ and $\partial_x\WCT\in X_{\delta,\kappaVecC+\muVecC,\infty}$. All the above conditions then suffice to show that each rescaled operator is $o(1)$ owing to (i)  
the control of the difference operators in the brackets provided by \Lemref{lem:princpartGN}, \Lemref{lem:princpartFN} 
and \Lemref{lem:hoc} together with the fact that the principal part matrices do not depend on $\UGcomp{6,-1}$, $\UGcomp{6,0}$, $\UGcomp{6,1}$, and (ii) the fact that the principal part matrices commute with $\RR{\kappaVecG+\muVecG}$ and $\RR{\kappaVecF+\muVecF}$, respectively.
\item [$\OTG\cdot$ and $\OTF\cdot$:]
These operators are $o(1)$ if,  in addition to the above, we have \[0<\nu_1<1-\max\{\muGcomp1,\muGcomp4,\muGcomp5,\muFcomp1\}\quad\text{and}\quad
  0<\nu_2<1-\muGcomp6.\]
  This follows from \Lemref{lem:princpartGN} and \Lemref{lem:princpartFN} and in particular from \Eqsref{eq:xiG} and  \eqref{eq:xiF}. Moreover we use that $\partial_x\UCT\in X_{\delta,\kappaVecC-\tilde\epsilon,\infty}$ for any\footnote{We require $\tilde\epsilon>0$ to control logarithms$\log t$-terms which arise since $k$ is not constant.} and that the matrices $\PPMatrixGnullcomp1$ and $\PPMatrixFnullcomp1$ commute with $\RR{\kappaVecG}$ and $\RR{\kappaVecF}$, respectively.
\end{description}
The final task is to check that the definitions of the exponents in terms of $\epsilon$ in the hypothesis of \Propref{prop:ModSIVP} are consistent with all of the above inequalities. Since this is the case, this completes the proof.

\paragraph{Step 3. The original mixed second-first order system of evolution equations}
Steps~1 and 2 together yield solutions of the first-order evolution system \Eqref{eq:EinsteinEuleroperatorred} and thereby of \eqref{eq:EinsteinEuleroperator}. Recall that \Eqref{eq:EinsteinEuleroperator} was derived from the original Einstein-Euler evolution equations (\Eqsref{eq:wgEinstEqs}, \eqref{eq:asympwavegauge}, \eqref{eq:choiceconstraintmultiples}, \eqref{eq:AAA1} and \eqref{momentum3}), which is a mixed second-first order system, by introducing the first-order variables \Eqsref{eq:firstordervariablesG1}--\eqref{eq:firstordervariablesLast}.
In \cite{Ames:2016uy} we have discussed in detail under which conditions solutions of the first-order system give rise to  solutions of the original system (the Euler equations are not discussed in \cite{Ames:2016uy}); the same arguments apply here. Under the hypotheses of Propositions~\ref{prop:SIVPtruncated} and \ref{prop:ModSIVP}, in particular,
 if $\alpha$ is sufficiently negative, we can show that the unknown variables are not independent:
\begin{equation*}
\UGcomp{i,0}=D\UGcomp{i,-1}-\alpha\UGcomp{i,-1},\quad
\UGcomp{i,1}=t\partial_x\UGcomp{i,-1},
\end{equation*}
for all $i=1,\ldots,6$. Given this, one can show that
\begin{gather}
  \label{eq:secondordermetric}
  g_{00}=\UGcomp{1,-1},\quad\!\!
  g_{11}=\UGcomp{2,-1},     \quad\!\!
  g_{01}=\UGcomp{3,-1},\quad\!\!
  R=\UGcomp{4,-1},\quad\!\!
  E=\UGcomp{5,-1},\quad\!\!
  Q=Q_*+\UGcomp{6,-1},\\
  \label{eq:secondorderfluid}
  v^0=\UFcomp1,\quad\!\!
  v^1=\UFcomp2, 
\end{gather}
is a solution to the original mixed second-first order system. 

\paragraph{Step 4. Better shift decay} 
So far the results from Step~3, \Propref{prop:SIVPtruncated} and \Propref{prop:ModSIVP}, imply the existence of some $\tau>0$ such that
\be
  \label{eq:weakshiftdecay}
  g_{01}, Dg_{{01}}, \partial_x g_{01}\in X_{\delta, (k^2-1)/2+1-\tau,\infty}.
\ee
In fact, it follows that this holds for \textit{any} $\tau>0$. 
As in the vacuum case \cite{Ames:2016uy}, this knowledge turns out to be insufficient to control the propagation of constraint violations in the next subsection. 
Using the same arguments as in \cite{Ames:2016uy} we find that there exists some (possibly different) $\tau>0$ such that the stronger estimate
\be
  \label{eq:improvedshift}
  g_{01}, Dg_{{01}}, \partial_x g_{01}\in X_{\delta,(k^2-1)/2+1+\tau,\infty}
\ee
holds, provided that, in addition to the hypotheses of \Propref{prop:SIVPtruncated} and \Propref{prop:ModSIVP}, the data satisfy
\be
  \label{eq:diffconstr}
  \frac{\Lambda_*'}{\Lambda_*}=
  -k\frac{ E_*'}{E_*}+2 k E_*^2 Q_{**} Q_{*}' -\frac{2\gamma v^1_*(\Lambda_*)^{-\frac{2-\gamma}{2(\gamma-1)}} (v^0_*)^{\frac{1-2\gamma}{\gamma-1}}}{\gamma -1}
\ee
in the case $\Gamma>0$,
and
\be
\label{eq:diffconstr-critical}
  \frac{\Lambda_*'}{\Lambda_*}=
  -k\frac{ E_*'}{E_*}
  -\frac{2\gamma v^0_* v^1_*(\Lambda_*)^{-\frac{2-\gamma}{2(\gamma-1)}} ((v^0_*)^2-(v^1_*)^2)^{\frac{2-3\gamma}{2(\gamma-1)}}}{\gamma -1}
\ee
in the case $\Gamma=0$.  
Observe that this is an important step to establish statement~\ref{statement:improvedshift} of \Theoremref{th:sub-criticalcoupledEstimates} and of \Theoremref{th:criticalcoupledEstimates}.

\subsection{Solving the constraint equations}
\label{sec:constraints}
The next step is to study the propagation of constraint violations and thereby to derive conditions under which the solutions of the evolution equations constructed in the previous subsection satisfy $\mathcal D_\alpha\equiv 0$; recall \Sectionref{sec:waveformalism}. Since the arguments now are very similar to the ones in \cite{Ames:2016uy} due to the fact that the matter variables do not enter the constraint propagation equations directly, we only give a short summary and point out the major differences.
Let us choose any solution to the evolution equation constructed in the previous subsection. The corresponding constraint violation terms $\mathcal D_\alpha$ can then in principle be calculated from \Eqsref{eq:defD} and \eqref{eq:asympwavegauge}. In general these terms will not be zero but and their non-trivial evolution is described by the subsidiary system \Eqsref{eq:ConstraintPropagationEquation} and  \eqref{eq:choiceconstraintmultiples}; observe that the matter variables do not enter this system.

The techniques in  \cite{Ames:2016uy} establish that the hypotheses of Propositions~\ref{prop:SIVPtruncated} and \ref{prop:ModSIVP}, together with \Eqref{eq:diffconstr} for $\Gamma>0$ (or \Eqref{eq:diffconstr-critical} in the case $\Gamma=0$), suffice to show that
\be
  \label{eq:constraintspaces}
  \mathcal D_0, D\mathcal D_0 \in X_{\delta,-1+\tau,\infty},\quad
  \mathcal D_1, D\mathcal D_1 \in X_{\delta,\tau,\infty}
\ee
for some $\tau>0$ (which is not necessarily the same $\tau$ as in \Eqref{eq:improvedshift}) while $\mathcal D_2\equiv\mathcal D_3\equiv 0$; as before we write $\delta=\tilde\delta$ to simplify the notation. In fact, \Eqref{eq:diffconstr} (or \Eqref{eq:diffconstr-critical}, respectively) is also the condition that guarantees that the constraint violation terms vanish \textit{in leading order} in the limit $t\searrow 0$. The task is now to show that this is sufficient to establish that $\mathcal D_\alpha$ vanish \textit{identically}.

To this end we consider the subsidiary system \Eqsref{eq:ConstraintPropagationEquation} and \eqref{eq:choiceconstraintmultiples}.
Since $\mathcal D_0\equiv \mathcal D_1\equiv\mathcal D_2\equiv\mathcal D_3\equiv 0$ is the trivial solution to this homogeneous system, the task is to show that this trivial solution is the \textit{unique solution} in the spaces given by \Eqref{eq:constraintspaces}  under the hypotheses of \Propref{prop:SIVPtruncated} and \Propref{prop:ModSIVP}  together with \Eqref{eq:diffconstr} for $\Gamma>0$ (or \Eqref{eq:diffconstr-critical} in the case $\Gamma=0$). 

Uniqueness can  in principle be obtained by applying \Theoremref{th:smoothexistenceN} to \Eqref{eq:ConstraintPropagationEquation} for the spaces given by \Eqref{eq:constraintspaces}. 
As in vacuum,  
 however it turns out also that this does {not} work because \Theoremref{th:smoothexistenceN} requires spaces with larger exponents than in \Eqref{eq:constraintspaces}. 
Technically, the obstacle here is the block diagonal condition of  \Theoremref{th:smoothexistenceN} 
which requires that the exponents for all the terms $\mathcal D_0, D \mathcal D_0, \mathcal D_1, D \mathcal D_1$ must all be the same. If we were able to show that $\mathcal D_0$, $D\mathcal D_0$ would rather be in the space $X_{\delta,\tau,\infty}$ (as opposed to $X_{\delta,-1+\tau,\infty}$), the argument would go through. 

The trick to establish this introduced in \cite{Ames:2016uy} is to pick \textit{any} function $\mathcal D_1$ in $X_{\delta,\tau,\infty}$ with $D\mathcal D_1 \in X_{\delta,\tau,\infty}$ and then to consider this quantity as a given quantity in the evolution equations for $\mathcal D_0$ given by \Eqref{eq:ConstraintPropagationEquation}. When we apply the Fuchsian theorem to this subsystem, the block diagonal condition becomes far less restrictive and one can show that it has a unique solution $\mathcal D_0, D\mathcal D_0 \in X_{\delta,\tau,\infty}$.

When we combine this now with the argument above, we indeed establish that the constraint violation terms vanish identically under the conditions above and hence under the hypothesis of Theorems~\ref{th:sub-criticalcoupledExistence} and \ref{th:criticalcoupledExistence}.


The ``asymptotic constraint'' \Eqref{eq:diffconstr} for $\Gamma>0$ (or \Eqref{eq:diffconstr-critical} in the case $\Gamma=0$) has played a crucial role in this argument. 
Finally now we seek conditions for which this differential equation has a smooth solution and hence for which a complete set of smooth consistent asymptotic data exist.
Let us start with the case $\Gamma>0$.
In contrast to the vacuum case \cite{Ames:2016uy}, the data $\Lambda_*$ also appears on the right hand side of \eqref{eq:diffconstr} and hence this equation cannot be integrated directly to determine $\Lambda_*$. However, if we replace the free data $v^1_*$ by another free data $\hat v^1_*$ defined by
\be
  \label{eq:newfreefluiddata}
  v^1_*(x):= \hat v^1_*(x) (\Lambda_*(x))^{\frac{2-\gamma}{2(\gamma-1)}},
\ee
then \Eqref{eq:diffconstr} becomes
\begin{equation*}
  \frac{\Lambda_*'}{\Lambda_*}=
  -k\frac{ E_*'}{E_*}+2 k E_*^2 Q_{**} Q_{*}' -\frac{2\gamma \hat v^1_* (v^0_*)^{\frac{1-2\gamma}{\gamma-1}}}{\gamma -1}.
\end{equation*}
We can now determine $\Lambda_*$ by integration, and the global smoothness condition reduces to 
\begin{equation*}
  0=
  \int_0^{2\pi}\left(-k\frac{ E_*'}{E_*}+2 k E_*^2 Q_{**} Q_{*}' -\frac{2\gamma \hat v^1_* (v^0_*)^{\frac{1-2\gamma}{\gamma-1}}}{\gamma -1}\right)dx.
\end{equation*}
Finally now we can swap the roles of $v^1_*$ and $\hat v^1_*$ and thereby fully establish \Theoremref{th:sub-criticalcoupledExistence}. 

In the case $\Gamma=0$ and $Q_*=\textrm{const}$, the asymptotic constraint takes the form \eqref{eq:diffconstr-critical}. Since $k$ is a constant here, it is now possible to consider the data $v^0_*$, $v^1_*$, and $\Lambda_*$ as free, and to determine $E_*$ by integration of \eqref{eq:diffconstr-critical}. This gives rise to \Eqref{eq:ESconstr}. 

The proofs of \Theoremref{th:sub-criticalcoupledExistence} and \Theoremref{th:criticalcoupledExistence} are now complete. We have also established part~\ref{statement:improvedshift} of  \Theoremref{th:sub-criticalcoupledEstimates} and \Theoremref{th:criticalcoupledEstimates}. We will not say much about part~\ref{statement:singular}, but next focus on parts~\ref{statement:coupledsub2} and \ref{statement:coupledsub3}  in \Sectionref{sec:vtd}.

\subsection{``Velocity term dominance'' and ``matter does not matter''}
\label{sec:vtd}

\paragraph{Velocity term dominance}
Consider any solution $(g_{\alpha\beta}, v^\alpha)$ of the Einstein-Euler equations asserted by \Theoremref{th:sub-criticalcoupledExistence} (again we mainly focus on the case $\Gamma>0$ here). \Propref{prop:ModSIVP} establishes that there exists a solution of the partially truncated evolution equations for which the associated metric given by \Eqref{eq:secondordermetric} agrees with $g_{\alpha\beta}$ at order $(1,1,1,1,1,\min\{1,2(1-k)\})$ 
and the associated fluid vector given by \Eqref{eq:secondorderfluid} agrees with $v^\alpha$ at order $(1,1-\Gamma)$. Part~\ref{statement:coupledsub2} of \Theoremref{th:sub-criticalcoupledEstimates} (and analogously for \Theoremref{th:criticalcoupledEstimates}) is therefore a consequence of the following result.
\begin{proposition}
  \label{prop:VTD}
  Choose  functions $k$, $\Lambda_*$, $E_*$, $Q_*$, $Q_{**}$, $v^0_*$ and $v^1_*$ in $C^\infty(T^1)$ such that $\Lambda_*,E_*,v^0_*>0$ and $1>k>0$, and a constant $\gamma\in (1,2)$. Choose  smooth functions $\muGcomp{i}$, $\muFcomp{i}$ and $\eta$ such that
  \begin{center}
    \begin{tabular}{rcccl}
      $\max\{\Gamma,2k-1\}$ & $<$ & $\eta$ & $<$ & $1$,\\
      $0$ & $<$ & $\muGcomp5$ & $<$ & $\min\{2k,2(1-k),1-\eta\}$,\\
      $0$ & $<$ & $\muGcomp4$ & $<$ & $1-\Gamma$,\\
      $0$ & $<$ & $\muGcomp1=\muGcomp6$ & $<$ & $\min\{\muGcomp4,\muGcomp5\}$,\\
      $0$ & $<$ & $\muFcomp1$ & $<$ & $\min\{\Gamma,\muGcomp1\}$,\\
      $\Gamma$ & $<$ & $\muFcomp2$ & $<$ & $\Gamma+\muFcomp1$.
    \end{tabular}
  \end{center}
  Let $\widehat\UC=\USC+\widehat\WC$ be the solution of the partially truncated equations, \Eqref{eq:EinsteinEulerPartialTruncated}, with $\widehat{\WC}\in X_{\tilde\delta,\kappaVecC+\muVecC,\infty}$ asserted by \Propref{prop:SIVPtruncated} (identifying $\delta$ and $\tilde\delta$) and $\UCT=\USC+\WCT$ be the solution of the fully truncated equations, \Eqref{eq:EinsteinEulerPartialTruncated} where the term $\FredGV{\WG}$ is replaced by $\FredGVOne{\WG}$, with $\WCT\in X_{\tilde\delta,\kappaVecC+\muVecC,\infty}$ (see the remark after \Propref{prop:SIVPtruncated}). Let $g_{\alpha\beta}$ be the metric associated with $\widehat\UC$ and $g_{\truncated,\alpha\beta}$ be the metric associated with $\UCT$ via \Eqref{eq:secondordermetric}. Analogously let $v^\alpha$ be the fluid vector associated with $\widehat\UC$ and $v^\alpha_\truncated$ associated with $\UCT$ via \Eqref{eq:secondorderfluid}. Then the two metrics agree at order $(2-2k,2-2k,2-2k,2-2k,2-2k,2-2k)$ and the two fluid vectors agree at order $(2-2k,2-2k-\Gamma)$.
\end{proposition}

In order to prove this proposition, let us set
$\WC:=\widehat\WC-\WCT$
so that $\UC=\USC+\WCT+\WC$. Observe that $\WCT$ and $\WC$ are different terms than the terms with the same names in Step~2 of \Sectionref{sec:applFuchsian}. Nevertheless, the reason why we choose the same variable names is that they will play exactly the same roles as the corresponding terms before. This is so because we can show that the partially and the fully truncated equations imply evolution equations for $\WC$ which are very similar to  \Eqsref{eq:modeqs1} and \eqref{eq:modeqs2}:
\be
  \label{eq:modeqs11}
  \begin{split}
  &\LG{\WGT+\WG}{\WG}-\PPMatrixFcomp1{\WGT+\WG} t\partial_{x} \WG\\
&=            \OGVOneOne{\WG}+{\OGVOneTwo{\WG}}
+\FGVTwoOne{\WGT+\WG}
+\FGVTwoTwo{\WGT+\WG}\\
  &+{\OGF{\WC}}
-{\OGPZero{\WG}}+\OOGPOne{\WG}
\end{split}
\ee
and
\be
  \label{eq:modeqs21}
  D\WF+\NNFT\WF=\OOF{\WC}  
\ee
with
\begin{align}
  \OOGPOne{\WG}&:=(t\PPMatrixGcomp1 {\WGT+\WG}-t\PPMatrixGcomp1 {\WGT})\partial_{x} \USG\\
  \OOF{\WC}&:=\FredTF{\WCT+\WC}-\FredTF{\WCT}.
\end{align}
We will now solve the singular initial value problem of this system of equations for $W\in X_{\tilde\delta,\kappaVecC+\muVecC+\nuVecC,\infty}$ 
where all data, the exponent vector $\muVecC$ and the exponent $\eta$ satisfy the hypothesis of \Propref{prop:VTD}. Moreover, we assume that $\nuVecC$ is of the form \Eqsref{eq:defnu1} for some so far unspecified $\nu_1,\nu_2>0$. This discussion follows the proof of \Propref{prop:ModSIVP} very closely. The proof of \Propref{prop:VTD} however is simpler because we can rely on the fact (from the proof of \Propref{prop:SIVPtruncated})  that the exponents $\muGcomp{i}$, $\muFcomp{i}$ and $\eta$ satisfy the correct inequalities. Hence we can focus our attention on $\nu_1$ and $\nu_2$. As before we study this singular initial value problem by means of \Theoremref{th:smoothexistenceN}. The block diagonal and eigenvalue conditions of this theorem are satisfied for any $\nu_1,\nu_2>0$ provided the other exponents satisfy the hypothesis of \Propref{prop:VTD}. We remark that we use \Eqref{eq:truncatedsol} for the solutions of both the partially and the fully truncated systems without further notice. 
It remains to establish the following results.

\begin{description}[labelsep=1ex, align=left,%
  labelwidth=0.08\textwidth,labelindent=0.02\textwidth,%
  font=\normalfont,leftmargin=0.112\textwidth]
\item [$\OGVOneOne\cdot$:] This operator does not depend on $\UGcomp{6,-1}$, $\UGcomp{6,0}$ and $\UGcomp{6,1}$. As a consequence of \Lemref{lem:sourceG1} and \Lemref{lem:hoc}, the rescaled operator is $o(1)$ provided
  \be
    \label{eq:fffff1}
    \nu_1\ge\nu_2.
  \ee
\item [$\OGVOneTwo\cdot$:] This \textit{does} depend on $\UGcomp{6,-1}$, $\UGcomp{6,0}$ and $\UGcomp{6,1}$ and therefore the generalized version of \Lemref{lem:hoc} above must be used together with \Lemref{lem:sourceG1}. The image exponent of the rescaled operator is
  \begin{align*}
  (&\infty,2 k-{\muGcomp1}+\nu_2-\nu_1,\infty,
\infty,2 \eta +2 k+{\muGcomp1}+\nu_2-\nu_1,\infty,
\infty,2 k+\nu_2-\nu_1,\infty,\\
&\infty,\infty,\infty,
\infty,2 k-{\muGcomp5}+\nu_2-\nu_1,\infty,
\infty,\min\{2 \eta   +{\muGcomp1},\muGcomp5-\muGcomp6\},{\muGcomp1}).
\end{align*}
  This is positive and hence the rescaled operator is $o(1)$ if
  \be
    \label{eq:fffff2}
    \nu_1-\nu_2<2k-\muGcomp5.
  \ee
\item [$\WG\mapsto\FGVTwoOne{\WGT+\WG}$:] 
  The image exponent of the rescaled operator is
  \begin{align*}
  (&\infty,\infty,\infty,
\infty,2 (1-k)-{\muGcomp1}-\nu_1,\infty,
\infty,\infty,\infty,\\
&\infty,\infty,\infty,
\infty,2 (1-k)-{\muGcomp5}-\nu_1,\infty,
\infty,1+\eta -2 k-\nu_2,\infty ).
\end{align*}
  This follows from \Lemref{lem:sourceG1}.
  This is positive and hence the rescaled operator is $o(1)$ if
  \be
    \label{eq:fffff3} 
    2 (1-k)-{\muGcomp5}>\nu_1,\quad 1 -2 k+\eta>\nu_2.
  \ee
\item [$\WG\mapsto\FGVTwoTwo{\WGT+\WG}$:] 
  The image exponent of the rescaled operator is
  \begin{align*}
  (&\infty,\infty,\infty,
\infty,\eta +1-\nu_1,\infty,
\infty,1-\eta  -{\muGcomp6}-\nu_1,\infty,\\
&\infty,\infty,\infty,
\infty,1+\eta -{\muGcomp5}+{\muGcomp6}-\nu_1,\infty,
\infty,\infty, \infty).
\end{align*}
  This follows from \Lemref{lem:sourceG1}.
  This is positive and hence the rescaled operator is $o(1)$ if
  \be
    \label{eq:fffff4} 
    1+\eta-{\muGcomp5}>\nu_1.
  \ee
\item  [$\OGF\cdot$:]  This operator does not depend on $\UGcomp{6,-1}$, $\UGcomp{6,0}$ and $\UGcomp{6,1}$. The rescaled operator is $o(1)$ as a consequence of \Lemref{lem:sourceG2} and \Lemref{lem:hoc} if $\nu_1\ge\nu_2$.
\item [$\OGPZero\cdot$ and $\OOGPOne\cdot$:] Here we make use of the fact that $D\WCT\in X_{\delta,\kappaVecC+\muVecC,\infty}$ and $\partial_x\USG\in X_{\delta,\kappaVecC-\tilde\epsilon,\infty}$ for any $\tilde\epsilon>0$. All the above conditions then suffice to show that each rescaled operator is $o(1)$ owing to (i)  
the control of the difference operators in the brackets provided by \Lemref{lem:princpartGN}, \Lemref{lem:princpartFN} and \Lemref{lem:hoc} together with the fact that the principal part matrices do not depend on $\UGcomp{6,-1}$, $\UGcomp{6,0}$, $\UGcomp{6,1}$, and (ii) the fact that the principal part matrices commute with $\RR{\kappaVecG}$, $\RR{\kappaVecG+\muVecG}$, and, $\RR{\kappaVecF}$, $\RR{\kappaVecF+\muVecF}$, respectively.
\item [$\OOF\cdot$:] This operator does not depend on $\UGcomp{6,-1}$, $\UGcomp{6,0}$ and $\UGcomp{6,1}$. The 
  rescaled operator is $o(1)$ as a consequence of \Lemref{lem:sourceF} and \Lemref{lem:hoc}.
\end{description}
Hence, we have established that for any choice of data and exponents consistent with the hypothesis of \Propref{prop:VTD},  \Eqsref{eq:modeqs11} and \eqref{eq:modeqs21} (determined by the functions $\widehat\WC$ and $\WCT$) has a unique solution $\WC\in  X_{\tilde\delta,\kappaVecC+\muVecC+\nuVecC,\infty}$ for any choice of $\nu$ consistent with
\Eqsref{eq:fffff1}, \eqref{eq:fffff2}, \eqref{eq:fffff3} and \eqref{eq:fffff4}. First we want to argue that this quantity $\WC$ is indeed the sought function $\widehat\WC-\WCT$ for which we only know so far that it is in $X_{\tilde\delta,\kappaVecC+\muVecC,\infty}$. To this end we first observe by small modifications of the above arguments that  \Eqsref{eq:modeqs11} and \eqref{eq:modeqs21} also have a unique solution $\WC$ in the slightly larger space  $X_{\tilde\delta,\kappaVecC+\muVecC+\nuVecC-\epsilon,\infty}$ for any choice of $\nu$ consistent with
\Eqsref{eq:fffff1}, \eqref{eq:fffff2}, \eqref{eq:fffff3} and \eqref{eq:fffff4} and any sufficiently small $\epsilon>0$. Since \Eqsref{eq:fffff1}--\eqref{eq:fffff4} allow us to pick $\nu_1$ and $\nu_2$ arbitrarily small we can therefore achieve that 
\[\kappaVecC+\muVecC+\nuVecC-\epsilon\le \kappaVecC+\muVecC+\nuVecC.\]
Uniqueness therefore confirms that the uniquely determined solution $\WC$ indeed agrees with $\widehat\WC-\WCT$. In order to establish \Propref{prop:VTD} now we need to check that we can choose $\nu_1$ and $\nu_2$ sufficiently large. Without loss of generality we can now assume specific values for the exponents. In particular we can choose $\muGcomp{5}$ is so small and $\eta$ so close to $1$ (in consistency with the hypothesis of \Propref{prop:VTD}) that \Eqsref{eq:fffff1}, \eqref{eq:fffff2}, \eqref{eq:fffff3} and \eqref{eq:fffff4} allow us to pick $\nu_1$ and $\nu_2$ arbitrarily close to $2-2k$. 
The final step is to use  \Eqsref{eq:secondordermetric} and \eqref{eq:secondorderfluid} and thereby to establish that we have
\[D\UGcomp{i,-1}=\UGcomp{i,0}+\alpha\UGcomp{i,-1}\]
as a consequence of both the partially and the fully truncated equations.

\paragraph{Matter does not matter}
Finally we are concerned with part~\ref{statement:coupledsub3} of \Theoremref{th:sub-criticalcoupledEstimates} (and analogously \Theoremref{th:criticalcoupledEstimates}). In full analogy to our comparison of solutions of the partially and the fully truncated systems with the same data in the previous paragraph, we now compare a solution of the Einstein-Euler evolution equations $\widehat\WC$ with a solution of the vacuum Einstein evolution equations  $\WCV$ determined by the same data.  Let $\WC$ be given by  \Propref{prop:SIVPtruncated} and \Propref{prop:ModSIVP} for some consistent choice of data and exponents. Since the vacuum evolution equations are obtained from the Einstein-Euler evolution equations by deleting the term $\FGV{\WC}$ and by ignoring the Euler equations one can convince oneself easily that the analogous singular initial value problem for this simpler system has a solution $\WCV$ for precisely the same data and the same exponents. 

In analogy to the previous paragraph we write $\widehat\WC=\WCV+W$. The equation for $\WC$ can now be written in the form of \eqref{eq:modeqs1} (or \Eqref{eq:modeqs11}) using the same operator names where we only need to replace $\WCT$ by $\WCV$:
\be
  \label{eq:mdm}
  \begin{split}
  \LG{\WG}{\WG}
&=      {\OGVOneOne{\WG}}+{\OGVOneTwo{\WG}}+{\OGVTwoOne{\WG}}+{\OGVTwoTwo{\WG}}\\
  &-{\OGPZero{\WG}}-{\OGPOne{\WG}}+\FGF{\WCV+\WC}
\end{split}
\ee
This equation is very similar to \Eqref{eq:modeqs1} and we now attempt to analyze it under precisely the same conditions.  In the same way as in the proof of \Propref{prop:ModSIVP}, the conditions given by \Eqsref{eq:bddddd1}, 
\eqref{eq:ccccccond2}, \eqref{eq:ccccccond3}, \eqref{eq:ccccccond4}, \eqref{eq:ccccccond5}, \eqref{eq:ccccccond6}, \eqref{eq:ccccccond7} and \eqref{eq:ccccccond8} must hold also here. 
Since \Eqref{eq:weakshiftdecay} holds for any $\tau>0$ (we do not require \Eqref{eq:diffconstr} or \Eqref{eq:diffconstr-critical} for this) and since $\partial_x E\in X_{\delta,-k-\tau,\infty}$ for any $\tau>0$, \Eqref{eq:badterm} implies that the inequality \eqref{eq:ccccccond6} can be relaxed slightly
\be
  2 -2 k-{\muGcomp6}-\nu_2>0.
\ee
Only the last term in \Eqref{eq:mdm} still has to be analyzed. For that we find the following. If we assume
  \[\muGcomp{4}<1-\Gamma,\quad 0<\muFcomp1<\min\{2\Gamma,\muGcomp1\}\]
  in addition to the above, then the image exponent of the rescaled operator is 
\begin{align*}
    (&\infty,1-\Gamma -{\muGcomp1}-\nu_1,\infty,
\infty,1-\Gamma  -{\muGcomp1}-\nu_1,\infty,\\
&\infty,\min\{1-\Gamma,1 -{\muGcomp1},1-\Gamma
   +{\muFcomp2}-{\muGcomp1}\}-\nu_1,\infty,\\
&\infty,1-\Gamma
   -{\muGcomp4}-\nu_1,\infty,
\infty,\infty,\infty,
\infty,\infty,\infty )
  \end{align*}
as a consequence of \Lemref{lem:sourceG2}. This is positive and hence the rescaled operator is $o(1)$ if in addition to the above
\[\nu_1<1-\Gamma
   -{\muGcomp4}.\]
If we choose the same quantity $\epsilon$ as in \Propref{prop:ModSIVP} and choose the exponents in exactly the same way, our singular initial value problem for $\WC$ has a unique solution provided
\[0<\nu_1<\min\{1-\Gamma,2k+\nu_2\}-3\epsilon,\quad
0<\nu_2<\min\{2(1-k)-2\epsilon,\nu_1\}.\]
We can now finalize the proof of part~\ref{statement:coupledsub3} of \Theoremref{th:sub-criticalcoupledEstimates} with the same arguments as in the proof of \Propref{prop:VTD}.

\small 


\section*{Acknowledgments} 

This paper was initiated in 2013 during a visit of the second author (PLF) at the University of Otago with the financial support of the first author's ``Divisional assistance grant''. This paper was completed during a visit of the first author (FB) to the Universit\'e Pierre et Marie Curie with the support from the Agence Nationale de la Recherche via the grant 06-2--134423. This material is also based upon work supported by the National Science Foundation under Grant No.~0932078 000, while both authors were in residence at the Mathematical Science Research Institute in Berkeley, California. The first author (FB) was also partly funded by a University of Otago Research Grant in 2013. 
The second author also gratefully acknowledges support from the Simons Center for Geometry and Physics (SCGP) at Stony Brook University, where this paper was completed during the one-month Program ``Mathematical Method in General Relativity', organized by M. Anderson, S. Klainerman, P.G. LeFloch, and J. Speck in January 2015.

\end{document}